\documentclass[a4paper,10pt]{article}

\pdfoutput=1
\usepackage{jheppub}
\usepackage{array}
\usepackage{multirow}
\usepackage{amsmath}
\usepackage{makecell}
\newenvironment{rcases}
  {\left.\begin{aligned}}
  {\end{aligned}\right\rbrace}

\definecolor{green}{rgb}{0.1,0.8,0.2}

\makeatletter
\newcommand{\footnoteref}[1]{\protected@xdef\@thefnmark{\ref{#1}}\@footnotemark}
\makeatother

\title{An entropy current and the second law in higher derivative theories of gravity}
 
\author[a]{Sayantani Bhattacharyya}
\affiliation[a]{School of Physical Sciences, National Institute of Science Education and Research, HBNI, Bhubaneswar, Khurda 752050, Odisha, India}

\author[b]{, Prateksh Dhivakar}
\affiliation[b]{Department of Physics, Indian Institute of Technology Kanpur, Kalyanpur, Kanpur 208016, India}

\author[a]{, Anirban Dinda}
 
\author[b]{, Nilay Kundu}

\author[a]{, Milan Patra}

\author[a]{and Shuvayu Roy}

\emailAdd{sayanta@niser.ac.in, prateksh@iitk.ac.in, anirban.dinda@niser.ac.in, nilayhep@iitk.ac.in, milan.patra@niser.ac.in, shuvayu.roy@niser.ac.in}

\abstract{We construct a proof of the second law of thermodynamics in an arbitrary diffeomorphism invariant theory of gravity working within the approximation of linearized dynamical fluctuations around stationary black holes. We achieve this by establishing the existence of an entropy current defined on the horizon of the dynamically perturbed black hole in such theories. By construction, this entropy current has non-negative divergence,  suggestive of a mechanism for the dynamical black hole to approach a final equilibrium configuration via entropy production as well as the spatial flow of it on the null horizon. This enables us to argue for the second law in its strongest possible form, which has a manifest locality at each space-time point. We explicitly check that the form of the entropy current that we construct in this paper exactly matches with previously reported expressions computed considering specific four derivative theories of higher curvature gravity. Using the same set up we also provide an alternative proof of the physical process version of the first law applicable to arbitrary higher derivative theories of gravity.}
 

\begin{document}

\begin{flushright} 
\end{flushright}

\def\Kt{\mathcal{K}}
\def\Kn{\overline{\mathcal{K}}}
\maketitle
\flushbottom
%

\section{Introduction and summary}
\label{sec:intro}
In Einstein's classical theory of gravity, black holes are interesting solutions that behave like thermodynamic objects \cite{Hawking:1971tu, Bardeen:1973gs, Bekenstein:1973ur, Hawking:1974sw} and one can associate notions like temperature and entropy to them. It is believed that there is also a consistent statistical picture of black hole thermodynamics. This allows us to consider black holes as basically a thermodynamic ensemble of micro-states in an underlying quantum theory of gravity. 

Any UV complete theory of gravity, in its low energy limit, will typically generate several higher derivative corrections to the classical two derivative theory of Einstein gravity. Black hole solutions should continue to exist in these theories, at least when the higher derivative coupling is treated perturbatively. The fact that black holes are indeed a collection of a large number of micro-states will remain so even with the higher derivative corrections being added to the two derivative gravity action. Therefore, they should also maintain the laws of thermodynamics provided we could correctly identify the thermodynamic properties with the geometric properties of black holes. This identification is well-known in two-derivative Einstein gravity. However, in higher derivative theories, we still do not have a complete understanding of how to do it for dynamical black holes.

More precisely, we know that in two-derivative theories of gravity, the area of a time-slice of the horizon plays the role of entropy of the black hole that satisfies both the first and the second law of thermodynamics. In particular, the second law follows from the area increase theorem for black holes \cite{Hawking:1973uf, waldbook}. However, when one considers higher derivative theories of gravity, the same horizon area does not work as the definition of black hole entropy. We need to modify it to account for the higher derivative corrections. In \citep{PhysRevD.48.R3427, Iyer:1994ys} a geometrical notion of black hole entropy, called the Wald entropy, was provided for an arbitrary diffeomorphism invariant theory of gravity (including theories with higher derivative corrections to Einstein's theory) in such a way that the first law of thermodynamics is satisfied. However, once a dynamical black hole solution is considered in general higher derivative theories of gravity, Wald entropy's construction becomes ambiguous, meaning a bunch of terms could be added to Wald entropy without affecting the first law \cite{Jacobson:1993xs, Jacobson:1993vj, Jacobson:1995uq}. These are generally known as the JKM ambiguities in the literature, and they are non-zero only for dynamical configurations.

The concept of an equilibrium black hole configuration in a theory of gravity is associated with a space-time metric admitting a Killing vector that becomes null on the event horizon. Such a geometric configuration is also known as the stationary black hole solution. As we know, the first law of thermodynamics is essentially a comparison between the black hole parameters (like mass, charge, etc.) for two slightly different but nearby stationary configurations. The second law of thermodynamics, however, necessarily refers to dynamics. It states that the black hole entropy compared between final and initial stationary configurations (not necessarily nearby) can never decrease, i.e., entropy is always produced in every dynamical process. Although the Wald entropy satisfies the first law by construction, there is no clear proof that it will obey the second law in any arbitrary diffeomorphism invariant theory of gravity. In other words, unlike the two-derivative theory, we still do not have a completely satisfactory geometric construction of entropy that satisfies both the first and the second law of thermodynamics for dynamical black hole solutions in higher derivative theories of gravity. 

Over the years, various attempts have been made to understand if indeed there is a notion of the second law for black hole thermodynamics in higher derivative theories of gravity, \cite{Iyer:1994ys, Jacobson:1993xs, Jacobson:1993vj, Jacobson:1995uq, Sarkar:2013swa, Bhattacharjee:2015yaa, Wall:2015raa, Bhattacharjee:2015qaa, Wall:2011hj}\footnote{See the recent reviews of black hole thermodynamics in higher derivative theories of gravity \cite{Wall:2018ydq, Sarkar:2019xfd} and references therein, for a detailed discussion on this topic.}. A natural approach is to start from Wald entropy as the equilibrium definition of black hole entropy and explore the possibilities of extending it to dynamical situations such that the second law is satisfied. It essentially means fixing the JKM ambiguities related to the Wald entropy for dynamical configurations by imposing the second law as a principle. This has so far been the basic theme of various approaches in this context.

In any such attempt, one would, however, naturally have to devise an algorithm to analyze dynamical black hole space-times, which by itself is very difficult to execute even in general relativity. Therefore we need to adapt a perturbative approach around the exactly known stationary solutions. A standard procedure, which we will also follow in our analysis in this paper, is to consider some stationary black hole space-time which then gets slightly perturbed due to an external source and then finally settles down to another stationary configuration. The amplitude of the dynamical fluctuation is small enough in the sense that we could always perform a perturbative expansion in its amplitude and work with only the linearized term in this expansion, ignoring all non-linear terms. We will call this approximation the linearized amplitude expansion\footnote{One should note that with this linearized amplitude expansion, we cannot study violent time-dependent processes like black hole mergers. Although we would have nothing to say about those situations within this paper's premise, see \cite{Liko:2007vi, Sarkar:2010xp, Chatterjee:2013daa} where a similar problem has been studied with exciting conclusions.}

\paragraph{The context and the backdrop of our current work:} Our analysis in this paper is significantly motivated and based on the important results that were reported recently in \cite{Wall:2015raa} and subsequently in a follow-up paper \cite{Bhattacharya:2019qal}. Therefore, to better understand the context of this current paper, it is useful and informative that we discuss them a priori\footnote{For a detailed discussion the reader is requested to see section $2$ of \cite{Bhattacharya:2019qal} where a comprehensive review of \cite{Wall:2015raa} is given.}. 

Working within the approximation of linearized amplitude expansion of small dynamical perturbation to a stationary black hole configuration, in \cite{Wall:2015raa}, it was shown that these out-of-equilibrium JKM  ambiguities of Wald entropy could uniquely be fixed up to a particular order by demanding that it must satisfy the linearized version of the second law. Additionally, in \cite{Wall:2015raa}, the statement of the second law was satisfied in a more robust sense: not only the difference between the total Wald entropy of the final and the initial equilibrium configuration was non-negative; instead, the entropy was monotonically increasing at every instant of ``time" throughout the evolution from the initial to the final stationary point. This work, thereby, provides an explicit construction of an entropy function - an out-of-equilibrium extension of the Wald entropy - satisfying the second law. 

The general strategy that was followed in \cite{Wall:2015raa} was to show that the ``time" derivative of this entropy function is non-negative at every instant along the evolution\footnote{The reason we are putting ``time" within the quotation mark is the fact that on horizon rather than having a `time' coordinate as the component of a time-like vector, we will have a parameter running along the curve generated by a null vector, the generator of the horizon. The horizon being a null surface, we cannot actually have such a time like coordinate in true sense.}. This way, the basic working principle essentially resembles the same that is responsible for the area increase theorem in Einstein's theory of gravity. After choosing a particular metric gauge suitable for the case at hand and working within the approximation of linearized amplitude expansion, a specific `time-time' component of the rank-two equation of motion was studied. More precisely, a particular sub-class of the residual symmetry transformation of the chosen metric gauge (named as the ``boost transformation") was used to classify how different quantities transform with definite weights (termed as the ``boost-weight") under that symmetry of the stationary background. Eventually, using this classification based on boost-weights, a specific expression of that ``time-time" component of the equation of motion was obtained, maintaining the second law up to linearized order in amplitude. Finally, it was argued that the zero boost-weight part of that expression determines the equilibrium Wald entropy. In contrast, the higher boost sector fixes the JKM ambiguities of the entropy (which is the same as the out-of-equilibrium part of it).

In essence, the importance of the work \cite{Wall:2015raa} lies in the fact that it not only provides an algorithm to fix the JKM ambiguities but also to reproduce the Wald entropy of the stationary configuration itself. As a check of its consistency, this entropy function defined locally in time should reduce to the known expression of Wald entropy for stationary configurations once the dynamics are switched off. However, in \cite{Bhattacharya:2019qal}, a subtle technical issue of this construction regarding its implementation was pointed out. Actually, this constructive algorithm of \cite{Wall:2015raa} stumbles upon a road-block at the very leading order in amplitude expansion. The zero boost-weight sector of the ``time-time" component of the equations of motion doesn't acquire the desired form, which is responsible for reproducing the Wald entropy of the stationary black holes. However, this should not be a problem since, for stationary black holes, we already know that Wald's construction works. In \cite{Wall:2015raa}, this trickier set of terms were dealt with by taking recourse to the physical process version of the first law \cite{Jacobson:1995uq, Gao:2001ut, Amsel:2007mh, Bhattacharjee:2014eea, Chakraborty:2017kob, Chatterjee:2011wj, Kolekar:2012tq}. It was argued that if the physical process version of the first law is to be valid, then things should work out nicely to reproduce the correct expression for the Wald entropy at equilibrium.    

The main goal of the analysis carried out in \cite{Bhattacharya:2019qal} was to perform a brute force calculation focussing on only four derivative theories of gravity to check if the arguments presented in \cite{Wall:2015raa} are indeed true. By explicitly working out the ``time-time" component of the equations of motion for those theories, it was shown that there are terms, in that zero boost-weight sector of it, which cannot be cast in the form required to identify them as the Wald entropy density. However, it was also noticed that the structure of every such anomalous term could be rearranged as the spatial divergence of a specific spatial vector, named the entropy current. In \cite{Wall:2015raa} the appearance of such additional terms in the particular components of the equations of motion with the specific structures mentioned above was not paid its due attention, but the physical process version of the first law was called for the rescue. 

The physical process version of the first law is actually formulated along the same lines as the linearized version of the second law\footnote{It should be noted that although the necessary formalism to address both of them is along the same lines, the implications of them are pretty different. One needs to fix the JKM ambiguities to prove a linearized version of the second law; however, they have vanishing contribution for the first law.}. A stationary black hole is driven out of its equilibrium state by a small stress tensor associated with some matter falling in the black hole from asymptotic infinity, and finally, it settles down to another stationary configuration. A relation between the changes in the parameters (like entropy, mass, etc.) of the two equilibrium black holes follows accordingly. The presence of the additional anomalous terms in the form of a spatial divergence of a spatial current, which was pointed out in \cite{Bhattacharya:2019qal}, didn't affect the statement of the physical process version of the first law. This is because, in the physical process version of the first law, one compares the total entropy integrated over the spatial slices of the horizon, and thereby the entire spatial divergence terms drop out. In turn, this also justifies the arguments presented in \cite{Wall:2015raa} that, once the dynamical evolution of a black hole between two equilibrium configurations satisfies the physical process version of the first law, the zero boost-weight sector of the ``time-time" component of the equations of motion reproduces the correct expression of the Wald entropy for stationary black holes. 

Although in \cite{Wall:2015raa} there was a temporal locality in the version of the second law, that the entropy was considered as an integrated expression over the spatial slices of the horizon (for the reasons mentioned above) signifies that there was no locality in the spatial directions. In other words, all the statements that were made in \cite{Wall:2015raa} involved an entropy function that was local in time but could only be associated to each globally complete spatial section of the horizon but not to any infinitesimal elements of it. On the other hand, the analysis of \cite{Bhattacharya:2019qal} explicitly showed that, although only for specific four derivative theories of gravity, introducing the notion of an entropy current, the second law for black holes could be formulated in its most potent possible form, i.e., in an ultra-local version which is local in both time and space. In fact, this statement has been recast even in a more robust sense as follows: in a general four derivative theory of gravity, working up to linear order in the amplitude of the dynamical fluctuations, the spatial components of the entropy current has to be accounted for if one aims to prove an ultra-local version of the second law (as described above).

The entropy current, constructed in \cite{Bhattacharya:2019qal}, is a $(d-1)$-vector defined on the horizon, with coordinates running along with the `time' and the spatial coordinates that span the horizon. The `time' component of the entropy current gives us the instantaneous entropy density, whereas the spatial components of the same account for a flow of entropy across different local segments of the spatial slice of the horizon. The existence of a non-negative divergence of this entropy current is definitely the statement that entropy is being produced locally at every point of the dynamical space-time. However, due to the non-zero spatial components of the entropy current, it also suggests that there is another simultaneous mechanism by which redistribution of entropy is taking place between different adjacent spatial regions\footnote{The existence of such an entropy current is well known in theories where there is no dynamical gravity, e.g., in fluid dynamics \cite{Bhattacharyya:2008xc, Bhattacharyya:2012nq, Bhattacharyya:2013lha, Bhattacharyya:2014bha}.}. This physical picture highlighting the significance of having an entropy current or the idea of an ultra-locality for the second law is undoubtedly consistent in Einstein's theory of gravity. It is well known that the area increase theorem in general relativity is a local statement that works for every instance of time as well as at each spatial point of a time slice of the horizon. It is therefore not at all absurd to expect that a similar physical picture of the same kind would prevail even when higher derivative theories of gravity are considered, at least for dynamical fluctuations with small enough amplitudes. Actually, the results of \cite{Bhattacharya:2019qal} very strongly favour such an interpretation, at least for the theories explicitly studied there.

It is, therefore, of utmost importance to know whether the lessons that we have learned so far in the discussions above are applicable much more generally and to other generic theories of gravity as well. In this present work, our primary goal is to address this question. Rather than focussing on specific theories as was done in \cite{Bhattacharya:2019qal}, here in this work, we will be considering a general diffeomorphism invariant theory of gravity. However, our methodology will still be the same. For example, we will extensively use the same boost-symmetry of the horizon geometry to constrain the possible structural form of the ``time-time" component of the equations of motion.

Additionally, we will also restrict ourselves only to those dynamical situations where the approximation of a small amplitude of dynamical perturbation is good enough. In other words, a linearized expansion in the amplitude of the non-stationary fluctuations is allowed. Interestingly, our final result turns out to be an affirmative answer to the question we posed above: in an arbitrary diffeomorphism invariant theory of gravity, construction of an entropy current to account for the ultra-local version of the second law of thermodynamics, at least for dynamical processes with small amplitudes, is indeed possible. 

It is worth highlighting that one crucial aspect of the construction of an entropy current in this paper has been to prove a linearized version of the second law without ever invoking the physical process version of the first law. It has been instrumental in enabling us to maintain strict locality throughout our work. As we have mentioned before, this is in contrast to \cite{Wall:2015raa}, where the physical process version of the first law was used to take care of the zero boost-weight sector in specific components of equations of motion, and hence the spatial locality was lost. On the other hand, we have been able to design an independent proof of the first law's physical process version, using the same basic formalism that was formulated to prove a linearized version of the second law. 

Before we proceed, let us also mention that in the literature, the concept of an entropy current in the context of dynamical black holes has already been introduced; see \cite{Guedens:2011dy, Chapman:2012my, Eling:2012xa, Dandekar:2019hyc, Saha:2020zho}. Although they share the common primary goal in terms of constructing a local entropy current, the methodology and the working principles that we are using in our paper are very different from theirs. For example, in \cite{Guedens:2011dy} the context was to interpret the field equations of gravity as an equation of state, the primary motivation behind  \cite{Chapman:2012my, Eling:2012xa} has mainly been the celebrated fluid gravity correspondence \cite{Bhattacharyya:2008jc}. On the other hand, a certain duality between membrane dynamics and black holes called the membrane-gravity duality had been the driving principle for \cite{Dandekar:2019hyc, Saha:2020zho}.

The outline of this paper is as follows. We start in \S\ref{statementofproblem} with making a precise statement of the problem at hand and a summary of the main result of our paper. In \S\ref{sec:setup} we will then discuss the basic set up of our construction. This will include an extensive discussion of various basic concepts which play an important role in our analysis. After introducing the coordinate system that we will work with, the idea of a boost transformation and the weights of different quantities under this specific symmetry of the horizon geometry in our metric gauge will be explained. We will also discuss how these are related to the concepts of stationarity and a slight deviation from it. 

Furthermore, an important identity relating the equations of motion and the Noether charges under diffeomorphism invariance will be established. We will end this section with a discussion of the strategy that will be followed. In the next section, i.e., \S\ref{sec:proof}, we will present detailed proof of how to extract out the entropy current. This will be the central technical part of our paper. In the following section \S\ref{sec:check_proof}, we will show that the expression of the entropy current obtained from the abstract proof in this paper exactly reproduces the results known in four derivative theories of gravity which were reported in \cite{Bhattacharya:2019qal}. Next, in \S\ref{sec:PPVFL_proof} we use the same technical set up and present an independent proof of the physical process version of the first law for any arbitrary diffeomorphism invariant theory of gravity. Finally, in \S\ref{sec:conclusion}, we conclude with a brief discussion of the consequences of our results and possible future directions. We have also summarised the notations, conventions, and useful definitions in Appendix-(\ref{app:notations}) for the convenience of the reader. The other appendices provide useful technical details for our computations. 

\subsection{Statement of the problem and summary of the final result} \label{statementofproblem}
With the motivations and the context of our paper being discussed so far, in this sub-section, we will write down a more precise and explicit statement of the problem at hand as well as the final result that we have been able to achieve. This will help us organize our presentation in the following sections in a much more coherent manner. 

We will work with the most general diffeomorphism invariant theory of gravity in $d$ space-time dimensions with coordinates denoted by $x^\mu$, with the action given by\footnote{We are using a convention such that the Newton's constant $G = 1/4$, such that the area of the horizon gives the entropy of a black hole in Einstein's gravity.} 
\begin{equation} \label{action}
I = {1 \over 4\pi} \int d^d x \, \sqrt{-g} \, \left( L_{grav} + L_{mat}\right) \, .
\end{equation}
In eq.\eqref{action}, $L_{grav}$ is the Lagrangian for the degrees of freedom corresponding to the gravity part\footnote{One can argue that, see section $2$ of \cite{Iyer:1994ys}, any diffeomorphism invariant Lagrangian will have the specific functional dependence as mentioned here.},
\begin{equation} \label{diff_inv_L}
L_{grav} = L_{grav}(g_{\alpha\beta}, \, R_{\alpha\beta\rho\sigma}, \, D_{\alpha_1}R_{\alpha\beta\rho\sigma}, \cdots) \, .
\end{equation}
Moreover, $L_{mat}$ is the Lagrangian for the matter fields present in the theory\footnote{All possible couplings between the matter and the gravity sector are contained within the Lagrangian $L_{mat}$. It is also possible that $L_{mat}$ has to contain higher spin fields to make the entire higher derivative theory consistent with causality concerning processes such as graviton scattering, as was pointed out in \cite{Camanho:2014apa}.}. In this paper, we will be focussing on the gravity part of the Lagrangian $L_{grav}$. For the matter sector, all that we will need is to consider that it gives us a stress tensor satisfying the null energy condition. In what follows, we will make this more precise. 

We will first make the following choice of gauge for the space-time metric with coordinates $x^\mu =\{r,\, v, \, x^i \}$, for $i= 1, \, \cdots, \, (d-2)$, that represents a dynamical black hole solution with a regular event horizon at $r=0$,
\begin{equation}\label{met0}
\begin{split}
ds^2 = 2 \,  dv \, dr - g_{vv}(r,v,x)~dv^2 + 2 g_{vi}(r,v,x) \, dv \, dx^i + h_{ij}(r,v,x) \, dx^i \, dx^j \, , \\ \text{such that}~ g_{rv} =1\, , ~g_{rr} =0\, , ~g_{ri} =0\, , ~ g_{vv} \big\vert_{r=0} = 0 \, ,~ \partial_r g_{vv} \big\vert_{r=0} = 0 \, ,~ g_{vi} \big\vert_{r=0} = 0 \, . 
\end{split}
\end{equation}
The null hyper-surface of the event horizon, which is spanned by the coordinates $\{v, \, x^i \}$, will be denoted by $\cal H$. The constant $v$-slices of the horizon will be represented by ${\cal H}_v$. The coordinate $v$ is an affine parameter corresponding to the null generator  $\chi^\mu \partial_\mu \equiv \partial_v$ of the horizon\footnote{For the reasons mentioned in footnote-$4$, we have been loosely calling the parameter $v$ as the ``time" coordinate, although it is actually a null direction.}. In \S\ref{sec:coord} we will explain in more detail the choice of our coordinates. 

This paper will only focus on scenarios where the dynamics can be treated as small fluctuations around a stationary background. In operational terms, the full space-time metric can be decomposed as follows 
\begin{equation}
g_{\mu\nu} = g_{\mu\nu}^{eq} + \epsilon \, \delta g_{\mu\nu} \, ,
\end{equation}
where $g_{\mu\nu}^{eq}$ is the metric for the stationary black hole space-time, and $\epsilon$ is a parameter denoting the amplitude of small fluctuations around that stationary background. The smallness of the fluctuation allows us to perform a perturbative expansion in the amplitude and work only up to linear order in that amplitude expansion, i.e., neglecting terms with $\mathcal{O}(\epsilon^2)$ and higher orders. 

As we have emphasized before, in this current work, the basic idea behind constructing a proof of the linearized version of the second law is to constrain the structure of a particular component of the equations of motion, namely the $vv$-component of it denoted by ${\cal E}_{vv}$. It should be noted that by ${\cal E}_{\mu\nu}$ we are denoting the equations of motion which follow from varying the complete Lagrangian, including both the gravitational and the matter sector, with respect to the space-time metric. The part of ${\cal E}_{\mu\nu}$ that is derived only from the gravitational part of the Lagrangian will be denoted by $E_{\mu\nu}$. Therefore, we will have the following relation 
\begin{equation} \label{EgravTvv}
{\cal E}_{\mu\nu}=E_{\mu\nu}+T_{\mu\nu} \, ,
\end{equation}
where $T_{\mu\nu}$ is the stress-energy tensor obtained from the matter part of the Lagrangian \footnote{We are working with the following definition of the stress tensor. While obtaining the equations of motion ${\cal E}_{\mu\nu}$ by varying the total Lagrangian with respect to the fluctuations in the metric, there will be one type of terms involving only the metric components and their derivatives. We denote them as $E_{\mu\nu}$. However, in ${\cal E}_{\mu\nu}$, there will be another type of terms that involve both the metric and the matter fields. They will all be included in $T_{\mu\nu}$.}. In \S\ref{Evv_2nd_law} we have discussed in detail why the structure of this particular component of the equations of motion is so crucial to validate the second law\footnote{Remembering that we are writing the components of equations of motion, we will always have ${\cal E}_{vv}=0$ when thinking of it in an on-shell manner. However, this is not how we would consider ${\cal E}_{vv}$; rather, we would like to emphasize that our purpose here is to constrain the off-shell structure of ${\cal E}_{vv}$.}. 

The primary content of our analysis in this paper can be precisely summarised as the following statement:

\textit{ Using arguments based on diffeomorphism invariance and a certain boost symmetry of the horizon geometry to classify possible structures that can appear in ${\cal E}_{vv}$, when evaluated on the horizon (i.e. at $r=0$), we have been able to show that
\begin{equation}\label{eq:main1a}
   \begin{split}
   {\cal E}_{vv} \big\vert_{r=0} = &\, \partial_v \bigg[{1\over\sqrt h}\partial_v \left({\sqrt h} \, {\cal J}^v\right) +{1\over\sqrt h}\partial_i \left({\sqrt h} \, {\cal J}^i\right)\bigg] + T_{vv} + \mathcal{O} (\epsilon^2)
   \\= & \, \partial_v \bigg[{1\over\sqrt h}\partial_v \left({\sqrt h} \, {\cal J}^v\right) +\nabla_i{\cal J}^i\bigg] + T_{vv} + \mathcal{O} (\epsilon^2) \, ,
   \end{split}
   \end{equation} 
   where $h$ is the determinant of the induced spatial metric, i.e. $h_{ij}$, on the co-dimension two constant $v$-slices of the horizon, and $\nabla_i$ is the covariant derivative compatible with $h_{ij}$. Also, ${\cal J}^v$ and ${\cal J}^i$ will get contribution only from the gravitational part of the Lagrangian and eq.\eqref{eq:main1a} can equivalently be stated as 
   \begin{equation}\label{eq:main1}
   \begin{split}
   E_{vv} \big\vert_{r=0} = &\,  \partial_v \bigg[{1\over\sqrt h}\partial_v \left({\sqrt h} \, {\cal J}^v\right) +\nabla_i{\cal J}^i\bigg] + \mathcal{O} (\epsilon^2) \, ,
   \end{split}
   \end{equation} 
which follows from eq.\eqref{EgravTvv}.}

Furthermore, we will only focus on situations where the dynamics is initiated due to some perturbation coming from a matter stress tensor such that it satisfies the null energy condition. In our metric gauge this becomes $T_{vv} \geq 0$. Using this and also the fact that for on-shell processes the RHS of eq.\eqref{eq:main1a} should vanish, we immediately obtain
\begin{equation}\label{del_v_entcur}
\partial_v \bigg[{1\over\sqrt h}\partial_v \left({\sqrt h}~{\cal J}^v\right) +\nabla_i{\cal J}^i\bigg]\leq 0 \, .
\end{equation} 
Additionally, since we also require that the dynamics settles down to some stationary black hole configuration at late future as $v\rightarrow\infty$,  where $\partial_v$ becomes proportional to the Killing direction, we get\footnote{The reason for this is the following: by construction ${\cal J}^i$ includes at least one $\partial_v$-derivative and when evaluated on a stationary metric all the $\partial_v$'s should vanish.} 
\begin{equation}\label{eqlm_v_infty}
\bigg[{1\over\sqrt h}\partial_v \left({\sqrt h}~{\cal J}^v\right) +\nabla_i{\cal J}^i\bigg]\rightarrow 0~~\text{ as}~~ v\rightarrow\infty \, .
\end{equation}
Therefore, from eq.\eqref{del_v_entcur} and eq.\eqref{eqlm_v_infty} it follows that 
\begin{equation} \label{cond3}
\bigg[{1\over\sqrt h}\partial_v \left({\sqrt h}~{\cal J}^v\right) +\nabla_i{\cal J}^i\bigg]\geq 0~~\text{for all finite $v$}\, .
\end{equation}

From eq.\eqref{cond3}, we can also conclude that our procedure of arriving at eq.\eqref{eq:main1} enables us to construct a local entropy current on the horizon with components ${\cal J}^v$ and ${\cal J}^i$. Consequently, we can immediately identify its $v$-component given by ${\cal J}^v$ as the local entropy density, which not only reproduces the equilibrium expression of Wald entropy but also determines the out-of-equilibrium extension of it in the form of fixing the JKM ambiguities. On the other hand, the spatial components, i.e., ${\cal J}^i$, are identified with the entropy current density signifying a possible spatial flow of entropy locally at each point of the constant $v$-slices of the horizon. Also, we can see that with these identifications, eq.\eqref{cond3} signifies a local statement of entropy production at each space-time point on the horizon. Interestingly, we have also been able to show that when the stationary limit is taken by switching off the dynamical perturbations, ${\cal J}^v$ reduces to the expression of Wald entropy known for equilibrium black holes and ${\cal J}^i$ identically vanishes\footnote{This actually justifies eq.\eqref{eqlm_v_infty} above}. It is worth highlighting that obtaining  eq.\eqref{eq:main1}, therefore, not only justifies a consistent out-of-equilibrium extension of the Wald entropy but also provides us an algorithm to obtain ${\cal J}^v$ and ${\cal J}^i$ as entirely geometric quantities in the sense that they are constructed solely out of the metric components and their derivatives. 
However, an important point worth emphasizing is the fact that the choice of our gauge for the space-time metric, eq.\eqref{met0}, plays a vital role in the definitions of ${\cal J}^v$ and ${\cal J}^i$ via eq.\eqref{eq:main1a} and eq.\eqref{eq:main1}. Consequently, we must admit that our construction of the local entropy current is not a covariant one, and it relies on the choice of the constant $v$-slices of the horizon. 

Finally, we conclude our introduction with the following comments regarding the novelty of the results reported in this paper. The key result of this paper is to prove that eq.\eqref{eq:main1} is valid for arbitrary diffeomorphism invariant theories of gravity, up to linearized order in the amplitude of the non-stationary perturbations. However, we must note that the rest of the arguments connected to the linearized version of the second law have been used before, e.g., see \cite{Wall:2015raa}. Even in the case of two derivative Einstein gravity, the area increase theorem for the second law of black hole mechanics is very similar. On the other hand, a case by case and explicit computation of $E_{vv}$ in \cite{Bhattacharya:2019qal} has already revealed the appearance of the terms involving ${\cal J}^i$ in eq.\eqref{eq:main1}, but only for four derivative theories of gravity. In this context, the originality of our results, therefore, lies in the fact that we have been able to justify the ultra-local version of the second law via the entropy current on the horizon with an analysis that is based on general principles like Noether charge for diffeomorphism invariance and most importantly it makes a statement applicable to any higher derivative theory of gravity.

\section{Basic setup and key conceptual elements}\label{sec:setup}
In this section, we will explain the basic setup of our problem. We will introduce a coordinate system adapted to the horizon. The choice of this coordinate system is akin to choosing a gauge for the metric, and we will work with this throughout this paper. Subsequently, we will also elaborate on the approximations that will be used. A specific symmetry, called the boost-invariance, will be introduced. By knowing how different terms transform under this boost transformation, we will see how any generic tensor quantity, built out of metric functions and their derivatives, in a theory of gravity can be constrained. After we have discussed various elements of the basic setup \footnote{We shall be very brief in explaining our setup here. For details, we refer the readers to \citep{Wall:2015raa}, \citep{Bhattacharyya:2016xfs}, \citep{Bhattacharya:2019qal}.}, we will also describe the main strategy, which will be followed in the subsequent section to prove the existence of an entropy current for a general diffeomorphism invariant theory of gravity.
    
\subsection{The coordinate system adapted to the horizon} \label{sec:coord}
We are considering a classical theory of gravity, and black holes are specific solutions described by a space-time that is endowed with a horizon - a null hyper-surface separating the black hole singularity from the asymptotic infinity. Our goal is to choose a coordinate system to describe a space-time with dynamical black holes in this subsection. We will work in $d$-dimensional space-times, and let us suppose that on the horizon, a co-dimension one hyper-surface, $\partial_v$ is the null generator with $v$ being an affine parameter. We will choose $v$ to be one of the coordinates along the horizon. Let $\{x^i\}$, for $i = 1 ,2,\cdots, (d-2),$ be the other $(d-2)$ spatial coordinates along the horizon. Thus, $\{v, x^i\}$ together construct a coordinate system on the horizon. Next, we consider null geodesics generated by $\partial_r$, emanating out of the horizon. We further demand that these geodesics, generated by $\partial_r$, make an angle such that, on the horizon, the inner product of $\partial_r$ and $\partial_v$ is one. Additionally, the inner product of $\partial_r$ and $\partial_i$ vanishes for every $i$. The origin of the $r$ coordinate is chosen to lie on the horizon, i.e., $r=0$ will always give the horizon's location. Finally, we also restrict the parameter $r$ to be an affine parameter along these geodesics, denoting the coordinate away from the horizon. The full $d$ dimensional coordinates will be denoted by $x^\mu \, = \, \{r,v,x^i\}$. In this coordinate system, the metric takes the following form
\begin{equation}\label{met1}
\begin{split}
ds^2 = 2 \,  dv \, dr - r^2 \, X(r,v,x)~dv^2 + 2 r \, \omega_i(r,v,x) \, dv \, dx^i + h_{ij}(r,v,x) \, dx^i \, dx^j \, . 
\end{split}
\end{equation}
The detailed arguments justifying that the most general form of the metric describing a black hole space-time can always be written as in eq.\eqref{met1} are given in Appendix-A of \citep{Bhattacharyya:2016xfs} \footnote{The particular form of the metric in which we have extracted out an explicit factor of $r^2$ in $g_{vv}$ and a factor of $r$ in $g_{vi}$ components of the metric has previously been used in \cite{Bhattacharya:2019qal}.}. It is interesting to note that the metric written in eq.\eqref{met1} should not be considered as the global coordinates. However, the space-time sufficiently close to the horizon will always obtain the form mentioned above. It can be explicitly checked that for Schwarzschild black holes the full space-time can be written in the form given in eq.\eqref{met1}. On the other hand, for rotating Kerr black holes the near-horizon metric can be written in the form of eq.\eqref{met1} but these coordinates cannot be used as global coordinates. For the arguments used in this paper it is sufficient that the local patch near the horizon can always be described by the metric in eq.\eqref{met1}. 

However, one should note that the metric structure given in eq.\eqref{met1} does not completely fix the coordinates on constant $r$ and $v$ slices. The following two types of coordinate redefinitions are still allowed, which preserve the gauge choice in eq.\eqref{met1}:
\begin{enumerate}
\item We are allowed to do a coordinate transformation as  
\begin{equation} \label{residual1}
v \rightarrow \tilde{v} = f_1 (x^i)\,  v + f_2(x^i), 
\end{equation}
along with an appropriate redefinition of the coordinate $r$ such that the form of the metric in eq.\eqref{met1} remains invariant. Here $f_1$ and $f_2$ are arbitrary functions of the $x^i$ coordinates. This coordinate transform essentially redefines the constant $v$-slicing of the horizon and hence, will be important in determining the form of the entropy current. We will actually consider a special sub-class of this residual freedom of coordinate transformation in the following sub-section extensively for our analysis in this paper. 
\item We can also perform a relabelling of the null generators of the horizon by 
\begin{equation} \label{residual2}
x^i \rightarrow \tilde{x}^i = g^i(x^j) \, ,
\end{equation}
which also keeps our metric gauge in eq.\eqref{met1} intact. Additionally, this does not change the constant $v$-slicing of the horizon. It rather mixes only the $x^i$ coordinates among themselves. Therefore, it represents the diffeomorphism invariance for the metric component $h_{ij}$, the induced metric on a co-dimension two constant $r$ and $v$ hyper-surface. This would, for example, enable us to convert all partial derivatives with respect to $x_i$ coordinates to covariant derivatives $\nabla_i$ compatible with the metric $h_{ij}$. 
\end{enumerate}

We are interested in obtaining an expression for the entropy density and entropy current density, both of which are defined on the horizon (i.e., on the $r=0$ hyper-surface). Hence, the entropy density and entropy current should be a scalar and a vector, respectively, under the diffeomorphism that mixes only the $x^i$ coordinates among themselves as in eq.\eqref{residual2}. Also, in our chosen gauge eq.\eqref{met1}, both  are constructed out of the metric functions $X$, $\omega_i$, $h_{ij}$ and their derivatives. Therefore, in our attempt to construct an entropy current for a diffeomorphism invariant theory of gravity, we identify the following quantities as our basic building blocks listed in Table-\ref{table:basicblock}.
\begin{table} [h!] 
\centering 
 \begin{tabular}{| c | c|} 
 \hline
 1. & \makecell{ The metric coefficients: a scalar function $X(r,v,x)$, a vector  \\ function $\omega_i(r,v,x)$ and a tensor function $h_{ij}(r,v,x)$.}\\ 
 \hline
 2. & \makecell{Three differential operators: $\{ \partial_r$, $\partial_v$, $\nabla_i \}$ acting on  \\ $X(r,v,x), \, \omega_i(r,v,x)$ and $ h_{ij}(r,v,x) $.}\\
 \hline
\end{tabular}
\caption{The basic building blocks}
\label{table:basicblock}
\end{table}
One should note that in the list given in Table-\ref{table:basicblock}, the scalar, vector, and tensor properties of different quantities are determined with respect to the spatial diffeomorphism that mixes only the $x^i$ coordinates among themselves.
   
\subsection{Stationarity and small deviation from it by dynamical fluctuations} \label{sec:stationarity}
As we have mentioned before, our analysis in this paper is limited to the perturbative approximation of working only up to linear order in the amplitude of fluctuations around a stationary black hole configuration. In other words, we should be able to decompose the full metric as a sum of a stationary metric and a dynamical fluctuation with a small amplitude around it \footnote{We only consider fluctuations that preserve our metric gauge eq.(\eqref{met1})}. Working with the horizon adapted coordinates $\{r,\, v,\, x\}$ described above in \S\ref{sec:coord}, in this subsection, our aim is to make these notions more precise, namely what it means for space-time to be stationary and also what a non-stationary dynamical perturbation is. 
   
It is well known that for a stationary black hole, we can always find one Killing vector such that it is null on the event horizon, a co-dimension one null hyper-surface. In other words, the event horizon becomes a Killing horizon for that Killing vector\footnote{However, for a static black hole, we have a stronger requirement, i.e., the Killing vector, which becomes null on the event horizon, is the time translations at the asymptotic infinity.}. With the knowledge of this, we can learn about what specific restrictions does the requirement of stationarity imposes on the space-time metric given in eq.\eqref{met1}. 

As was discussed before, we consider a special sub-class of the residual coordinate transformation freedom mentioned in eq.\eqref{residual1}  with $f_1(x^i) = \lambda$, an arbitrary constant parameter and $f_2(x^i) = 0$. To be more precise, we get the following rescaling of the $v$ and $r$ coordinate 
 \begin{equation} \label{boosttransf}
 r\rightarrow\tilde r = \lambda ~r,~~v\rightarrow\tilde v = {v\over \lambda} \, ,
 \end{equation}
 with $\lambda$ being some arbitrary constant. Following \citep{Wall:2015raa} and \citep{Bhattacharya:2019qal}, we will call this scaling transformation as the boost transformation in the rest of this paper. The infinitesimal version of this transformation is generated by the following vector \footnote{The norm of $\xi$ is given by
$\xi^\mu \xi^\nu g_{\mu\nu} = -2rv - r^2v^2 X(r, v, x)$.
So,  $\xi$  becomes null at the horizon located at $r=0$, where it is proportional to the null generator ($\partial_v$) of the horizon.}
\begin{equation} \label{BT_generator}
\xi = \xi^\mu\partial_\mu = v\partial_v - r\partial_r \, .
\end{equation}
  Under this coordinate transformation, the metric eq.\eqref{met1} remains almost invariant apart from the scaling of the arguments ($r$ and $v$) of the metric functions $X$, $\omega_i$ and $h_{ij}$, 
\begin{equation}
ds^2 = 2 d\tilde v~d\tilde r - \tilde r^2 \tilde X(\tilde r,\tilde v,x)~d\tilde v^2 + 2\tilde r~\tilde \omega_i(\tilde r,\tilde v,x)~d\tilde v ~dx^i + \tilde h_{ij}(\tilde r,\tilde v,x) dx^i~dx^j \, ,
\end{equation} 
  where $\tilde X(\tilde r,\tilde v,x) = X\left({\tilde r\over \lambda} , \lambda\tilde v, x\right), ~\tilde \omega_i(\tilde r,\tilde v,x)=\omega_i\left({\tilde r\over \lambda} , \lambda\tilde v, x\right),~\tilde h_{ij}(\tilde r,\tilde v,x) = h_{ij}\left({\tilde r\over \lambda} , \lambda\tilde v, x\right)$.
It is then obvious that the metric remains invariant if all the metric functions $X$, $\omega_i$ and $h_{ij}$ depend on the coordinates $r$ and $v$ only through their product $rv$.  More precisely, when evaluated on the horizon, the vector field $\xi$, which reduces to the null generator of the horizon, will be a Killing vector for any space-time with the following metric
\begin{equation}\label{metequl} 
ds^2 = 2 \, dv \, dr - r^2 \, X(rv,x)\, dv^2 + 2 \, r \,\omega_i(rv,x)\, dv \, dx^i + h_{ij}(rv,x) \, dx^i \, dx^j \, .
\end{equation}
Therefore, we learn that in our horizon adapted coordinates $\{ r, \, v, \, x^i\}$ any metric of the form given above in eq.\eqref{metequl} describes a stationary space-time\footnote{It might seem strange to see that the stationary metric in eq.\eqref{metequl} has $v$ dependent components. As explained in detail in Appendix-A of \citep{Bhattacharyya:2016xfs}, we should remember that $v$ is an affine parameter along $\partial_v$, the null generator of the horizon. However, the metric functions for a stationary black hole should be independent of the Killing coordinate. The latter is generated along the Killing vector $(v\partial_v - r\partial_r)$. This is not exactly equal to, but rather proportional to the $\partial_v$, even at the horizon $r=0$.}. In \citep{Bhattacharya:2019qal} (see Appendix-A in it), it was shown that up to possible coordinate transformations, this is the most general form of stationary space-time with a Killing horizon.

Having understood the most general form of any stationary metric, our next job will be to quantify the departure from stationarity due to a dynamical perturbation with small amplitude. We remember that our basic  building blocks are the metric functions $X$, $\omega_i$ and $h_{ij}$ and their $\partial_r$, $\partial_v$ and $\nabla_i$ derivatives given in Table-\ref{table:basicblock}. Let us now consider a generic covariant tensor with an expression of the form
\begin{equation} \label{eqlmTens}
{\cal A} \sim (\partial_r)^{m_r}(\partial_v)^{m_v}{\cal B} \, ,
\end{equation}
 where ${\cal B}$ collectively denotes the metric coefficients: $X, \, \omega_i, \, h_{ij} $ or any other covariant tensor constructed out of them using the action of $\nabla_i$ only \footnote{It should be noted that the possible appearances of $\partial_r$ and $\partial_v$ in $\cal A$ are explicitly shown in eq.\eqref{eqlmTens}, i.e. ${\cal B}$ does not contain any factors involving $\partial_r$ or $\partial_v$ acting on $X, \, \omega_i, \, h_{ij} $. To avoid clutter of indices, we have also suppressed the components of the tensors ${\cal A}$ and ${\cal B}$.}. If we evaluate ${\cal B}$ for any equilibrium or stationary space-time configuration given by the metric eq.\eqref{metequl}, it will be a function of the product $rv$ and $x$, i.e. 
${\cal B} (r,v,x)\, \vert_{equil.} \sim {\cal B} (rv,x)$. 
Therefore, it can be argued that whenever $m_v>m_r$, the expression of ${\cal A}$ given in eq.\eqref{eqlmTens} will vanish on the horizon ($r=0$) as it will have $(m_v - m_r)$ factors of $r$. However, this will not be true if we are evaluating ${\cal A}$ for a general dynamical metric where the functional dependence of metric coefficients ($X, \, \omega_i, \, h_{ij} $) on $r$ and $v$ is not restricted only through the product of them. Hence, we conclude that whenever any expression of the form given in eq.\eqref{eqlmTens} with $m_v>m_r$ evaluates to a non-zero value, the corresponding space-time metric is dynamical,
\begin{equation} \label{defEQvsNEQ}
\begin{split}
\text{Equilibrium configuration} \rightarrow ~& (\partial_r)^{m_r}(\partial_v)^{m_v}{\cal B}(rv,x) \vert_{r=0}  =0 ~\text{(for $m_v>m_r$)} \, , \\
\text{Non-equilibrium configuration} \rightarrow ~& (\partial_r)^{m_r}(\partial_v)^{m_v}{\cal B}(r,v,x) \vert_{r=0} \neq 0 ~\text{(for $m_v>m_r$)} \, .
\end{split}
\end{equation}
We will use this as a criterion for stationary and dynamical space-times. 
 
 To explain the linearity of dynamical fluctuations to a stationary configurations, we decompose the full space-time metric $g_{\mu\nu}$ as a sum of two parts
\begin{equation}\label{break_stationarity_0}
g_{\mu\nu} = g_{\mu\nu}^{eq} + \epsilon \, \delta g_{\mu\nu} \, ,
\end{equation}
where $g_{\mu\nu}^{eq}$ is a stationary metric and $\delta g_{\mu\nu}$ captures the dynamics away from equilibrium. The small parameter $\epsilon$ denotes the amplitude of the dynamical perturbation. In the coordinate system that we are working with, such a decomposition will imply that all the metric functions could also be written accordingly as a sum of two contributions
\begin{equation} \label{break_stationarity}
\begin{split}
X=& \,  X^{eq}(rv,x) + \epsilon \, \delta X(r,v,x) \, , \\
\omega_i =& \,  \omega^{eq}_i(rv,x) + \epsilon \, \delta\omega_i(r,v,x) \, ,\\
 h_{ij}= & \,  h_{ij}^{eq}(rv,x) + \epsilon \, \delta h_{ij}(r,v,x) \,.
\end{split}
\end{equation}
Furthermore, as a consequence, we can also argue that for any ${\cal B}$, some covariant expression constructed out of the metric functions and their $\nabla_i$ derivative(s), can always be written as 
\begin{equation} \label{Beq+neq}
{\cal B}(r,v,x) ={\cal B}^{eq}(rv,x) + \epsilon \, \delta{\cal B}(r,v,x) \, .
\end{equation}
Following our arguments leading to eq.\eqref{defEQvsNEQ}, it is obvious that if we operate $\left(\partial_r^{m}\partial_v^{m +k}\right)$ (for $k>0$) on such a ${\cal B}$ given in eq.\eqref{Beq+neq} and evaluate it on the horizon, it is only $\delta{\cal B}$ that will contribute, rendering the expression linear in dynamics
\begin{equation} \label{linA}
{\cal A}\, \vert_{r=0} \sim\left(\partial_r^{m}\partial_v^{m +k}\right) {\cal B} \vert_{r=0} = \epsilon \, \left(\partial_r^{m}\partial_v^{m +k}\right)\delta{\cal B} \vert_{r=0} \, \sim \, {\mathcal O} (\epsilon) \, ,~~~ (\text{for $k>0$}) .
\end{equation}
Now, let us consider an expression of the form 
\begin{equation} \label{B1B2eq+neq}
{\cal A}_1 \sim \left(\partial_r^{m_1}\partial_v^{m_1 +k_1}{\cal B}_1\right)\left(\partial_r^{m_2}\partial_v^{m_2 +k_2}{\cal B}_2\right)\, ,~~~(\text{for $k_1, k_2>0$}),
\end{equation}
where both ${\cal B}_1$ and ${\cal B}_2$ are two different but arbitrary covariant tensor expressions admitting similar decomposition as ${\cal B}$ in eq.\eqref{Beq+neq}. Following the similar set of arguments presented above, we could now see that both of the factors in the RHS of eq.\eqref{B1B2eq+neq} will have non-zero contributions only when ${\cal B}_1$ and ${\cal B}_2$ are replaced by $\delta{\cal B}_1$ and $\delta{\cal B}_2$ respectively,
\begin{equation}\label{nonlinA}
\begin{split}
{\cal A}_1\,\vert_{r=0} \sim \left(\partial_r^{m_1}\partial_v^{m_1 +k_1}\right) {\cal B}_1  \, \left(\partial_r^{m_2}\partial_v^{m_2 +k_2}\right) {\cal B}_2 \,\vert_{r=0}  \, \sim \, {\mathcal O} (\epsilon^2) \, ,~~~ (\text{for $k_1, k_2>0$}) \, ,
\end{split}
\end{equation}
leading us to the conclusion that the expression ${\cal A}_1$ will be non-linear (quadratic) in the amplitude of dynamical fluctuations.

We will summarise the main lessons that we have learnt regarding the classification of generic tensor structures built out of the metric coefficient functions and their derivatives, in terms of stationarity and dynamical fluctuations (linear or non-linear) around it in Table-\ref{table:sum_dv_ep}
\begin{table} [h!] 
\centering 
 \begin{tabular}{| c| m{13cm} | } 
 \hline
 1. & Any expression that has more $\partial_v$-derivatives than $\partial_r$-derivatives could be non-zero solely because of dynamics. \\
  \hline
2. &  The expression will be linear in the amplitude of dynamics when the extra $\partial_v$ derivatives (i.e. those $\partial_v$'s which are not compensated or paired with $\partial_r$'s) are all acting on a single covariant expression of the form of ${\cal B}$, eq.\eqref{linA}. \\
 \hline
3. &  Whenever the extra $\partial_v$ derivatives are distributed between different factors, the resulting expression (e.g. ${\cal A}_1$ in eq.\eqref{nonlinA}) would be non-linear in the amplitude of dynamics. \\
 \hline
\end{tabular}
\caption{Distribution of $\partial_v$-derivatives and non-linearities in amplitude expansion.}
\label{table:sum_dv_ep}
\end{table}

\subsection{Killing symmetry and boost weight of quantities} \label{KillSymBW}
As we have mentioned in the introduction, we would be analyzing the equation of motion for the metric, and our goal is to write its most general form in terms of our building blocks given in Table-\ref{table:basicblock}. The equation of motion in a theory of gravity is a rank two tensor with respect to the full set of diffeomorphism that mixes all the space-time coordinates. In this subsection, we shall briefly describe how we can relate a particular component of a tensor (when evaluated on the horizon) with the number and distribution of $\partial_r$ and $\partial_v$ derivatives acting on different factors.

In our horizon adapted coordinate system $\{r, \, v, \, x\}$, the most general stationary black hole solution as in eq.\eqref{metequl}, admits a Killing vector 
\begin{equation} \label{killingvector}
\xi = \xi^\mu\partial_\mu =( v\partial_v - r\partial_r) \, .
\end{equation}
As was mentioned in the previous subsection (see eq.\eqref{BT_generator}), this is also the generator of the boost transformation eq.\eqref{boosttransf}, or more precisely of the infinitesimal version of it. 

In order to study small dynamical fluctuations away from a stationary black hole configuration, it will be useful to see how various quantities transform under this scaling or boost transformation that generates the Killing symmetry of the stationary background. Furthermore, since we are viewing every quantity to be built out of the basic building blocks (Table-\ref{table:basicblock}), it is sufficient to know the transformation property of them. 

Let us define the power of $\lambda$ by which any quantity transform under the boost transformation eq.\eqref{boosttransf}, as the boost weight of that quantity
\begin{equation} \label{defBW0}
{\cal A} \rightarrow \widetilde{\cal A} = \lambda^w {\cal A} \, , ~~\text{under}~~ \{r\rightarrow\tilde r = \lambda ~r,~v\rightarrow\tilde v = \lambda^{-1} v\} ~~ \Rightarrow ~~ \text{boost weight of ${\cal A}$ is $w$} \, .
\end{equation}
The first thing we should note following this definition, is that the coordinates $r$ and $v$ have boost weights $+1$ and $-1$ respectively. Similarly all the metric functions: $X$, $\omega_i$ and $h_{ij}$ have zero boost weight and only $\partial_v$ and $\partial_r$ are the operators that transform non-trivially under eq.\eqref{boosttransf}
 \begin{equation}
\partial_r \rightarrow \partial_{\tilde{r}} = \lambda^{-1} \, \partial_r \, , ~~ \partial_v \rightarrow \partial_{\tilde{v}} =\lambda  \, \partial_v \, .
\end{equation}
Hence, the operators $\partial_r$ and $\partial_v$ have boost weights $-1$ and $+1$ respectively, and the operator $\nabla_i$ has zero boost weight when operated on $X$, $\omega_i$ and $h_{ij}$. 

In other words, since $\xi$ given in eq.\eqref{killingvector} generates Killing symmetry of the stationary background, the Lie derivative with respect to $\xi$ (denoted as ${\cal L}_\xi$) when operated on any covariant tensor that is constructed out of the stationary metric eq.\eqref{metequl}, will vanish. Therefore, a non-zero value of this ${\cal L}_\xi$ operator acting on any covariant tensor built out of metric components, is indicative of non-stationary or out-of-equilibrium dynamics. The action of ${\cal L}_\xi$ on any covariant tensor $S_{\alpha_1 \alpha_2\cdots \alpha_k}$ is given by
\begin{equation}\label{Lxi}
\begin{split}
{\cal L}_\xi S_{\alpha_1 \alpha_2\cdots \alpha_k}= \, &\xi^\beta\partial_\beta S_{\alpha_1 \alpha_2\cdots \alpha_k}+ \left(\partial_{\alpha_1}\xi^{\beta}\right)S_{\beta \alpha_2\cdots \alpha_k}+ \left(\partial_{\alpha_2}\xi^{\beta}\right)S_{\alpha_1 \beta \alpha_3 \cdots \alpha_k}+\cdots\\
& + \left(\partial_{\alpha_k}\xi^{\beta}\right)S_{\alpha_1 \alpha_2 \cdots  \beta}\, ,
\end{split}
\end{equation}
and when we evaluate this in our coordinate system, eq.\eqref{met1}, with  $\xi =( v\partial_v - r\partial_r) $, we will obtain the following general form
\begin{equation}\label{Lxi1}
\begin{split}
{\cal L}_\xi S_{\alpha_1 \alpha_2\cdots \alpha_k} =\,  \left[ w +(v\partial_v -r\partial_r) \right]S_{\alpha_1 \alpha_2\cdots \alpha_k} \, 
\end{split}
\end{equation}
where we identify the quantity $w$ as the boost weight of the covariant tensor $S_{\alpha_1 \alpha_2\cdots \alpha_k}$ as 
\begin{equation}\label{defBW}
\begin{split}
w\, \equiv \, &\text{net boost weight of a covariant tensor (with all indices lowered)}\\
= \, & \text{number of lower $v$ indices - number of lower $r$ indices.} 
\end{split}
\end{equation}
A more detailed explanation regarding the idea of boost weight of different quantities is provided in Appendix-(\ref{app:BW}). Following eq.\eqref{defBW}, we can check that under ${\cal L}_\xi$, the operators $\partial_v$ and $\partial_r$ transform like the $v$ and $r$ components of a lower-indexed vector $\partial_{\mu}$ and therefore have boost weights $+1$ and $-1$ respectively, as expected. We also learn that if we evaluate a particular component of a covariant tensor on the horizon $r=0$, the number of lower $v$ indices corresponds to the number of $\partial_v$'s (and the number of lower $r$ indices corresponds to the number of $\partial_r$'s) acting on metric functions ($X$, $\omega_i$, and $h_{ij}$) with boost weight zero\footnote{If we allow ourselves to go away from the horizon, positive boost weights could also be absorbed by factors of $r$. For example, the $vv$ component of the metric, $g_{vv}$, which is one example of a  rank-2 covariant tensor component with boost weight $2$, has a factor of $r^2$ multiplying a boost-invariant function $X$. The same is true for $g_{vi}$ and $g_{ij}$ the $vi$ and $ij$ components of the metric respectively. In fact, this is why all the metric functions ($X$, $\omega_i$ and $h_{ij}$) are of zero boost weights.}.
 
Before we proceed, let us summarise the main points of this section in Table-\ref{table:sum_bas_setup}, which we are going to use in the rest of the paper.
\begin{table} [h!] 
\centering 
 \begin{tabular}{| c| m{13cm} | } 
 \hline
 1. & We have chosen a special coordinate system adapted to the dynamical null horizon. The coordinates on the horizon are $v$ (parameter along the null generator) and $\{x^i\}$ (all the spatial coordinates), whereas the coordinate away from the horizon is denoted by $r$, see eq.\eqref{met1}. The coordinates $v$ and $r$ are also affinely parametrized. \\
  \hline
2. &  In this coordinate system, if we evaluate any component of a covariant tensor restricted on the horizon, schematically, it will have a structure with some number of $\partial_r$, $\partial_v$ and $\nabla_i$ operators acting on the functions appearing in the metric eq.\eqref{met1}:($X$, $\omega_i$, and $h_{ij}$) or product of such structures. \\
 \hline
3. &  For an expression, we defined the boost weight $w$ of it to be the difference between the number of $\partial_v$'s and $\partial_r$'s, and then we have argued that for a tensor component, this is the same as the difference between the number of lower $v$ (or upper $r$) and the lower $r$ (or upper $v$) indices, see eq.\eqref{defBW0} and eq.\eqref{defBW}. \\
 \hline
4. &  Furthermore, the boost weight of a tensor component is nothing but the multiplicative factor it will have under the action of Lie derivative ${\cal L}_\xi$ along the vector $\xi \left(= v\partial_v - r\partial_r\right)$, the general Killing vector field whenever the metric has a Killing horizon, see eq.\eqref{Lxi1}. \\
 \hline
5. &  Finally, any expression with positive boost weight is non-zero only when evaluated on a dynamical, non-stationary metric. It is linear in the amplitude of the dynamical fluctuation provided all the excess $\partial_v$'s are acting on a single function, and it is not a product of more than one expression, each having positive boost weight, see eq.\eqref{defEQvsNEQ}, eq.\eqref{linA} and eq.\eqref{nonlinA}. \\
 \hline
\end{tabular}
\caption{Summary of the important lessons so far that will serve as guidelines for later computations.}
\label{table:sum_bas_setup}
\end{table}

\subsection{Significance of the structure of certain components of the equations of motion for a proof of the second law} \label{Evv_2nd_law}
Until now, we have summarized the basic concepts in relevance to our work. In the following, we will briefly review the arguments required to justify a local version of the second law of thermodynamics for linearized dynamical perturbations around a stationary black hole configuration. As we have mentioned before, to achieve this, we will be looking at the structure of a particular component of the equations of motion, to be more specific, the `$vv$'-component of it, i.e., $E_{vv}$. Thereby, we will also explain how significant the role of the structure of $E_{vv}$ is to establish the second law. 

The notion of entropy for stationary black holes in arbitrary diffeomorphism invariant theory of gravity is associated with the Wald entropy defined as
\begin{equation} \label{def_Wald_ent}
S_W = -2 \pi \int_{{\cal H}_v} d^{d-2}x \,  \sqrt{h} \, {E^{\alpha_1\alpha_2\alpha_3\alpha_4}_R} \epsilon_{\alpha_1 \alpha_2} \epsilon_{\alpha_3 \alpha_4} \, = \int_{{\cal H}_v} d^{d-2}x \,  \sqrt{h} \, \left(1 + s^{\text{HD}}_w\right) ,
\end{equation}
where $E^{\alpha_1\alpha_2\alpha_3\alpha_4}_R$ is given by eq.\eqref{def_E_abcd}, $\epsilon_{\alpha_1 \alpha_2}$ are the bi-normal to the constant $v$-slices of the horizon, $\sqrt{h}$ is the determinant of the induced metric $h_{ij}$ on the horizon and $s^{\text{HD}}_w$ is the contribution to the Wald entropy density from the higher derivative part of the Lagrangian leaving the Einstein-Hilbert part aside. This definition of the Wald entropy is, by construction, consistent with the first law of thermodynamics. From the last expression on the RHS above it is clear that for Einstein's gravity $S_W$ reduces to the well known formula of computing area of the spatial slices of the horizon.  

Starting from the Wald entropy formula in eq.\eqref{def_Wald_ent}, one can then write down an entropy function which is the out-of-equilibrium extension of it as the following
\begin{equation} \label{def_S_tot}
S_{\text{total}} = S_W + \int_{{\cal H}_v} d^{d-2}x \,  \sqrt{h} \, s_{\text{cor}} = \int_{{\cal H}_v} d^{d-2}x \, \sqrt{h} \,\left( 1 + s^{\text{HD}}_w + s_{\text{cor}} \right) \, , 
\end{equation}
where $s_{\text{cor}}$ are the corrections to the Wald entropy density due to non-stationary dynamics, and it certainly includes the JKM ambiguities that we mentioned before. Once the dynamics is switched off, $s_{\text{cor}}$ vanishes and therefore $S_{\text{total}}$ reduces to $S_W$, i.e. 
$$ s_{\text{cor}} \big\vert_{equilibrium} = 0.$$
The main goal of \cite{Wall:2015raa} was to constrain the form of $s_{\text{cor}}$, and hence to fix the JKM ambiguities, by requiring $S_{\text{total}}$ to satisfy a linearized version of the second law. More precisely, the strategy would be to determine $s_{\text{cor}}$ such that $\partial_v S_{\text{total}} \geq 0$. One defines a local expansion parameter $\vartheta$, \footnote{The corresponding quantity in Einstein's gravity, say $\vartheta_E$, is the expansion parameter for the null congruence $\partial_v$. The evolution of this parameter $\vartheta$ with $v$ is governed by the Raychaudhuri equation and this plays a crucial role in proving the area increase theorem in general relativity.} as follows
\begin{equation} \label{deftheta}
{\partial S_{\text{total}} \over \partial v} \equiv \int_{{\cal H}_v} d^{d-2}x \,  \sqrt{h} \, \vartheta \, .
\end{equation}
Now, one way to prove $\partial_v S_{\text{total}} \geq 0$ is to show that $\vartheta \geq 0$ for all $v \geq 0$. The trick one uses for this is to argue that $\vartheta$ is a monotonically decreasing function of $v$ (i.e. $\partial_v \vartheta \leq 0$), along with the restriction that at asymptotic future it approaches to zero, meaning $\vartheta \rightarrow 0$ for $v \rightarrow \infty$. This is actually motivated by the physical situation that we are examining here. After getting slightly perturbed by a small dynamical fluctuation, the initial stationary black hole solution finally settles down to another stationary black hole solution in the future. 

Now calculating $\partial_v \vartheta$ in our chosen metric gauge by using eq.\eqref{deftheta} as the definition of $\vartheta$ and $S_{\text{total}}$ from eq.\eqref{def_S_tot}, we obtain\footnote{We have used here the fact that $\partial_v \vartheta_E = -R_{vv} + \mathcal{O} (\epsilon^2)$, where $\vartheta_E$ is the corresponding quantity for $\vartheta$ when computed from just the Einstein-Hilbert term in the action.}
\begin{equation}\label{Dv_theta}
\partial_v \vartheta = - R_{vv} + \partial_v \left({1 \over \sqrt{h}} \partial_v \left(\sqrt{h} \left(s^{\text{HD}}_w + s_{\text{cor}}\right) \, \right)\right) + \mathcal{O} (\epsilon^2) \, .
\end{equation} 
Next, we should remember that the $vv$-component of the equation of motion $E_{\mu\nu}$ can be written as
\begin{equation} \label{EOM_Evv}
-E_{vv} = R_{vv} + E^{\text{HD}}_{vv} = T_{vv} \, 
\end{equation}
where $E_{vv}$ is obtained from just the gravity part of the Lagrangian\footnote{The minus sign on the LHS is there to make it consistent with the convention used in eq.\eqref{EgravTvv}. With this choice of convention, for Einstein gravity we get $$E_{vv} = - R_{vv} = -T_{vv},$$ and this also defines the sign convention of the stress tensor such that the null energy condition works out as $T_{vv} \geq 0$. We should also remember that while writing the $vv$-component of the equations of motion in the form of eq.\eqref{EOM_Evv}, we have evaluated it on the horizon which is located at $r=0$. For example, from our choice of metric given in eq.\eqref{met1} it is clear that $g_{vv} \vert_{r=0}=0$. }, $E^{\text{HD}}_{vv}$ only contains the contribution from the higher derivative part of the gravity Lagrangian excluding the Einstein-Hilbert term, $T_{vv}$ is the $vv$-component of the stress energy tensor $T_{\mu\nu}$ coming from the matter sector. Substituting this for $R_{vv}$ in eq.\eqref{Dv_theta} we get
\begin{equation}\label{Dv_theta_1}
\partial_v \vartheta = E^{\text{HD}}_{vv} + \partial_v \left({1 \over \sqrt{h}} \partial_v \left(\sqrt{h} \left(s^{\text{HD}}_w + s_{\text{cor}}\right) \, \right)\right)-T_{vv} + \mathcal{O} (\epsilon^2) \, .
\end{equation} 
Furthermore, we also assume that the stress-energy tensor, for the matter part, satisfies the null energy condition, which, when translated to our choice of metric gauge and $\xi^\mu$ eq.\eqref{killingvector}, takes the form of $T_{vv} \geq 0$. Thus from eq.\eqref{Dv_theta_1} we see that the imposition of null energy condition on the stress-energy tensor for the matter perturbation actually contributes in favor of making $\partial_v \vartheta \leq 0$. In that sense, with the assumption of null energy condition, $T_{vv}$ trivially drops out from our analysis. 

We are considering dynamical perturbations to a stationary black hole configuration such that the metric of the equilibrium background gets corrected at $\mathcal{O} (\epsilon)$, see eq.\eqref{break_stationarity_0}. Consequently, we must also take into consideration the fact that terms linear in the amplitude of the dynamical fluctuation can come with both signs depending on the sign of $\epsilon$. On the other hand, on equilibrium configurations, i.e. $\epsilon \rightarrow 0$, we must get $\partial_v \vartheta \rightarrow 0$. Therefore, the only way in which $\partial_v \vartheta \leq 0$ can be met while working up to linear order $\mathcal{O} (\epsilon)$, is to show that 
\begin{equation} \label{conddvtheta}
\partial_v \vartheta = \mathcal{O} (\epsilon^2) \, ,
\end{equation}
or in other words, $\partial_v \vartheta$ should vanish up to linear order in the amplitude expansion. 

As a result it is obvious from eq.\eqref{Dv_theta_1} that eq.\eqref{conddvtheta} will be satisfied if the following holds true
\begin{equation}\label{cond2}
 E^{\text{HD}}_{vv} = -\, \partial_v \left({1 \over \sqrt{h}} \partial_v \left(\sqrt{h} \left(s^{\text{HD}}_w + s_{\text{cor}}\right) \, \right)\right)+\mathcal{O} (\epsilon^2) \, .
\end{equation}
This relation in eq.\eqref{cond2} highlights the significance of $E^{\text{HD}}_{vv}$ in arriving at a proof of the linearized version of the second law presented in the form of eq.\eqref{conddvtheta} .

At this point, it is worth making a comment regarding the dependance of $T_{vv}$ on the order of $\epsilon$ and its significance for eq.\eqref{conddvtheta}. As mentioned before in \S\ref{sec:stationarity} (see eq.\eqref{break_stationarity_0}), the correction to the stationary background metric due to dynamical perturbations is of $\mathcal{O} (\epsilon)$. Let the matter fields that induce the perturbation are of order $\epsilon$. Now, following our assumptions the stress tensor due to the matter perturbation satisfies null energy condition $T_{vv} \geq 0$, and vanishes in the equilibrium limit $\epsilon \rightarrow 0$. As a consequence, based on general arguments given above for $\partial_v \vartheta$, we learn that $T_{vv} \sim \mathcal{O} (\epsilon^2)$; see section-(2.1) and Appendix-B of \cite{Bhattacharya:2019qal} for a detailed discussion on this point. Therefore, up to our working precision of $\mathcal{O} (\epsilon)$, $T_{vv}$ can be dropped from the RHS of eq.\eqref{Dv_theta_1} and the equality in eq.\eqref{conddvtheta} is justified once the restriction on $E_{vv}$ in eq.\eqref{cond2} is met \footnote{We can instead have a situation, possibly a more physically appealing one, where the corrections to the space-time geometry are generated due to backreactions from a $\mathcal{O} (\epsilon^2)$ stress tensor representing the non-stationary perturbation. The metric which is sourced by this stress tensor would then receive its first correction at $\mathcal{O} (\epsilon^2)$ and not at $\mathcal{O} (\epsilon)$. The corrections to the metric at $\mathcal{O} (\epsilon)$ would trivially vanish. In other words, the dynamical fluctuations of the metric will get non-trivial contributions starting at $\mathcal{O}(\epsilon^2)$. This is certainly true for minimal coupling of the matter sector, which we will be considering in this paper. However, as it has been discussed in section-(2.1) of  \cite{Bhattacharya:2019qal}, even in that situation, the statement made in eq.\eqref{cond2} would be valid. In that case eq.\eqref{conddvtheta} would become trivial at $\mathcal{O} (\epsilon)$. The first non-trivial contribution for this would occur at $\mathcal{O} (\epsilon^2)$ as the following: $\partial_v \vartheta \vert_{\mathcal{O} (\epsilon^2)} \leq 0$.}. 

In other words, we can view eq.\eqref{cond2} as a constraint on the off-shell structure of $E_{vv}$, which can be summarised as follows:

\textit{ In any diffeomorphism invariant theory of gravity, working in the chosen metric gauge eq.\eqref{met1} and up to linear order in the amplitude expansion, if we can show that the structure of the $vv$-component of equations of motion can always be written as the RHS of eq.\eqref{cond2}, we will be able to construct a proof of the linearized version of the second law by satisfying eq.\eqref{conddvtheta}.}

In view of the statement made above, we are now convinced that the primary goal of this current paper is indeed justified. Once we are able to show that in any arbitrary higher derivative theory of gravity $E_{vv}$ can always be written as eq.\eqref{eq:main1}, which we write here again for convenience
\begin{equation}\label{eq:main1_repeat}
   \begin{split}
   E_{vv} \big\vert_{r=0} = & \, \partial_v \bigg[{1\over\sqrt h}\partial_v \left({\sqrt h} \, {\cal J}^v\right) +\nabla_i{\cal J}^i\bigg] + \mathcal{O} (\epsilon^2) \, .
   \end{split}
   \end{equation}
Once this is compared with eq.\eqref{cond2}, one can straightforwardly obtain the components of entropy current (i.e. ${\cal J}^v$ and ${\cal J}^i$). Furthermore, we are also convinced that when the equilibrium limit is taken ${\cal J}^v$ should reproduce the equilibrium Wald entropy density\footnote{Since $E_{vv}$ also includes the contributions from the Einstein-Hilbert term, in ${\cal J}^v$ we get an extra additive factor of one in addition to $s^{\text{HD}}_w$. This extra factor of one is actually the Area of the horizon piece.}
\begin{equation}
{\cal J}^v\big\vert_{equilibrium} = (1+s^{\text{HD}}_w) \, .
\end{equation}

\subsection{An important identity relating the equations of motion with the Noether charge for diffeomorphism} \label{rel_Evv_Q} 
In the previous sub-sections, we have established the importance of a certain desired structure of the $vv$-component of the equations of motion, denoted by $E_{vv}$, for our work. In the following, we will write down a key equation that will help us construct the entropy current from $E_{vv}$ by relating the latter to the Noether charge under diffeomorphism. This relation is, in fact, an identity that is true for any diffeomorphism invariant theory of gravity. We will follow the construction of Noether currents and charges corresponding to the diffeomorphism invariance in such a theory of gravity as prescribed in \cite{Iyer:1994ys}\footnote{Before we proceed, let us emphasize that this is a well-known identity in gravitational theories and here we are just stating and reproducing it for the sake of completeness. We refer the readers to Appendix-(\ref{app:ReviewIyerWald}) where a review of \cite{Iyer:1994ys} is provided in component notation.}.
 
In a diffeomorphism invariant theory of gravity with an action $S= \int\sqrt{-g} \, L$, the variation of the Lagrangian due to an arbitrary variation $\delta g_{\mu\nu}$ of the metric, is given by
\begin{equation}\label{eq:hijibiji-1}
\begin{split}
\delta\left[\sqrt{-g} ~L\right]&= \sqrt{-g} \, E^{\mu\nu} \, \delta g_{\mu\nu}  + \sqrt{-g} \, D_\mu \Theta ^\mu [\delta g] \, ,
\end{split}
\end{equation}
where $E^{\mu\nu}$ is the equation of motion, $\Theta^\mu [\delta g]$ is the total derivative term generated in the variation of the Lagrangian and following our convention in this paper, $D_\mu$ is the covariant derivative with respect to the metric $g_{\mu\nu}$. 

Next we consider that $\delta g_{\mu\nu}$ is produced due to an infinitesimal coordinate transformation generating the diffeomorphism $x^\mu \rightarrow x^\mu + \zeta^\mu$, 
\begin{equation}
\delta g_{\mu\nu} = D_\mu \zeta_\nu + D_\nu \zeta_\mu, 
\end{equation}
for which we obtain 
\begin{equation}
\delta\left[\sqrt{-g} \, L\right] = \sqrt{-g}\, D_{\mu} \left( \zeta^\mu L\right).
\end{equation}
Substituting this back in eq.\eqref{eq:hijibiji-1} and cancelling the $\sqrt{-g}$ factor on both sides we get
\begin{equation}\label{hijibiji-2}
\begin{split}
D_\mu \left(\zeta^\mu L-2E^{\mu\nu} \zeta_\nu -\Theta^\mu[\partial\zeta]\right) =-2\zeta_\nu D_\mu E^{\mu\nu} \, .
\end{split}
\end{equation}
Now we make a choice for $\zeta^\mu$ such that it is non-zero only in a small region ${\cal R}$. With this, if one integrates the both sides of eq.\eqref{hijibiji-2} over full space-time, the LHS will vanish since it would integrate to a pure boundary term at infinity where $\zeta$ vanishes. Hence, we obtain
\begin{equation}
\int_\text{full space-time} \, \zeta_\nu \, D_\mu E^{\mu\nu}=0 \, ,
\end{equation}
which is true for any $\zeta$ as long as it is has a non-zero support over a finite region $\mathcal{R}$. Therefore, we see that the following relation identically holds 
\begin{equation}\label{bianchi1}
 D_\mu E^{\mu\nu}=0 \, .
 \end{equation}
We should note that this is essentially the Bianchi identity corresponding to any diffeomorphism invariant theory.

Let us now substitute the identity eq.\eqref{bianchi1} in eq.\eqref{hijibiji-2} and allow $\zeta^\mu$ to be any vector without any restriction on its support. We obtain that the vector $\zeta^\mu L-2E^{\mu\nu} \zeta_\nu -\Theta^\mu(\partial\zeta)$ is identically conserved  on any geometry for any $\zeta$
\begin{equation}
D_\mu \left(\zeta^\mu L-2E^{\mu\nu} \zeta_\nu -\Theta^\mu[\partial\zeta] \right) =0 \, .
\end{equation}

Next, we use the result that an identically conserved vector can itself be always written as the divergence of an arbitrary antisymmetric rank-$2$ tensor. Hence one can argue that locally it is always true that 
\begin{equation}\label{hijibiji-3}
\begin{split}
\zeta^\mu L-2E^{\mu\nu} \zeta_\nu -\Theta^\mu = - D_{\nu} Q^{\mu\nu}\, ,
\end{split}
\end{equation}
for some arbitrary antisymmetric tensor $ Q^{\mu\nu}$\footnote{The negative sign on the RHS of eq.\eqref{hijibiji-3} is due to our convention and could be understood as a part of the definition of $Q^{\mu\nu}$ through this equation. In Appendix-(\ref{app:ReviewIyerWald}), see eq.\eqref{defDQ}, where we have derived this equation while reviewing \cite{Iyer:1994ys}.}. 

We can contract the free index on both sides of eq.\eqref{hijibiji-3} with another $\zeta_\mu$ to get 
\begin{equation}\label{hijibiji-4}
\begin{split}
& \zeta^2 L -\Theta^\mu \zeta_\mu+ \zeta_\mu \, D_\nu  Q^{\mu\nu} =2 \,  \zeta^\mu\zeta^\nu E_{\mu\nu} \, .
\end{split}
\end{equation}
Although eq.\eqref{hijibiji-4} is true for any $\zeta^\mu$, for our purpose, we will choose $\zeta^\mu$ to be $\xi = v\partial_v - r\partial_r$ and evaluate both sides of eq.\eqref{hijibiji-4} on the horizon $r=0$, 
\begin{equation}\label{hijibiji-5}
\begin{split}
 \left(2\xi^\mu\xi^\nu E_{\mu\nu}\right)\vert_{r=0}=\, &2 v^2 E_{vv} \vert_{r=0}\\
 \left(\xi^2 L -\Theta\cdot \xi+ \xi_\mu\,  D_\nu  Q^{\mu\nu}\right)\vert_{r=0} = \,& v\left(-\Theta^r+  D_\nu Q^{r \nu} \right)\vert_{r=0} \, .
\end{split}
\end{equation}
Finally, cancelling one factor of $v$ from both sides, we arrive at
\begin{equation}\label{keyeqn}
\begin{split}
2\, v \, E_{vv}\big\vert_{r=0}=~&\big(-\Theta^r+ D_\mu Q^{r \mu}\big)\big\vert_{r=0}\, .
\end{split}
\end{equation}  
Equation \eqref{keyeqn} constitutes the main relation that we will use in \S\ref{sec:proof} to prove the existence of an entropy current in a diffeomorphism invariant theory of gravity.

\section{A proof of the existence of an entropy current}\label{sec:proof}
Following the fundamental concepts developed in \S\ref{sec:setup}, in this section, we will carry out an extensive analysis to derive an expression of the entropy current for dynamical black holes in any diffeomorphism invariant theory of gravity up to linearized order in the amplitude of the out-of-equilibrium fluctuations. In practice, we aim to establish that a specific component of the equations of motion, namely $E_{vv}$, in such a theory can always be recast in a form similar to the RHS of eq.\eqref{eq:main1}, and thereby obtaining the form of ${\cal J}^v$ and ${\cal J}^i$. As we have already discussed in \S\ref{Evv_2nd_law}, showing eq.\eqref{eq:main1} is equivalent to argue for a linearized version of the second law involving black hole dynamics. 

The starting point of our calculation would be to consider the important relation, eq.\eqref{keyeqn}, that was derived in \S\ref{rel_Evv_Q}. Here we re-write that equation again for the sake of convenience, 
\begin{equation}\label{keyeqn_again}
\begin{split}
2\, v \, E_{vv}\big\vert_{r=0}=~&\big(-\Theta^r+ D_\mu Q^{r \mu}\big) \big\vert_{r=0}\, .
\end{split}
\end{equation}
With this at hand, our main job now is to establish that the expression for $E_{vv}$ on the RHS of eq.\eqref{keyeqn_again} can always be rearranged as the following
\begin{equation}\label{eq:main1_again}
   \begin{split}
   E_{vv} \big\vert_{r=0} = &\,  \partial_v \bigg[{1\over\sqrt h}\partial_v \left({\sqrt h} \, {\cal J}^v\right) +\nabla_i{\cal J}^i\bigg] + \mathcal{O} (\epsilon^2) \, ,
   \end{split}
   \end{equation} 
which is the same as eq.\eqref{eq:main1}.

For the sake of convenience of the reader, in the following, we will separate the material to be presented into two sub-sections. To start with, in \S\ref{ssec:short_prf} we will provide a brief schematic overview of the proof. The primary purpose of this would be to convey the essential ingredients of the analysis postponing the details of technical computations to the following sub-section \S\ref{ssec:main_prf}.

\subsection{A brief and schematic sketch of the proof highlighting the basic strategy} \label{ssec:short_prf}
In this sub-section, our primary goal is to provide an illustrative sketch of the arguments justifying the construction of a local entropy current in an arbitrary higher derivative theory of gravity. Before we dive into the technical computations presented in the following sub-section, it will be beneficial to develop an overall and intuitive understanding of the line of logical sequence. 

\paragraph{Basic strategy:} We start with a brief explanation of the basic strategy as follows. Since a particular coordinate system, as in eq.\eqref{met1}, has been chosen for our analysis, it is worth noting the explicit appearance of the coordinate $v$ on the LHS of eq.\eqref{keyeqn_again}. The off-shell structure of $E_{vv}$ cannot involve any factor of $v$ except through the metric functions $(X, \, \omega_i, \, h_{ij})$ or their derivatives. Therefore, our primary task going ahead would be to figure out the explicit $v$-dependence in each of the two terms appearing on the RHS of eq.\eqref{keyeqn_again} so that we can compare the terms with equal powers of $v$ on both sides of it. This, in turn, would produce the desired structure of $E_{vv}$ that we are looking for.

We notice that on the RHS of eq.\eqref{keyeqn_again} two quantities determine the structure of $E_{vv}$ when evaluated on the horizon. One of them is the $r$-component of $\theta^\mu$, which is generated as a total derivative term in the variation of the Lagrangian under a diffeomorphism, eq.\eqref{eq:hijibiji-1}. The second term involves $Q^{\mu\nu}$, which is defined in eq.\eqref{hijibiji-3}. In order to extract out the explicit $v$-dependence in both of these terms, we will use their corresponding definitions outlined in the construction of the Noether charge for diffeomorphism invariant gravity Lagrangians, \cite{Iyer:1994ys}. These definitions are given in eq.\eqref{theta_expr} and eq.\eqref{theta_p_expr} for $\Theta^\mu$, and in eq.\eqref{expr_Q_ab} for $Q^{\mu \nu}$.

Furthermore, we will also need to consider the Killing symmetry of the stationary background geometry, eq.\eqref{metequl}, generated by the null generator of the horizon $\xi^\mu$ eq.\eqref{BT_generator}, and its breaking by a non-stationary perturbation in the linearised amplitude approximation (see eq.\eqref{break_stationarity}). Since $\xi^\mu$ also generates the boost transformation eq.\eqref{boosttransf}, we will have to figure out how each of the two terms on the RHS of eq.\eqref{keyeqn_again} transforms under it, i.e., the boost weights for each of them. 

One crucial element for our computations would be determining the general structure of an arbitrary covariant tensor quantity with a given boost weight. At this point, it should be emphasized that every tensorial quantity appearing in our analysis should be considered as if they are built out of the basic building blocks as given in Table-\ref{table:basicblock}. In \S\ref{KillSymBW} we have already learned about how to assign boost weights for the basic building blocks as well as for any generic tensor built out of them by counting the number of $r$ and $v$ components of that particular tensor, see eq.\eqref{defBW}. 

Once we know how to assign boost weight for a generic covariant tensor, we can easily determine its generic structure. For that, we need to analyze the distribution of $\partial_v$-derivatives carefully. The fact that we will be working in the linearised amplitude approximation will play a crucial role in executing this. We should also remember another related and equally important aspect, that is, eq.\eqref{keyeqn_again} has to be evaluated on the horizon located at $r=0$. The guiding principles which are important in this context are summarised in Table-\ref{table:sum_dv_ep} and Table-\ref{table:sum_bas_setup}. Following these guidelines, we can compute the generic structure of any covariant tensor with a positive boost weight, and the result is given in eq.\eqref{sq01}. The detailed arguments for deriving this relation would be discussed in \S\ref{BW_cov_Ten}. We will call this relation as ``Result:\,$1$" and it will be repeatedly used in our calculations.

Next, we would directly implement the strategy outlined above for both the terms on the RHS of eq.\eqref{keyeqn_again} one at a time. Using the definitions of $\Theta^\mu$ and $Q^{\mu \nu}$ as given in eq.\eqref{theta_expr} and eq.\eqref{expr_Q_ab} respectively, we will analyze every term that contributes to each of them. Finally, substituting the resulting expressions in eq.\eqref{keyeqn}, we will derive the desired structure for $E_{vv}$ as mentioned in eq.\eqref{eq:main1}.

\paragraph{Schematic illustration of the proof:}
As written above in eq.\eqref{eq:main1_again}, the principal quantity that we are going to examine is $E_{vv}$, the $vv$ component of the equation of motion. Following the criteria set in eq.\eqref{defBW} and Table-\ref{table:sum_bas_setup}, $E_{vv}$ is a tensor component with boost weight $+2$. Following the arguements presented in \cite{Wall:2015raa} and also reviewed in section-(2.2.2) of \cite{Bhattacharya:2019qal}, we have seen that $E_{vv}$ being a term with boost weight $+2$ could be expressed, up to linear order in the amplitude of the dynamical fluctuations, as 
\begin{equation}\label{eq:Evv_start}
E_{vv} \sim  \partial_v^2 C_{(0)} + A_{(0)}\, \partial_v^2 B_{(0)} + \mathcal{O} (\epsilon^2) \, .
\end{equation}
Here we have used the convention that the subscripts in $A_{(0)}$, $B_{(0)}$ and $C_{(0)}$ denote the boost weights of these quantities, and this will be followed throughout in this paper. It is also essential to keep in mind that the purpose of writing eq.\eqref{eq:Evv_start} is to highlight the structural form on the RHS, ignoring specific numerical factors, possible index contraction, and factors of $\sqrt{h}$ to avoid the clutter.

Furthermore, for a more ellaborate derivation of eq.\eqref{eq:Evv_start}, one should compare it with eq.(2.22) in \cite{Bhattacharya:2019qal}. This would also emphasise an important difference in the structures of the two terms on the RHS of eq.\eqref{eq:Evv_start}. In general, for a given theory of higher derivative gravity the quantities $A_{(0)}$, $B_{(0)}$ and $C_{(0)}$ should be thought of as they are built out of the metric functions $(X, \, \omega_i, \, h_{ij})$ or their derivatives with possible index contractions such that the resulting boost weights are zero for all of them. However, $C_{(0)} $ is always a product of two terms with non-zero boost weights equal in magnitude but opposite in signs as shown below
\begin{equation}
C_{(0)} \sim \sum_{k}  \widetilde T_{(-k)}~\partial_v^{(k)}T_{(0)} \, .
\end{equation}
It is obvious from this equation that $C_{(0)}$ has boost weight zero, but it is product of two quantities $\widetilde T_{(-k)}$ and $\partial_v^{(k)}T_{(0)}$ which have individual boost weights as $-k$ and $+k$ respectively. However, the other term on the RHS of eq.\eqref{eq:Evv_start} involves $A_{(0)}$ and $B_{(0)}$ which could not be written like this. We name the terms of the type $C_{(0)}$ as `JKM-type' terms and  the terms of the form $A_{(0)}\partial_v^2 B_{(0)}$ as the `zero-boost' terms\footnote{The fact that the second term on the RHS of eq.\eqref{eq:Evv_start} has been given the name `zero-boost' terms should not be confused with the fact that this term $A_{(0)}\, \partial_v^2 B_{(0)}$ actually has a non-zero boost weight. To see the origin of this nomenclature we refer the reader to section-(2.2.2) and section-(2.2.3) in \cite{Bhattacharya:2019qal}. Both the terms $A_{(0)}$ and $B_{(0)}$ do not have any $v$ or $r$ derivatives in them which is not the case for $C_{(0)}$.}. Therefore, it follows from eq.\eqref{eq:Evv_start} that any tensor component with boost weight $+2$, such as $E_{vv}$, must have a set of such `zero-boost' terms plus $\partial_v^2$ acting on `JKM type' terms. The statements mentioned above actually follows more generally from the important result that determines the general structure of any covariant tensor quantity with positive boost-weight, as given in eq.\eqref{sq01} named as ``Result:\,$1$". 

Our goal in this paper is to show that the `zero-boost' terms in $E_{vv}$ could also be written as\footnote{This is very schematic, and there have to be relevant factors of $\sqrt{h}$ to make everything consistent, but in this sub-section, since we are aiming at a heuristic argument we shall be ignoring all these subtleties.}
\begin{equation}
A_{(0)}\partial_v^2 B_{(0)} ~\sim ~\partial_v^2 {\cal Q}_{(0)} +\partial_v \left(\nabla_i {\cal Q}_{(1)}^{i}\right) + {\cal O}\left(\epsilon^2\right) \, ,
\end{equation} 
where ${\cal Q}_{(0)}$ is a quantity with vanishing boost weight and ${\cal Q}_{(1)}^{i}$ is a quantity with boost weight $+1$ with a spatial index $i = 1,2,\cdots, D-2$. Also, $\nabla_i$ is the covariant derivative associated with the induced metric $h_{ij}$ on the co-dimension two spatial slices of the horizon. With this eq.\eqref{eq:Evv_start} becomes
\begin{equation}\label{eq:Evv_result}
E_{vv} \sim  \partial_v^2 \big(  C_{(0)} + {\cal Q}_{(0)}\big) + \partial_v \left(\nabla_i {\cal Q}_{(1)}^{i}\right) + \mathcal{O} (\epsilon^2) \, ,
\end{equation}
In the following sub-section, we will carry out a detailed analysis to arrive at eq.\eqref{keyrel3} which is actually a more precise version of eq.\eqref{eq:Evv_result}. It should also be noted that the presence of the first term on the RHS of eq.\eqref{eq:Evv_result} was already explained in \cite{Wall:2015raa}. The actual new result in this current paper is to argue that in addition to the first term on the RHS, the second term on the RHS can also, in general, be there in $E_{vv}$. This new term was identified with an entropy current in \cite{Bhattacharya:2019qal}. 

In the following we will present a rough sketch for the arguments to argue that eq.\eqref{eq:Evv_result} is indeed true. For this we will use equation eq.\eqref{keyeqn_again} and analyze both the terms on the RHS of this equation individually. 
\begin{itemize}
\item From the construction of Noether's charge for diffeomorphism as pointed out in \cite{Iyer:1994ys}, we know that both $\Theta^r$ and $Q^{r\mu}$ are linear in $ \xi^\mu \partial_\mu = v\partial_v - r\partial_r$. The corresponding expressions justifying this can be found in eq.\eqref{schemetheta} for  $\Theta^r$ and eq.\eqref{expr_Q_ab} for $Q^{r\mu}$.
\item Since $\xi$ is the only source of explicit dependence on $v$, both $\Theta^r$ and $Q^{r\mu}$, being linear in $\xi^\mu$, will have the following structure 
\begin{equation}\label{eq:struct1}
\begin{split}
\Theta^r\vert_{r=0} &= \Theta_{(1)} + v \, \Theta_{(2)} \, ,~~~
Q^{rv}\vert_{r=0} = Q_{(0)} + v \, Q_{(1)} \, ,~~~
Q^{ri}\vert_{r=0} = J^i_{(1)} + v \,  \partial_v\tilde J^{i}_{(1)} \, .
\end{split}
\end{equation}
Here also, the subscripts in the RHS denote the boost weights of the corresponding quantities. In the last equation, we have also used the fact that any spatial current with boost weight $+2$ could always be expressed as $\partial_v$ acting on a current term with boost weight $+1$, up to corrections non-linear in the amplitude of the dynamics. We have suppressed some index structures to avoid cluttering in the presentation.
\item Using the fact that $\Theta^\mu$ could be expressed in terms of $\mathcal{L}_\xi$ (and a single derivative of it)\footnote{Here $\mathcal{L}_\xi$ denotes Lie derivative along the vector $\xi^\mu$.} acting on covariant tensors (see the arguments leading to eq.\eqref{schemetheta} in \S\ref{subsec:Theta_r}), one could give further structure to $\Theta^r$. Actually, as discussed in Appendix-(\ref{App_eqShrtPrf}), one can show that
 \begin{equation}\label{eq:fstruct}
\Theta_{(2)} =\partial_v \Theta_{(1)} + \partial^2_v {\cal M}_{(0)} \, ,
 \end{equation}
 where ${\cal M}_{(0)}$ is a JKM-type term, i.e. a boost invariant term  and also a product of two factors each of which has opposite boost weights.
\item Now, substituting eq.\eqref{eq:struct1} and eq.\eqref{eq:fstruct} in eq.\eqref{keyeqn_again} we find
\begin{equation}\label{eq:struct2} 
\begin{split}
2 v \, E_{vv}\big\vert_{r=0} =&\left[ -\Theta_{(1)} + Q_{(1)}+\partial_v Q_{(0)}+ \nabla_i J^i_{(1)}\right] \\
&+ v\,  \partial_v\left[-\Theta_{(1)} - \partial_v{\cal M}_{(0)} + Q_{(1)}+ \nabla_i \tilde J^i_{(1)}\right]+ \mathcal{O} (\epsilon^2) \,.
\end{split}
\end{equation}
Since the commutator of $\partial_v$ and $\nabla_i$ is of boost weight $+1$, it has been neglected in the last line where it is acting on $\tilde J^i_{(1)}$. 
\item It is important to note that the $v$-dependence in eq.\eqref{eq:struct2} is manifest explicitly. In other words, it is an identity with no explicit factors of $v$ in $E_{vv}$, $\Theta$, $Q$ and $\tilde J^i$. Therefore, it follows that the coefficients of different powers of $v$ must vanish independently
\begin{equation}\label{eq:struct3a}
 \Theta_{(1)} - Q_{(1)}-\partial_v Q_{(0)}- \nabla_i J^i_{(1)}=0\\
\end{equation}
\begin{equation}\label{eq:struct3b}
\partial_v\left[-\Theta_{(1)} -\partial_v{\cal M}_{(0)}+Q_{(1)}+\nabla_i\tilde J^i_{(1)}\right]= 2E_{vv}\big\vert_{r=0}
\end{equation}

\item Finally, substituting eq.\eqref{eq:struct3a} in eq.\eqref{eq:struct3b} we get the desired result, 
\begin{equation}
2 E_{vv}\big\vert_{r=0} = -\partial_v\left[\partial_v\left(Q_{(0)} + {\cal M}_{(0)} \right) + \nabla_i\left(J^i_{(1)} - \tilde J^i_{(1)}\right)\right] + \mathcal{O} (\epsilon^2) \, .
\end{equation}
By comparison, we find this is to be of the same form as on the RHS of eq.\eqref{eq:main1_again}. In the following sub-section, we will derive this relation with much more technical rigour and arrive at eq.\eqref{keyrel3} with the expressions of ${\cal J}^v$ and ${\cal J}^i$ in eq.\eqref{keyrel3_JvJi}.
\end{itemize}

\subsection{An elaborate analysis of the proof with technical details} \label{ssec:main_prf} 
\begin{figure}[h]
\centering
\includegraphics[width=0.95\textwidth]{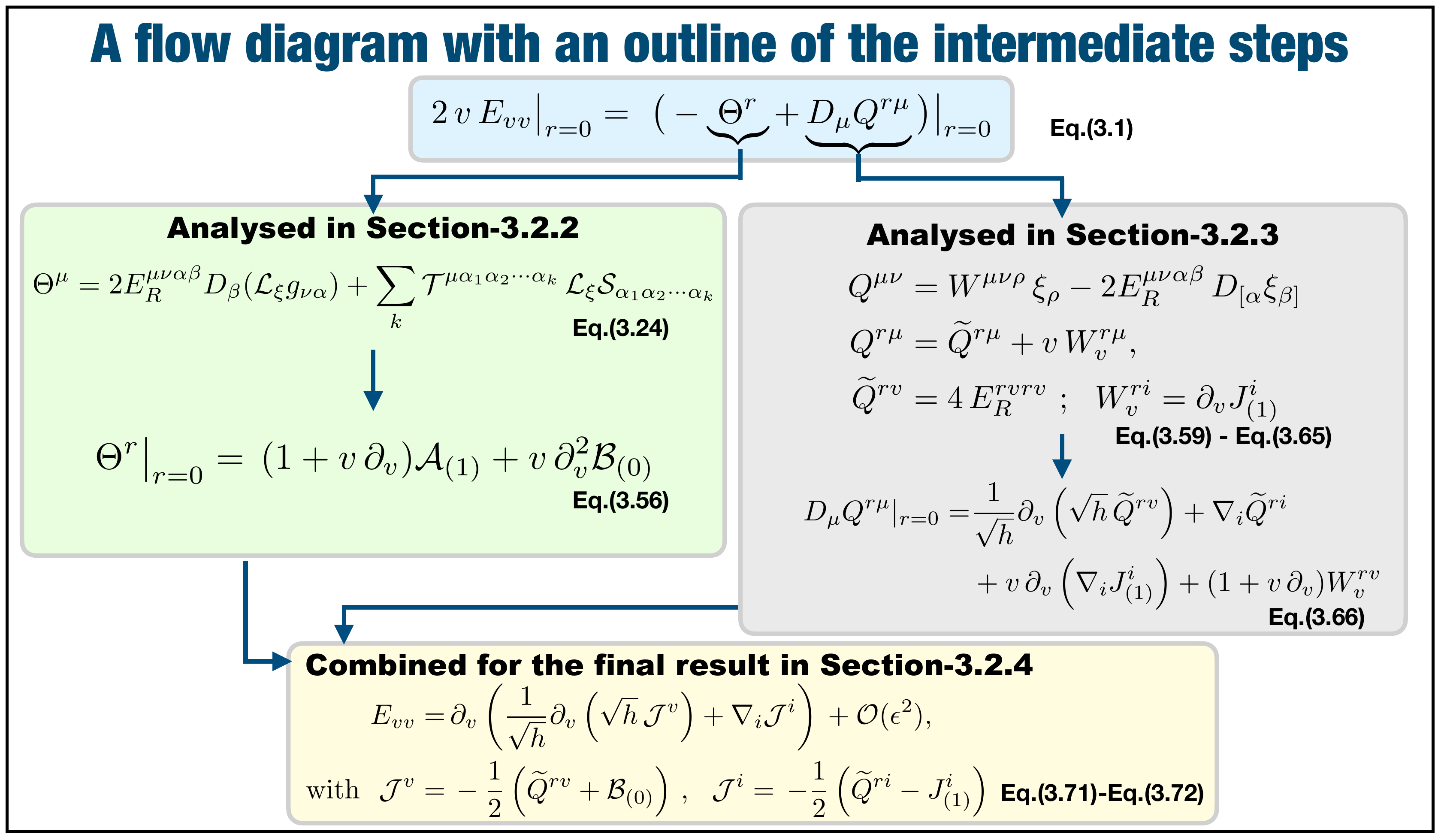}
\caption{A flow diagram summarising the main technical result in each sub-section.}
\label{chart}
\end{figure}

In this sub-section, our main goal is to methodically discuss in more detail how eq.\eqref{keyeqn_again} can lead us to the construction of an entropy current in any higher derivative theory of gravity. Simultaneously one also obtains the components ${\cal J}^v$ and ${\cal J}^i$ as given in eq.\eqref{eq:main1_again} such that the linearized second law is satisfied. For a better organization of the technical computations to be presented in the following, we will separate them into several sub-parts. In \S\ref{BW_cov_Ten} we will start with obtaining an important result regarding the generic structure of a general covariant tensor with any positive boost weight. Using this result, in \S\ref{subsec:Theta_r} and \S\ref{subsec:DQrmu}, the two terms on the RHS of eq.\eqref{keyeqn_again} will be analyzed one at a time. Finally, we will collect them all together in \S\ref{final_Evv}, to arrive at our final result for the structural form of $E_{vv}$.  In Fig:\ref{chart}, we provide a flow-diagram summarising the main results from each of the following sub-sections.

\subsubsection{The generic structure of a covariant tensor with positive boost weight}\label{BW_cov_Ten}
We shall first write down an important result involving the boost weight of a covariant tensor and possible distributions of partial differential operator $\partial_v$, evaluated on the horizon. The typical structure of any covariant tensor component with positive boost weight $w = a+1>0$, restricted to the horizon and up to linear order in the amplitude of the non-stationary perturbation, will be of the form
\begin{equation}\label{sq00}
t^{(k)}_{(a+1)} \big\vert_{r=0} = \widetilde T_{(-k)} \, \partial_v^{(k+a +1)}T_{(0)}  \big\vert_{r=0} + {\cal O}(\epsilon^2) \, .
\end{equation}
Here the subscripts denote the boost weight of the corresponding quantity and for convenience we have suppressed the index structure for $t^{(k)}_{(a+1)}, \, \widetilde T_{(-k)}$ and $T_{(0)}$. In writing eq.\eqref{sq00} we have followed the guidelines summarised in both Table-\ref{table:sum_dv_ep} and Table-\ref{table:sum_bas_setup}. More specifically, we made use of the fact that one $\partial_v$-derivative operator carries a boost weight of $(+ 1)$. The quantities $\widetilde T_{(-k)}$ and $T_{(0)}$, appearing on the RHS of eq.\eqref{sq00} are arbitrary tensor quantities with boost weight `$-k$' and zero respectively. The superscript `$k$' on $t^{(k)}_{(a+1)}$ signifies the fact that any covariant tensor with boost weight $w = a+1>0$, can be represented by a member in an one parameter family of terms denoted by $k= \, 0,\, 1,\, 2, \cdots$. 

Once a typical covariant tensor with positive boost weight has been written as in eq.\eqref{sq00}, one can rearrange the $\partial_v$-derivatives and recast it in the following form 
\begin{equation}\label{sq01}
\begin{rcases}
\begin{split}
t^{(k)}_{(a+1)} \big\vert_{r=0}= \, &  \widetilde T_{(-k)} \,  \partial_v^{(k+a +1)}T_{(0)} + {\cal O}(\epsilon^2)\\
= \, & \partial_v^{(a+1)}\left[\sum_{m=0}^{k-1} (-1)^m \, \left[{^{m+a}C}_{m}\right] \,  \widetilde T_{(-k+m)}~\partial_v^{(k-m)}T_{(0)} \right]~~~~\\
\, & + (-1)^k \left[ {^{k+a}C}_a\right]\widetilde T_{(0)}~\partial_v^{a+1} T_{(0)}  + {\cal O}(\epsilon^2) \, ,
\end{split}
\end{rcases}
\text{\, \, Result:\,$1$}
\end{equation}
where $\widetilde T_{(-k +m)} = \partial_v^m \widetilde T_{(-k)}$ and the numerical coefficient $\left[ {^{k+a}C}_a\right] = {(k+a)! \over k! \, a!}$.
The detailed arguments justifying that such a rearrangement is always possible allowing us to write $t^{(k)}_{(a+1)}$ as the RHS of eq.\eqref{sq01}, is presented in Appendix-(\ref{app:pro1}).

This is a significant result which we will use quite extensively in the rest of this section. For that reason, we have also identified eq.\eqref{sq01} as ``Result: $1$" and will refer to it accordingly whenever we use it.  

Before we proceed, let us mention a notable feature of the final expression in eq.\eqref{sq01}. In the first term on the RHS, the quantity within the square bracket is boost invariant, however, it is product of two terms which are individually not boost invariant: one of them has positive boost weight and the other one has equal boost weight but of negative sign,
\begin{equation}\label{1st_sq01}
\text{The first term on the RHS of eq.\eqref{sq01}} ~ \sim ~ \partial_v^{(a+1)}\left[\widetilde T_{(-k+m)}~\partial_v^{(k-m)}T_{(0)}\right] \, 
\end{equation}
where $\widetilde T_{(-k+m)}$ and $\partial_v^{(k-m)}T_{(0)}$ have boost weights equal to $(-k+m)$ and $(k-m)$ respectively. Following our set of guidelines presented in Table:\ref{table:sum_dv_ep}, we can therefore conclude that boost invariant terms of this kind are non-zero only due to dynamics and will vanish when evaluated on stationary configurations.

\subsubsection{Analyzing the structure of the first term (i.e. $\Theta^r$) in $E_{vv}$} \label{subsec:Theta_r}
In this subsection, our focus will be on studying the first term's structure on the RHS of eq.\eqref{keyeqn_again}, i.e., $\Theta^r$. 

For any diffeomorphism invariant theory of gravity, the functional dependence of the Lagrangian is given in eq.\eqref{diff_inv_L}. Under an arbitrary change of the space-time metric denoted by $g_{\mu\nu} \rightarrow g_{\mu\nu} + \delta g_{\mu\nu}$, the variation in the Lagrangian, written in eq.\eqref{eq:hijibiji-1}, contains a total derivative term, i.e. $D_{\mu} \theta^\mu$. Furthermore, following the Lemma $3.1$ of \cite{Iyer:1994ys},
(see also eq.\eqref{app_theta} and eq.\eqref{app_theta_p} in Appendix-(\ref{app:ReviewIyerWald})), 
the general structure of $\theta^\mu$ can be obtained as the following
\begin{equation} \label{theta_expr}
    \Theta^{\mu} [g_{\beta_1 \beta_2},  \delta g_{\beta_1 \beta_2}, \delta \phi] = 2 E^{\mu\nu\alpha\beta}_R D_{\beta} \delta g_{\nu\alpha} + \Theta'^{\mu} \, ,
\end{equation}
where
\begin{equation}\label{theta_p_expr}
\begin{split}
\Theta'^{\mu} = S^{\mu\alpha\beta}(g_{\beta_1 \beta_2},\phi)\,  \delta g_{\alpha\beta} & + \sum_{i=0}^{m-1} T^{\mu\alpha\beta\rho\sigma \alpha_1 \dots \alpha_i}_i (g_{\beta_1 \beta_2},\phi)\,  \delta D_{(\alpha_1} \dots D_{\alpha_i)} R_{\alpha\beta\rho\sigma}  
\\ & + \sum_{i=0}^{l-1} U^{\mu \alpha_1 \dots \alpha_i}_i (g_{\beta_1 \beta_2},\phi)\,  \delta D_{(\alpha_1} \dots D_{\alpha_i)} \phi 
\, ,
\end{split}
\end{equation}
where $E^{\mu\nu\alpha\beta}_R$ is given by
\begin{equation} \label{def_E_abcd}
    E^{\mu\nu\alpha\beta}_R = \dfrac{\partial \, L}{\partial R_{\mu\nu\alpha\beta}} - D_{\rho_1} \dfrac{\partial \, L}{\partial D_{\rho_1} R_{\mu\nu\alpha\beta}} + \dots + (-1)^m D_{(\rho_1} \dots D_{\rho_m)} \dfrac{\partial \, L}{\partial D_{(\rho_1} \dots D_{\rho_m)} R_{\mu\nu\alpha\beta} } \, ,
\end{equation} which is a covariant tensor with the same symmetry properties of $R_{\mu\nu\rho\sigma}$ and $\phi$ is the matter field that are also present in the Lagrangian. 

It should be noted that in the expression of $\theta^{\mu}$ on the RHS of eq.\eqref{theta_expr}, the variation that is denoted by $\delta$, can be put to the left of the covariant derivatives everywhere in $\Theta'^{\mu}$ except for the term $E^{\mu\nu\alpha\beta}_R D_{\beta} \delta g_{\nu\alpha}$. This distinction between the two terms on the RHS of eq.\eqref{theta_expr} will be crucial in the following. 

Let us rewrite eq.\eqref{theta_expr} schematically as the following
\begin{equation} \label{theta_mu_form1}
\Theta^{\mu} = 2 E^{\mu\nu\alpha\beta}_R D_{\beta} \delta g_{\nu\alpha} +\sum_{k} {\cal T}^{\mu \alpha_1 \alpha_2\cdots \alpha_k} \, \delta {\cal S}_{\alpha_1 \alpha_2\cdots \alpha_k} \, ,
\end{equation}
where on the RHS, the second term collectively includes all the contributions that are in $\theta'^\mu$, eq.\eqref{theta_p_expr}. Therefore, the quantity ${\cal S}_{\alpha_1 \alpha_2\cdots \alpha_k}$ is actually a representative member of the family of covariant tensors as indicated below
\begin{equation} \label{arg_S_theta}
\begin{split}
{\cal S}_{\alpha_1 \alpha_2\cdots \alpha_k}: \big\{ &\,  g_{\alpha\beta}, \, R_{\alpha\beta\rho\sigma}, \, D_{\alpha_1}R_{\alpha\beta\rho\sigma}, \cdots, D_{(\alpha_1} \dots D_{\alpha_m)} R_{\mu\nu\alpha\beta}, \cdots , \\
& \, \phi, \,  D_{\alpha_1} \phi, \cdots , D_{(\alpha_1} \dots D_{\alpha_i)} \phi, \cdots \big\} \, .
\end{split}
\end{equation}
It should be noted that in our discussion so far, there has been neither any reference to the diffeomorphism transformation nor have we used any invariance of the theory under diffeomorphism. Hence, eq.\eqref{theta_expr} is true for arbitrary variations $\delta g_{\alpha\beta}$. 

Now, if $\delta g_{\alpha\beta}$ is generated due to a diffeomorphism given by $x^\mu \rightarrow x^\mu + \zeta^\mu(x)$ where $\zeta^\mu$ is some vector field, we can write
\begin{equation}
\delta g_{\alpha\beta} = {\cal L}_{\zeta}g_{\alpha\beta} = D_{\alpha} \zeta_{\beta} + D_{\beta} \zeta_{\alpha} \, ,
\end{equation}
where ${\cal L}_{\zeta}$ is the Lie derivative along $\zeta^\mu$. If we substitute for $\delta g_{\alpha\beta}$ from the equation above in eq.\eqref{theta_expr}, the corresponding $\Theta^{\mu}$ for diffemorphism will be obtained, which we denote by $\Theta^{\mu}[g, \,{\cal L}_{\zeta}g]$. Implementing this in eq.\eqref{theta_mu_form1} leads us to the following
\begin{equation}\label{theta_mu_form2}
\Theta^{\mu} = 2 E^{\mu\nu\alpha\beta}_R D_{\beta} ({\cal L}_{\zeta} g_{\nu\alpha}) + \sum_{k} {\cal T}^{\mu \alpha_1 \alpha_2\cdots \alpha_k} \, \delta {\cal S}_{\alpha_1 \alpha_2\cdots \alpha_k}[\delta g_{\alpha\beta}\rightarrow {\cal L}_{\zeta} g_{\alpha\beta}] \, .
\end{equation}
In the second term on the RHS of eq.\eqref{theta_mu_form2}, the meaning of the substitution $\delta g_{\alpha\beta}\rightarrow {\cal L}_{\zeta} g_{\alpha\beta}$ in the argument of $\delta {\cal S}_{\alpha_1 \alpha_2\cdots \alpha_k}$ is obvious in the following sense: for each possible terms in $\delta {\cal S}_{\alpha_1 \alpha_2\cdots \alpha_k}$ listed in eq.\eqref{arg_S_theta}, one should first reduce it to a form involving   $\delta g_{\alpha\beta}$ and arbitrary number of derivatives on it, and then substitute for $\delta g_{\alpha\beta}\rightarrow {\cal L}_{\zeta} g_{\alpha\beta}$. 

Next, we adopt a specific choice for the vector field that generates diffeomorphism as $\xi^{\mu}$ in place of $\zeta^{\mu}$. In our coordinate system $\{r,v,x^i\}$ in eq.\eqref{met1}, $\xi^{\mu}$ is given by $\xi^{\mu} \partial_{\mu} = v \partial_v -r \partial_r$. From our discussions both in \S\ref{sec:stationarity} and \S\ref{KillSymBW}, we immediately identify this $\xi^\mu$ as the generator of the boost transformation given in eq.\eqref{boosttransf} as well as the Killing vector, eq.\eqref{killingvector}, for the most general stationary black holes in eq.\eqref{metequl}. 

Furthermore, with the input of diffeomorphism invariance of our theory, we can show that 
\begin{equation} \label{diffeo_covT_rel}
\delta {\cal S}_{\alpha_1 \alpha_2\cdots \alpha_k}[\delta g_{\alpha\beta}\rightarrow {\cal L}_{\xi} g_{\alpha\beta}] = {\cal L}_{\xi} {\cal S}_{\alpha_1 \alpha_2\cdots \alpha_k} \, ,
\end{equation}
and thus obtain the following expression for $\theta^{\mu}$
\begin{equation}\label{schemetheta}
\Theta^{\mu} = 2 E^{\mu\nu\alpha\beta}_R D_{\beta} ({\cal L}_{\xi} g_{\nu\alpha}) + \sum_{k} {\cal T}^{\mu \alpha_1 \alpha_2\cdots \alpha_k} \, {\cal L}_{\xi} {\cal S}_{\alpha_1 \alpha_2\cdots \alpha_k} \, .
\end{equation}
We refer the readers to Appendix-(\ref{app:diffeo_covT}) for a detailed clarification regarding eq.\eqref{diffeo_covT_rel}.

In what follows, we will consider each of the two terms on the RHS of eq.\eqref{schemetheta} one at a time and try to obtain the general structure of them. As explained in \S\ref{ssec:short_prf}, we would like to know the explicit $v$-dependence of each of the two terms that determines $\Theta^{\mu}$ in eq.\eqref{schemetheta}. In both of these quantities, the source of $v$-dependence is solely from the factor of $\xi^\mu$. We know that $\xi$ dependence in $\Theta$ comes from ${\cal L}_\xi$, which is linear in $\xi$ and therefore linear in $v$. From here, we learn that, when evaluated on the horizon, the $v$-dependence in $\Theta^\mu$ can at most be linear in $v$, and schematically we can write
\begin{equation} \label{theta_mu_form3}
\Theta^\mu \big\vert_{r=0} = \Theta_1^\mu + v \,\Theta_2^\mu \, ,
\end{equation}
where $\Theta_1^\mu$ and $\Theta_2^\mu$ are, as of now, components of covariant tensors without any explicit $v$ dependence. Focussing on $\Theta^r$, which is our primary object, and following our rules of boost weight counting, we learn that the corresponding weights of $\Theta_{(1)}^r$ and $\Theta_{(2)}^r$ are $(+1)$ and $(+2)$ respectively, as also denoted by their subscripts. With this knowledge of boost weights for $\Theta_{(1)}^r$ and $\Theta_{(2)}^r$, we can use 
Result-\,$1$ from eq.\eqref{sq01} to predict the following schematic structures for each of them (ignoring the explicit constant coefficients for now) 
\begin{equation}\label{th1th2}
\begin{split}
\Theta_{(1)}^r \big\vert_{r=0}\sim~& \partial_v{\cal X} +\widetilde {\cal X}_{(0)}\partial_v{\cal X}_{(0)} \\
\Theta_{(2)}^r \big\vert_{r=0}\sim~& \partial_v^2{\cal Y}+ \widetilde{\cal Y}_{(0)}~ \partial_v^2{\cal Y}_{(0)} \, ,
\end{split}
\end{equation}
where ${\cal X}, \, {\cal X}_{(0)}, \, \widetilde {\cal X}_{(0)},\, {\cal Y}, \, {\cal Y}_{(0)}, \, \widetilde {\cal Y}_{(0)}$ are all boost invariant quantities. Recall that according to Result-\,$1$ (in eq.\eqref{sq01}), although ${\cal Y}$ and ${\cal X}$ are both boost invariant terms, each of them could have pieces that are always product of two terms, one with strictly positive and the other with with strictly negative boost weight and therefore vanish on stationary horizons (see the discussions around eq.\eqref{1st_sq01}). 

We now readily obtain the following form for $\Theta^r$
\begin{equation}\label{scm}
\begin{split}
\Theta^r \big\vert_{r=0}
\sim \, \partial_v{\cal X} +\widetilde {\cal X}_{(0)}\, \partial_v{\cal X}_{(0)} +v\, \left[\partial_v^2{\cal Y}+ \widetilde{\cal Y}_{(0)}\,  \partial_v^2{\cal Y}_{(0)}\right] \, .
\end{split}
\end{equation}
Following our discussion so far, it might appear that $\Theta_{(1)}^r$ and $\Theta_{(2)}^r$ are two independent quantities and therefore their structures as written in eq.\eqref{th1th2} are not related to each other. However, this is not true and due to the fact that both of them are actually generated by an action of the operator ${\cal L}_\xi$ in eq.\eqref{schemetheta}, we can constrain the structure of $\Theta^r$ further. In particular we will explicitly show that one can relate the quantities appearing in eq.\eqref{th1th2} up to non-linear order corrections in the amplitude of the fluctuations,
\begin{equation}\label{key1}
\widetilde{\cal Y}_{(0)} \, \partial_v^2{\cal Y}_{(0)}+ {\cal O}(\epsilon^2) = \widetilde{\cal X}_{(0)} \,  \partial_v^2{\cal X}_{(0)} + {\cal O}(\epsilon^2) = \partial_v\left(\widetilde{\cal X}_{(0)} \,  \partial_v{\cal X}_{(0)}\right)+ {\cal O}(\epsilon^2) \, .
\end{equation}
Substituting this relation in equation eq.\eqref{scm} we find the final schematic structure for $\Theta^r$
\begin{equation}\label{thf}
\begin{split}
\Theta^r \big\vert_{r=0}=&\,  (1 + v~\partial_v) {\cal A}_{(1)}+ v~\partial_v^2{\cal B}_{(0)}+ {\cal O}(\epsilon^2) \\
\text{such that,}~~ {\cal A}_{(1)} =& \,\tilde{\cal X}_{(0)}~ \partial_v{\cal X}_{(0)} + \partial_v{\cal X}\, , ~~ \text{and}~~{\cal B}_{(0)} =\, {\cal Y}-{\cal X}\, .
\end{split}
\end{equation}
Therefore, in the following, our job is to justify the critical relation eq.\eqref{key1}, for both the terms in the expression of $\Theta^r$ as given in eq.\eqref{schemetheta}.

\subsubsection*{Analysis of the second term in $\Theta^r$ from eq.(\ref{schemetheta}):} \label{subThetaR1} 
For reasons that will be clear later, to start with, we focus first on the second term on the RHS of\eqref{schemetheta}. As we are considering the $r$-component of $\Theta^\mu$, it follows from our rule of assigning boost weights that $\Theta^r$ must have a weight of $(+1)$ and consequently both the terms on the RHS of eq.\eqref{schemetheta} will have boost weights equal to $(+1)$. With this knowledge let us now write down the schematic form of the second term in $\Theta^r$ for our convenience
\begin{equation}\label{theta1a}
\sum_{j} {\cal T}^{r \alpha_1 \alpha_2\cdots \alpha_j} \, {\cal L}_{\xi} {\cal S}_{\alpha_1 \alpha_2\cdots \alpha_j} 
= ~\sum_{k\geq 0}\widetilde T_{(-k)}{\cal L}_\xi T_{(k+1)} +\sum_{k\geq 0}S_{(k+1)} {\cal L}_\xi \widetilde S_{(-k)} \, ,
\end{equation}
where we have suppressed the details of the index structures of the covariant tensor components $T$, $\widetilde T$, $S$ and $\widetilde S$, and the subscripts denote the boost weights for each of them. One can see that the two terms appearing on the RHS of eq.\eqref{theta1a}, are two different cases where the Lie derivative ${\cal L}_\xi$ is acting on quantities with positive and negative boost weight respectively. It should be noted that here we are just doing an operation which is in principle possible without stating anything specific about the structures of the quantities on the RHS above. Thus all the quantities $T$, $\widetilde T$, $S$ and $\widetilde S$ on the RHS of the equation above are covariant tensors which are otherwise arbitrary apart from having the specific boost weights assigned to them. We can then perform further manipulations as shown below to bring it to a form as given in the following
\begin{equation}\label{theta1}
\begin{split}
\sum_{j} &{\cal T}^{r \alpha_1 \alpha_2\cdots \alpha_j} \, {\cal L}_{\xi} {\cal S}_{\alpha_1 \alpha_2\cdots \alpha_j} \\ 
= ~&\sum_{k\geq 0}\widetilde T_{(-k)}{\cal L}_\xi T_{(k+1)} +\sum_{k\geq 0}S_{(k+1)} {\cal L}_\xi \widetilde S_{(-k)}\\
 = ~&\sum_{k\geq 0}\widetilde T_{(-k)}{\cal L}_\xi T_{(k+1)}+{\cal L}_\xi \bigg( \sum_{k\geq 0}\left[S_{(k+1)}  \widetilde S_{(-k)}\right]\bigg) -\sum_{k\geq 0}\widetilde S_{(-k)} {\cal L}_\xi S_{(k+1)}\\
 = ~&\left(1 + v\partial_v\right)\bigg(\sum_{k\geq 0}\left[S_{(k+1)}  \widetilde S_{(-k)}\right]\bigg) +\sum_{k\geq 0}\widetilde T_{(-k)}{\cal L}_\xi T_{(k+1)}-\sum_{k\geq 0}\widetilde S_{(-k)} {\cal L}_\xi S_{(k+1)}\\
 = ~&\left(1 + v\partial_v\right){\mathbb S}_{(1)}+\sum_{k\geq 0}\widetilde{\mathbb T}_{(-k)}~{\cal L}_\xi {\mathbb T}_{(k+1)} \\
=~& \left(1 + v\partial_v\right){\mathbb S}_{(1)}+\sum_{k\geq 0}\widetilde{\mathbb T}_{(-k)}~\left(k+1+v\partial_v\right) {\mathbb T}_{(k+1)} \, .
 \end{split}
\end{equation}
In the second last line we have defined new quantities for brevity
\begin{equation}
{\mathbb S}_{(1)} \equiv \sum_{k\geq 0}\left(S_{(k+1)}  \widetilde S_{(-k)}\right),~~\text{and}~~\widetilde{\mathbb T}_{(-k)}{\cal L}_\xi {\mathbb T}_{(k+1)} \equiv\widetilde T_{(-k)}{\cal L}_\xi T_{(k+1)}-\widetilde S_{(-k)} {\cal L}_\xi S_{(k+1)} \, .
\end{equation}
Comparing with eq.\eqref{thf}, it is clear from the final expression on the RHS of eq.\eqref{theta1} that the first term there is already of the desired form. 

In the following we will process the second term in eq.\eqref{theta1}. Schematically, this has two different structures in it, namely $\widetilde{\mathbb T}_{(-k)} \, {\mathbb T}_{(k+1)}$ and $\widetilde{\mathbb T}_{(-k)} \, \partial_v{\mathbb T}_{(k+1)}$. Our strategy would  be to use Result:\,$1$ from eq.\eqref{sq01}  repeatedly for each of them. Firstly, we see that, according to Result:\,$1$, ${\mathbb T}_{(k+1)}$ will have the following schematic structure (ignoring the coefficients)
\begin{equation}
{\mathbb T}_{(k+1)}= \partial_v^{k+1} C_{(0)} + A_{(0)}~\partial_v^{(k+1)}B_{(0)} \, .
\end{equation}
Using this we define the quantities $t^{(k)}_{(1)}$ and $t^{(k)}_{(2)}$ as given by 
\begin{equation}\label{theta2ab}
\begin{split}
t^{(k)}_{(1)}  \, \equiv \,  &\widetilde{\mathbb T}_{(-k)}{\mathbb T}_{(k+1)}
=~\widetilde{\mathbb T}_{(-k)} \, \partial_v^{k+1}C_{(0)} + A_{(0)} \, \widetilde{\mathbb T}_{(-k)} \, \partial_v^{(k+1)}B_{(0)} \, ,\\
t^{(k)}_{(2)} \, \equiv  \, &\widetilde{\mathbb T}_{(-k)} \, \partial_v{\mathbb T}_{(k+1)}=~\widetilde{\mathbb T}_{(-k)} \, \partial_v^{k+2}C_{(0)} + A_{(0)} \, \widetilde{\mathbb T}_{(-k)} \, \partial_v^{(k+2)}B_{(0)} \, ,
\end{split}
\end{equation} 
which then leads us to the following form of the second term in eq.\eqref{theta1}
\begin{equation}\label{theta2a}
\begin{split}
\sum_{k\geq 0}\widetilde{\mathbb T}_{(-k)}  \,  \mathcal{L}_{\xi}  {\mathbb T}_{(k+1)}=\sum_{k\geq 0}\widetilde{\mathbb T}_{(-k)} \left(k+1+v\partial_v\right) {\mathbb T}_{(k+1)} =   \sum_{k\geq 0}\left((k+1)~t^{(k)}_{(1)} + v~t^{(k)}_{(2)}\right) \, .
\end{split}
\end{equation}
Furthermore, from eq.\eqref{theta2ab} we also learn that, in this case, both the terms that are present in $t^{(k)}_{(1)}$ has the general structure as 
\begin{equation} \label{form_tk1}
t^{(k)}_{(1)}\sim X_{(-k)}\, \partial_v^{k+1}Y_{(0)} \, ,
\end{equation}
with the following possibilities 
\begin{equation}
\begin{split}
\text{either}~~ X_{(-k)} \sim \, \widetilde{\mathbb T}_{(-k)} \,,& ~~ Y_{(0)} \sim \, C_{(0)}\, ;~~ 
\text{or} ~~ X_{(-k)} \sim \, A_{(0)} \,\widetilde{\mathbb T}_{(-k)}, ~~ Y_{(0)} \sim \, B_{(0)} \, .
\end{split}
\end{equation}
However, for both the cases mentioned above $t^{(k)}_{(2)}$ must also have the form given by
\begin{equation}\label{form_tk2}
t^{(k)}_{(2)}\sim X_{(-k)}\, \partial_v^{k+2}Y_{(0)}\,. 
\end{equation} 
We should remember that eq.\eqref{key1} is the key equation that we need to show to recast $\Theta^r$ into the desired structure given by eq.\eqref{thf}. In this case, for the choices of $t^{(k)}_{(1)}$ and $t^{(k)}_{(2)}$ given by eq.\eqref{form_tk1} and eq.\eqref{form_tk2} respectively in terms of $X_{(-k)}$ and $Y_{(0)}$, eq.\eqref{key1} results in the following statement: \\
{\it If we apply  Result:\,$1$ to $t^{(k)}_{(1)}$ in eq.\eqref{form_tk1}, it will generate a term of the form $X_{(0)}\,\partial_v Y_{(0)}$ and similarly $t^{(k)}_{(2)}$ in eq.\eqref{form_tk2}, will generate a term of the form $X_{(0)}\,\partial_v^2 Y_{(0)}$. The constant coefficient multiplying $X_{(0)}\partial_v Y_{(0)}$ in $t^{(k)}_{(1)}$ is $(k+1)$ times the constant coefficient multiplying $X_{(0)}\partial_v^2 Y_{(0)}$ in the expansion of $t^{(k)}_{(2)}$, see eq.\eqref{theta2a} .}
\\
The italicized statement written above follows trivially, as we show explicitly below, once we apply Result:\,$1$ to $t^{(k)}_{(1)}$ and $t^{(k)}_{(2)}$ and keep track of the coefficients of these two particular terms. This also recasts the term $\widetilde{\mathbb T}_{(-k)}\, {\cal L}_\xi {\mathbb T}_{(k+1)}$ in the following form, 
\begin{equation}\label{theta2b}
\begin{split}
\widetilde{\mathbb T}_{(-k)}\,&{\cal L}_\xi {\mathbb T}_{(k+1)}\\
=~&(k+1)\left(X_{(-k)} \partial_v^{k+1}Y_{(0)}\right) + v ~\left(X_{(-k)} \partial_v^{k+2}Y_{(0)}\right)\\
=~&(k+1)\bigg[\partial_v\left(\sum_{m=0}^{k-1}(-1)^{m}X_{(-k+m)}\partial_v^{k-m}Y_{(0)}\right) + (-1)^k X_{(0)}\partial_vY_{(0)}\bigg]\\
&+v~\partial_v^2\left(\sum_{m=0}^{k-1}(-1)^{m}(m+1)X_{(-k+m)}\partial_v^{k-m}Y_{(0)}\right)\\
&+v ~(-1)^k(k+1)\left(X_{(0)}\partial_v^2 Y_{(0)}\right)\\
=~&(k+1)\bigg[\partial_v\left(\sum_{m=0}^{k-1}(-1)^{m}X_{(-k+m)}\partial_v^{k-m}Y_{(0)}\right) + (-1)^k X_{(0)}\partial_vY_{(0)}\bigg]\\
&+v~\partial_v^2\left(\sum_{m=0}^{k-1}(-1)^{m}\left[(k+1)-(k-m)\right]X_{(-k+m)}\partial_v^{k-m}Y_{(0)}\right)\\
&+v ~(-1)^k(k+1)\left(X_{(0)}\partial_v^2 Y_{(0)}\right)\\
=~&(1+v\partial_v)\bigg[(k+1)\left(\sum_{m=0}^{k-1}(-1)^{m}\partial_v\left[X_{(-k+m)}\partial_v^{k-m}Y_{(0)}\right]+ (-1)^k X_{(0)}\partial_vY_{(0)}\right) \bigg]\\
&-v~\partial_v^2\left(\sum_{m=0}^{k-1}(-1)^{m}(k-m)X_{(-k+m)}\partial_v^{k-m}Y_{(0)}\right)+ {\cal O}\left(\epsilon^2\right) \, .
\end{split}
\end{equation}
In the last line we have simply rearranged the terms and used the fact that
$$X_{(0)}\,\partial_v^2Y_{(0)}= \partial_v\left(X_{(0)} \, \partial_v Y_{(0)}\right) + {\cal O}\left(\epsilon^2\right) \, .$$
Next, we substitute eq.\eqref{theta2b} in eq.\eqref{theta1} and obtain the following form for the second term in $\Theta^r$ eq.\eqref{schemetheta},
\begin{equation}\label{theta1_final}
\begin{split}
\sum_{j} {\cal T}^{r \alpha_1 \alpha_2\cdots \alpha_j} \, {\cal L}_{\xi} {\cal S}_{\alpha_1 \alpha_2\cdots \alpha_j} =~ & \left(1 + v\partial_v\right){\mathbb S}_{(1)}+\sum_{k\geq 0}\widetilde{\mathbb T}_{(-k)}~{\cal L}_\xi {\mathbb T}_{(k+1)} + {\cal O}\left(\epsilon^2\right)\\
=~& (1 + v~\partial_v){\cal M}_{(1)} + v~\partial_v^2 {\cal N}_{(0)}+ {\cal O}\left(\epsilon^2\right) \, ,
 \end{split}
\end{equation}
such that 
\begin{equation}\label{theta1_final_1}
\begin{split}
{\cal M}_{(1)} =& \, {\cal S}_{(1)}+\sum_{k\geq 0} (k+1)\left(\sum_{m=0}^{k-1}(-1)^{m}\partial_v\left[X_{(-k+m)}\partial_v^{k-m}Y_{(0)}\right]+ (-1)^k X_{(0)}\partial_vY_{(0)}\right) \, , \\
\text{and}~~{\cal N}_{(0)} =& \sum_{k\geq 0} \sum_{m=0}^{k-1}(-1)^{m+1}\, (k-m)\,  X_{(-k+m)}\,  \partial_v^{k-m}Y_{(0)} \, .
\end{split}
\end{equation}
It should be noted that the form in eq.\eqref{theta1_final} is consistent with eq.\eqref{thf}. 

Finally, we conclude this sub-section with the following comments regarding the structure of ${\cal N}_{(0)}$ that we have obtained in eq.\eqref{theta1_final_1}. It is evident that ${\cal N}_{(0)}$ is a boost invariant quantity of the type mentioned in eq.\eqref{1st_sq01}. From its expression, we note that it is a product of two terms that are not individually boost invariant, i.e. 
\begin{equation} \label{N_JKM}
{\cal N}_{(0)} \, \sim \, X_{(-k+m)}\,  \partial_v^{k-m}Y_{(0)} \, ,
\end{equation}
where $X_{(-k+m)}$ has boost weight of $(-k+m)$ but $ \partial_v^{k-m}Y_{(0)}$ has a boost weight of $(k-m)$. Following our guidelines listed in Table-\ref{table:sum_dv_ep} (see Rule:\,$2$) we can conclude that ${\cal N}_{(0)}$ is evaluated to be linear in the amplitude of the dynamical fluctuations, i.e. ${\cal N}_{(0)} \, \sim \, \mathcal{O}(\epsilon)$. Furthermore, using Rule:\,$1$ from the same Table-\ref{table:sum_dv_ep} we also learn that ${\cal N}_{(0)}$ will vanish when the stationarity limit is taken by turning the dynamics off. From these properties it is evident that ${\cal N}_{(0)}$ is a possible candidate to generate the JKM ambiguities associated with the Wald entropy, which will be used in our subsequent analysis.

\subsubsection*{Analysis of the first term in $\Theta^r$ from  eq.(\ref{schemetheta}):} \label{subThetaR2}
Next, we consider, for the $r$ component of $\Theta^\mu$, the first term on the RHS of eq.(\ref{schemetheta}) 
\begin{equation}\label{1stterm1}
\begin{split}
&2 \, E^{r\nu\rho\sigma}_R \, D_{\sigma}\left[{\cal L}_\xi g_{\nu\rho}\right]  = 2 \, D_{\sigma} \left(E^{r\nu\rho\sigma}_R \, {\cal L}_\xi  g_{\nu\rho}\right) - 2 \, D_{\sigma} E^{r\nu\rho\sigma}_R \, {\cal L}_\xi  g_{\nu\rho} \, .
\end{split}
\end{equation}
Now the second term on the RHS of the above equation has the same structure as that of the second term in eq.\eqref{schemetheta}. Therefore one can do a similar analysis for that term following the steps mentioned in \S\ref{subThetaR1}. 

\subparagraph{For the first term in eq.\eqref{1stterm1}:} In the following, we will only process the first term in  eq.\eqref{1stterm1}. 

Let us define the following quantity 
 \begin{equation}
 H^{\mu  \sigma}  \equiv E^{\mu\nu\rho\sigma}_R \, {\cal L}_\xi g_{\nu\rho}\, ,
 \end{equation}
and using the symmetries of $E^{\mu\nu\rho\sigma}_R$ under exchanging its indices, we note that $H^{\mu  \sigma}$ is a symmetric tensor, 
 \begin{equation}\label{chkstm}
 \begin{split}
H^{\sigma\mu} = E^{\sigma \nu \rho \mu}_R \, {\cal L}_\xi g_{\nu\rho} =E^{\sigma \rho \nu \mu}_R \,{\cal L}_\xi g_{\rho \nu} =E^{\nu \mu \sigma \rho}_R \,{\cal L}_\xi g_{\nu\rho} =E^{\mu \nu \rho \sigma}_R \, {\cal L}_\xi g_{\nu\rho} =H^{\mu\sigma}\, . 
 \end{split}
 \end{equation}
Next, we calculate $D_{\sigma} H^{r\sigma}$ and evaluate that at the horizon to obtain the following expression\footnote{We have used the following formula to compute the divergence of a symmetric tensor 
$$D_{\sigma} H^{\sigma}_{\mu} = {1\over\sqrt{g}}\partial_{\sigma}\left[\sqrt{g} H^{\sigma}_{\mu}\right] -{1\over 2} H^{\sigma \nu} \partial_{\mu}g_{\sigma \nu}$$}
\begin{equation}\label{1stterm2}
\begin{split}
D_{\sigma}&H^{r\sigma}\big\vert_{r=0}=g^{r\mu}D_{\sigma}H^{\sigma}_{\mu}\big\vert_{r=0} = D_{\sigma}H^{\sigma}_v\big\vert_{r=0}\\
&=\left[ {1\over\sqrt{g}}\partial_{\sigma}\left[\sqrt{g} H^{\sigma}_v\right] -{1\over 2}H^{\mu\rho}\partial_vg_{\mu\rho}\right]_{r=0}\\
&=\left[ {1\over\sqrt{g}}\partial_r\left[\sqrt{g} H^r_v\right] +{1\over\sqrt{g}}\partial_v\left[\sqrt{g} H^{v}_v\right] + {1 \over \sqrt{g}}\partial_i \left[\sqrt{g} H^i_v\right] -{1\over 2}H^{ij}\partial_vg_{ij}\right]_{r=0}\\
&=\left[ {1\over\sqrt{g}}\partial_r\left[\sqrt{g} H^{r\mu}g_{\mu v}\right] + {1\over\sqrt{g}}\partial_v\left[\sqrt{g} H^{v \rho}g_{\rho v}\right] +{1\over\sqrt{g}}\partial_i\left[\sqrt{g} H^{i \rho}g_{\rho v}\right] -{1\over 2}H^{ij}\partial_vg_{ij}\right]_{r=0}\\
&=\left[ {1\over\sqrt{g}}\partial_r\left[\sqrt{g} H^{rr}\right] +\omega_i H^{ri}+  {1\over\sqrt{g}}\partial_v\left[\sqrt{g} H^{v r}\right] + {1\over\sqrt{g}}\partial_i\left[\sqrt{g} H^{i r}\right] -{1\over 2}H^{ij}\partial_vg_{ij}\right]_{r=0} \, .
\end{split}
\end{equation}
To proceed from here we need to compute different components of $H^{\mu\sigma}$ for our metric gauge and evaluate that at the horizon. First we define $\Delta_{\mu\nu} \equiv {\cal L}_\xi g_{\mu\nu}$, and compute its non-zero components as given below
\begin{equation}\label{nonzerdelta}
\begin{split}
\Delta_{vv} = r^2(v\partial_v -r\partial_r) X,~~\Delta_{vi} = r(v\partial_v -r\partial_r) \omega_i,~~\Delta_{ij} = (v\partial_v -r\partial_r)h_{ij} \, .
\end{split}
\end{equation}
This enables us to obtain the following expressions for different components of $H^{\mu\sigma}$ 
\begin{equation}\label{hcomp}
\begin{split}
H^{rr} &=E^{rvvr}_R \, \Delta_{vv} + E^{rvir}_R \, \Delta_{vi} + E^{rijr}_R \, \Delta_{ij} =r\,v\, E^{rvir}_R\, \partial_v \omega_i + E^{rijr}_R (v\partial_v -r\partial_r) h_{ij} + {\cal O}\left(r^2\right)\\
H^{rv}&=H^{vr}= E^{rvjv}_R \, \Delta_{vj} + E^{rjkv}_R \, \Delta_{jk}=v \, E^{rjkv}_R \,  \partial_vh_{jk} + {\cal O}(r)  \\
H^{ri} &= H^{ir} =E^{rvvi}_R\Delta_{vv} + 2 E^{rvji}_R \, \Delta_{vj} + E^{rjki}_R\Delta_{jk}  = v \,  E^{rjki}_R\,  \partial_v h_{jk} + {\cal O}(r) \\
H^{ij} &= E^{ivvj}_R \, \Delta_{vv} + 2E^{ivkj}_R \, \Delta_{vk} + E^{ikk'j}_R \, \Delta_{kk'} =  v \, E^{ikk'j}_R \, \partial_vh_{kk'} + {\cal O} \left(r\right)
\end{split}
\end{equation}
Remembering the rule of assigning boost weights, we note that $H^{ri}\vert_{r=0} ={\cal O}(\epsilon^2)$ and $H^{ij}\vert_{r=0} = {\cal O}\left(\epsilon\right)$, and, consequently we conclude $H^{ij}\partial_vg_{ij}= H^{ij}\partial_v h_{ij} = {\cal O}(\epsilon^2)$. Implementing all of these results in eq.\eqref{1stterm2} we are left with 
\begin{equation} \label{1stterm3}
D_{\sigma} H^{r\sigma} = \left[ {1\over\sqrt{g}}\partial_r\left[\sqrt{g} H^{rr}\right] +{1\over\sqrt{g}}\partial_v\left[\sqrt{g} H^{v r}\right]
 \right]_{r=0} + {\cal O}\left(\epsilon^2\right) \, ,
\end{equation}
with $H^{rr}$, $H^{vr}$ and $H^{ir}$ being given in eq.\eqref{hcomp}. 
Next we evaluate the terms on the RHS of eq.\eqref{1stterm3}. The first term can be manipulated using eq.\eqref{hcomp} as
\begin{equation}\label{termh}
\begin{split}
{1\over\sqrt{g}}\partial_r\left[\sqrt{g} H^{rr}\right]\big\vert_{r=0}&=  E^{rijr}_R\, (v\partial_r\partial_v -\partial_r) h_{ij} + {\cal O}\left(\epsilon^2\right)= E^{rijr}_R\, (-1 +v\partial_v)(\partial_r h_{ij}) + {\cal O}\left(\epsilon^2\right)\\
&= \left[ E^{rijr}_R\, {\cal L}_\xi(\partial_rh_{ij})\right]_{r=0} + {\cal O}\left(\epsilon^2\right)\\
\end{split}
\end{equation}
It should be noted that $\partial_r h_{ij}$ cannot be identified as a component of any covariant tensor constructed out of metric. However, it has a definite boost weight $(=-1)$ and therefore, the action of ${\cal L}_\xi$, once it is expressed in terms of boost weight and $(\xi\cdot\partial)$ as in eq.\eqref{Lxi1}, is well-defined. We have used this in arriving at the last equality above in eq.\eqref{termh}. 

We can clearly see from the RHS of eq.\eqref{termh} that it has the same schematic structure as the second term in equation eq.\eqref{schemetheta}, analyzed in previous subsection \S\ref{subThetaR1}. Though $(\partial_r h_{ij})$ is not a covariant tensor component, the analysis of the previous subsection will go through here since it uses only the boost weight defined through the eq.\eqref{Lxi1}. Therefore following the same analysis as in \S\ref{subThetaR1} we get
\begin{equation}\label{termh1}
\begin{split}
{1\over\sqrt{g}}\partial_r\left(\sqrt{g} H^{rr}\right)\big\vert_{r=0}  =~ & E^{rijr}_R\, {\cal L}_\xi(\partial_rh_{ij})\big\vert_{r=0}\\
=~& (1 + v\,\partial_v){\cal C}_{(1)} + v\,\partial_v^2 {\cal D}_{(0)}+ {\cal O}\left(\epsilon^2\right) \, ,
 \end{split}
\end{equation}
where ${\cal C}_{(1)}$ is a tensor quantity with boost weight $(+1)$ and ${\cal D}_{(0)}$ is a boost invariant quantity. 

For the second term in eq.\eqref{1stterm3}, using the corresponding relation from eq.\eqref{hcomp}, we obtain the following form 
\begin{equation}\label{termh2}
\begin{split}
{1\over \sqrt{g}}\partial_v\left(\sqrt{g}\, H^{rv}\right)&= {1\over \sqrt{g}}\partial_v\left[\sqrt{g}  \,v\, E^{rjkv}_R\,\partial_v h_{jk}\right]= E^{rjkv}_R\,\partial_v h_{jk}+v\left({1\over \sqrt{h}}\partial_v\left[\sqrt{h} \, E^{rjkv}_R\,\partial_v h_{jk}\right]\right)\\
&=(1+v \, \partial_v){\cal E}_{(1)} +{\cal O}(\epsilon^2) \, ,
\end{split}
\end{equation}
where 
\begin{equation}\label{termh2a}
\begin{split}
{\cal E}_{(1)} =& E^{rjkv}_R\,\partial_v h_{jk} \, .
\end{split}
\end{equation}
\subparagraph{For the second term in  eq.\eqref{1stterm1}:} As we have mentioned before the second term on the RHS of eq.\eqref{1stterm1} is already in the form of the second term in eq.\eqref{schemetheta} which has been analyzed in detail in the previous sub-section \S\ref{subThetaR1} \, . Therefore following a similar method we can obtain the following form for this term as shown below
\begin{equation}\label{termE2}
\begin{split}
- 2 \, D_{\sigma} E^{r\nu\rho\sigma}_R \, {\cal L}_\xi  g_{\nu\rho}\, \big\vert_{r=0} 
=~ (1 + v \, \partial_v){\cal F}_{(1)} + v~\partial_v^2 {\cal G}_{(0)}+ {\cal O}\left(\epsilon^2\right) \, ,
 \end{split}
\end{equation}
where again, ${\cal F}_{(1)}$ is a tensor quantity with boost weight $(+1)$ and ${\cal G}_{(0)}$ is a boost invariant quantity.

Finally, collecting all the results derived in parts so far from eq.\eqref{termh1}, eq.\eqref{termh2} and eq.\eqref{termE2}, we obtain the expression for the first term in eq.(\ref{schemetheta}) as
\begin{equation}\label{termE3}
\begin{split}
2 \, E^{r\nu\rho\sigma}_R \, D_{\sigma}\left[{\cal L}_\xi g_{\nu\rho}\right]\, \big\vert_{r=0} 
=~ (1 + v \, \partial_v){\cal P}_{(1)} + v~\partial_v^2 {\cal Q}_{(0)}+ {\cal O}\left(\epsilon^2\right) \, ,\\
\text{such that} ~~ {\cal P}_{(1)} = \, {\cal C}_{(1)}+{\cal E}_{(1)}+{\cal F}_{(1)} \, , ~~\text{and}~~{\cal Q}_{(0)} =\, {\cal D}_{(0)} +{\cal G}_{(0)} \, .
 \end{split}
\end{equation}
which is, again, consistent with the expected form of eq.\eqref{thf}.

With all these results at our disposal, we are in a position to collect them together in eq.\eqref{schemetheta} for $\Theta^r$ as predicted in eq.\eqref{thf}, to write down
\begin{equation}\label{termfinal1}
\begin{split}
\Theta^r \, \big\vert_{r=0} 
=~ (1 + v \, \partial_v){\cal A}_{(1)} + v~\partial_v^2 {\cal B}_{(0)}+ {\cal O}\left(\epsilon^2\right) \, , \\
\text{such that} ~~ {\cal A}_{(1)} = \, {\cal M}_{(1)} +  {\cal P}_{(1)}\, , ~~\text{and}~~{\cal B}_{(0)} =\,{\cal N}_{(0)} +{\cal Q}_{(0)} \, .
 \end{split}
\end{equation}

Before we conclude, let us make the following comment regarding ${\cal B}_{(0)}$. The term ${\cal Q}_{(0)}$ in eq.\eqref{termE3} has the same origin as the term ${\cal N}_{(0)}$ in eq.\eqref{theta1_final_1}. To be more precise, both of them are produced from the first term on the RHS of eq.\eqref{sq01}, i.e. terms of the kind as given in eq.\eqref{1st_sq01}. Therefore following the discussion around eq.\eqref{N_JKM}, we can argue that 
${\cal B}_{(0)}$ is a boost invariant term, although it is product of two terms who are not individually boost invariant. Such terms will vanish in stationary configurations and are non-zero only because of dynamical fluctuations. Using the summary listed in Table-\ref{table:sum_dv_ep} we can also argue that 
\begin{equation} \label{B0_ep}
{\cal B}_{(0)} ~\sim ~\mathcal{O}(\epsilon)
\end{equation}
 As we will see later ${\cal B}_{(0)}$ will, therefore, contribute to the JKM ambiguities of the Wald entropy. 

\subsubsection{Analyzing the structure of the second term (i.e. $D_\mu Q^{r\mu }$) in $E_{vv}$} \label{subsec:DQrmu}
Next we turn our focus to the second term on the RHS of eq.\eqref{keyeqn_again} in the expression of $E_{vv}$. This term involves the quantity $Q^{\mu\nu}$, which was defined in eq.\eqref{hijibiji-3}. This $Q^{\mu\nu}$ is the Noether charge corresponding to diffeomorphism invariance. The general form for this $Q^{\mu\nu}$ in a diffeomorphism invariant theory of gravity was constructed in \cite{Iyer:1994ys}. Following the Proposition $4.1$ in \cite{Iyer:1994ys} 
(also see eq.\eqref{app_Q_1} and eq.\eqref{app_Q_2} in Appendix-(\ref{app:ReviewIyerWald})), 
we can obtain this as given below
\begin{equation}\label{expr_Q_ab_0}
Q^{\mu\nu} = W^{\mu\nu\rho}  \,  \xi_\rho - 2E^{\mu\nu\alpha\beta}_R  \,  D_{[\alpha}\xi_{\beta]} + Y^{\mu\nu}  + D_{\alpha}Z^{\mu\nu\alpha}  \, ,
\end{equation}
where we have suppressed the functional dependence of the various quantities, e.g. $W^{\mu\nu\rho}(g_{\beta_1 \beta_2},\phi)$, etc. and  $E^{\mu\nu\rho\sigma}_R$ is already defined in eq.\eqref{def_E_abcd}. It should be noted that $Y^{\mu\nu}$ and $Z^{\mu\nu\alpha}$ are the possible ambiguities in the definition of the Noether charge. They are also present in the definition of $\theta^\mu$, but, interestingly, get cancelled in the combination $(-\Theta^\mu + D_\nu Q^{\nu\mu})$ which is important for our calculations, see eq.\eqref{keyeqn_again}. Therefore, in whatever follows we will neglect them, focussing only on  
\begin{equation} \label{expr_Q_ab}
Q^{\mu\nu} = W^{\mu\nu\rho}  \,  \xi_\rho - 2E^{\mu\nu\alpha\beta}_R  \,  D_{[\alpha}\xi_{\beta]}  \, .
\end{equation}
It should be noted that the two terms present on the RHS above are linear either in $\xi^\mu$ or in $D_{\alpha}\xi_{\beta}$. Therefore, following the same argument as in the case of $\Theta^r$, we could decompose $Q^{r\mu}$, by using $\xi^\mu$ from eq.\eqref{killingvector} and explicitly working out the $v$ dependence as
\begin{equation} \label{Qrmu1}
Q^{r\mu} = \widetilde{Q}^{r\mu} + v\, W^{r\mu}_v \, , 
\end{equation}
where $\widetilde{Q}^{r\mu}$ and $W^{r\mu}_v$ have no explicit $v$ dependence\footnote{Since $Q^{\nu\mu}$ is anti-symmetric in its two indices, for $Q^{r\mu}$ we should note that possible entries for $\mu$ are $\mu = v, i, j, k \cdots $, except $r$.}. In the process of deriving eq.\eqref{Qrmu1} from eq.\eqref{expr_Q_ab}, we can also obtain the following expression for $\widetilde{Q}^{r\mu}$
\begin{equation} \label{Qrv1}
\widetilde{Q}^{rv} = 4 \, E^{rvrv}_R \, ,
\end{equation}
from which it can be concluded that $\widetilde{Q}^{r\mu}$ actually reproduces the equilibrium Wald entropy density. This will be important later in our analysis. 
 
Following our rules for determining boost weight that we have already declared in Table-\ref{table:sum_bas_setup} using index counting, $Q^{rv}$ is boost invariant but $Q^{ri}$ has boost weight $+1$. Next, substituting this decomposition for $Q^{r\mu}$ from eq.\eqref{Qrmu1} in the expression of $D_\mu Q^{r\mu }$ we find
\begin{equation}\label{Q1}
\begin{split}
D_\mu Q^{r\mu} =~& D_\mu \widetilde{Q}^{r\mu} + W^{rv}_v + v ~D_\mu W^{r\mu}_v\\
=~&{1\over \sqrt{h}}\partial_v\left(\sqrt{h}~\widetilde{Q}^{rv}\right) + \nabla_i \widetilde{Q}^{ri} +W^{rv}_v + v\left[{1\over \sqrt{h}}\partial_v\left(\sqrt{h}~W^{rv}_v \right) +\nabla_iW^{ri}_v\right] \, .
\end{split}
\end{equation}
Let us now process the last term on the RHS above. It should be noted that  $W^{rv}_v$ has boost weight $(+1)$ and therefore must be of order ${\cal O}(\epsilon)$. It follows that
\begin{equation} \label{DWrvv}
{1\over \sqrt{h}}\partial_v\left(\sqrt{h}~W^{rv}_v \right) = \partial_vW^{rv}_v + {\cal O}(\epsilon^2) \, .
\end{equation}
Following the same logic, we learn that $W^{ri}_v$ is a term of boost weight $(+2)$. Therefore, applying Result:\,$1$ from eq.\eqref{sq01} it could be written as\footnote{We are not careful with the coefficients dictated by Result:\,$1$.}
\begin{equation} \label{Wriv}
W^{ri}_v =\partial_v J^i_{(1)} + {\cal O}(\epsilon^2) \, 
\end{equation}
such that $J^i_{(1)}$ is some spatial vector with boost weight $(+1)$. 

It should be noted here that the form of $W^{ri}_v$ written above is a simple representation of its possible structure. For an arbitrary covariant tensor with boost weight equal to $+2$, we will generally obtain a much more complicated structure. However, here our purpose is to justify the existence of such a term in general. We should also point out that the first term on the RHS of Result:\,$1$ (in eq.\eqref{sq01}) will always generate a term like $\partial_v J^i_{(1)}$ and there is no obstruction to that in principle. As we will see later in this paper, this particular term will be crucial to give us the spatial components of an entropy current. Having said this, we should also mention that one might come up with a particular theory of gravity where such a term is not generated. Einstein's two derivative gravity or $R^2$ theory of higher derivative gravity are such examples where $J^i_{(1)}$ turns out to be zero. 

From eq.\eqref{Wriv} we can then derive
\begin{equation}\label{DWriv}
\nabla_iW^{ri}_v= \partial_v\left(\nabla_i J^i_{(1)}\right)+ {\cal O}(\epsilon^2)  \, .
\end{equation}
Finally, substituting eq.\eqref{DWrvv} and eq.\eqref{DWriv} in equation eq.\eqref{Q1} we arrive at the following expression for $D_\mu Q^{r\mu}$ 
\begin{equation}\label{Q2}
\begin{split}
D_\mu Q^{r\mu} = {1\over \sqrt{h}}\partial_v\left(\sqrt{h}\,\widetilde{Q}^{rv}\right) + \nabla_i \widetilde{Q}^{ri} +  v\,\partial_v \left(\nabla_iJ^i_{(1)}\right) +(1+v\,\partial_v) W^{rv}_v + {\cal O}(\epsilon^2) \, .
\end{split}
\end{equation}
For our purpose, we do not need to process this term any further.
\subsubsection{Final analysis for the structure of $E_{vv}$} \label{final_Evv}
In the previous sub-sections, we have obtained the expressions for both $\Theta^r$ and $D_\mu Q^{r\mu}$. For convenience, let us rewrite the final structures of them here again,
\begin{equation}\label{finstr}
\begin{split}
\Theta^r =& \, (1 + v~\partial_v){\cal A}_{(1)} + v~\partial_v^2 {\cal B}_{(0)}+ {\cal O}(\epsilon^2) \, ,
~ \text{and}\, , ~\\
D_\mu Q^{r\mu} =& {1\over \sqrt{h}}\partial_v\left(\sqrt{h}\,\widetilde{Q}^{rv}\right) + \nabla_i \widetilde{Q}^{ri} +  v~\partial_v \left(\nabla_iJ^i_{(1)}\right) +(1+v~\partial_v) W^{rv}_v + {\cal O}(\epsilon^2) \, ,
\end{split}
\end{equation}
where ${\cal A}_{(1)}$ and ${\cal B}_{(0)}$ are given in eq.\eqref{termfinal1}.
Next we collect them together and substitute on the RHS of eq.\eqref{keyeqn_again} to obtain 
\begin{equation}\label{Key2}
\begin{split}
2  \, v \, E_{vv} \big\vert_{r=0}=&\left( -\Theta^r  + D_\mu Q^{r\mu}\right)\big\vert_{r=0}\\
=&  -{\cal A}_{(1)}  +{1\over \sqrt{h}}\partial_v\left(\sqrt{h} \, \widetilde{Q}^{rv}\right) + \nabla_i \widetilde{Q}^{ri} + W^{rv}_v\\
&+v\,\partial_v \bigg[ -{\cal A}_{(1)}  +W^{rv}_v+ \nabla_iJ^i_{(1)}-\partial_v {\cal B}_{(0)}  \bigg] + {\cal O}(\epsilon^2)\, .
\end{split}
\end{equation}
Since the dependence on $v$ is explicit in the equation above, we can compare the coefficients of $v^0$ and $v$ on both sides of equation eq.\eqref{Key2} to find the following two relations
\begin{equation}\label{keyrel}
\begin{split}
 {\cal A}_{(1)} &= {1\over \sqrt{h}}\partial_v\left(\sqrt{h}\,\widetilde{Q}^{rv}\right) + \nabla_i \widetilde{Q}^{ri} + W^{rv}_v\, ,\\
2E_{vv} &= \partial_v \bigg[ -{\cal A}_{(1)}  +W^{rv}_v+ \nabla_iJ^i_{(1)} -\partial_v {\cal B}_{(0)}  \bigg] \, .
\end{split}
\end{equation}
Substituting ${\cal A}_{(1)}$ from the first relation into the second one of eq.\eqref{keyrel}, we find
\begin{equation}\label{keyrel2}
\begin{split}
2E_{vv}= \partial_v \bigg[-\left({1\over \sqrt{h}}\partial_v\left(\sqrt{h}\,\widetilde{Q}^{rv}\right) + \nabla_i \widetilde{Q}^{ri}\right)+\nabla_iJ^i_{(1)}-\partial_v {\cal B}_{(0)} \bigg]+ {\cal O}(\epsilon^2) \, .
\end{split}
\end{equation}
As it was explained after eq.\eqref{termfinal1}, we know that ${\cal B}_{(0)}$ is of boost weight zero and it is always of order ${\cal O}(\epsilon)$,  see eq.\eqref{B0_ep}. Therefore, we learn that 
$$\partial_v^2{\cal B}_{(0)} = {1\over\sqrt h}\partial_v\left[\sqrt{h}~ \partial_v{\cal B}_{(0)} \right] + {\cal O}\left(\epsilon^2\right)\, .$$ Substituting this in the expression of $E_{vv}$ in eq.\eqref{keyrel2}, we find
\begin{equation}\label{keyrel3}
2E_{vv} \big\vert_{r=0}= -\partial_v\bigg({1\over\sqrt{h}}\partial_v\left[\sqrt{h} \, \left(\widetilde{Q}^{rv} +  {\cal B}_{(0)}\right)\right]+\nabla_i\left[\widetilde{Q}^{ri} - J^i_{(1)}\right]\bigg) + {\cal O}\left(\epsilon^2\right)\, .
\end{equation}
This constitutes the main result of our paper. We have established that the general structure of $E_{vv}$ as given above in eq.\eqref{keyrel3} exactly matches with required expression mentioned in eq.\eqref{eq:main1}. By comparing them we obtain the following components of the entropy current
\begin{equation} \label{keyrel3_JvJi}
\begin{split}
{\cal J}^v = & - \dfrac{1}{2} \left(\widetilde{Q}^{rv} +  {\cal B}_{(0)}\right) \, ,~~\text{and} ~~{\cal J}^i = - \dfrac{1}{2}\left(\widetilde{Q}^{ri} - J^i_{(1)} \right)\, .
\end{split}
\end{equation}
In other words, we have also constructed an entropy current on the horizon, with components given by ${\cal J}^v$ and ${\cal J}^i$ for an arbitrary higher derivative theory of gravity. 

We also note that $\widetilde{Q}_{rv}$, as worked out in eq.\eqref{Qrv1}, is the Wald entropy density. Additionally, as argued immediately after eq.\eqref{termfinal1}, ${\cal B}_{(0)}$ will contribute to the JKM ambiguities that are associated with Wald entropy, and it will vanish on stationary horizons. Therefore, eq.\eqref{keyrel3} also highlights how these ambiguous terms are fixed in order to argue for a linearized second law, which was already pointed out in \cite{Wall:2015raa}.

\section{Verification of the proof for specific examples} \label{sec:check_proof}
In section \S\ref{sec:proof} we have constructed a proof of the existence of an entropy current in arbitrary higher derivative theories of gravity. In the following, we would like to verify that the expression of the entropy current suggested by the abstract proof, as given in eq.\eqref{keyrel3}, indeed matches with the explicit calculation of the same in some specific model of higher derivative gravity, namely the four derivative theories. The entropy currents for four derivative theories of gravity have already been worked out in \citep{Bhattacharya:2019qal}. By comparing them with our result reported in this paper, we would, therefore, have a conclusive check of the entropy currents derived via general Noetherian principles and the boost symmetry of the horizon. For the reader's convenience, we summarize the main equations that we would like to explicitly check. The Key equation that we work with is the following:
\begin{equation}
    2v \, E_{vv}|_{r=0} = [ -\Theta^r + D_{\mu} Q^{r \mu} ] |_{r=0} \, ,
\end{equation}
where $\Theta^r$ has been argued to have the following form, see eq.\eqref{termfinal1}, 
\begin{equation}
    \Theta^r = (1 + v ~\partial_v) \mathcal{A}_{(0)} + v ~ \partial^2_v \mathcal{B}_{(0)} \, .
\end{equation}
The Noether charge has the following form eq.\eqref{Qrmu1}: 
\begin{equation}
    Q^{r\mu} = \widetilde{Q}^{r\mu} + v~W^{r\mu}_v
\end{equation}
Finally we have the expression for the entropy currents from $E_{vv}$ eq.\eqref{keyrel3}
\begin{equation}
    2E_{vv} = -\partial_v\bigg({1\over\sqrt{h}}\partial_v\left[\sqrt{h} \,  (\widetilde{Q}^{rv} + {\cal B}_{(0)})\right]+\nabla_i\left[\widetilde{Q}^{ri} - J^i_{(1)})\right]\bigg) + {\cal O}\left(\epsilon^2\right)\, .
\end{equation}
From Proposition 4.1 of \cite{Iyer:1994ys}, we have the following general from for the Noether charge \footnote{We have chosen not to consider the ambiguities of the Noether charge here.}
\begin{equation}
    Q^{\mu\nu} = W^{\mu\nu\alpha} \xi_{\alpha} - 2 E^{\mu\nu\alpha\beta}_R D_{[\alpha} \xi_{\beta]} 
\end{equation}
If we evaluate the above expression in our gauge eq.\eqref{met1}, which we note below for convenience
\begin{equation}
\begin{split}
ds^2 = 2 \,  dv \, dr - r^2 \, X(r,v,x)~dv^2 + 2 r \, \omega_i(r,v,x) \, dv \, dx^i + h_{ij}(r,v,x) \, dx^i \, dx^j \, . 
\end{split}
\end{equation}
and using the inverse metric components 
\begin{equation}
    \begin{split}
        g^{rr} = r^2 X(r,v,x^i) + r^2 \omega^i (r,v,x^i) \omega_i (r,v,x^i) = r^2 X + r^2 \omega^2 \\
        g^{ri} = - r \omega^i \hspace{0.8cm} g^{rv} = 1 \hspace{0.8cm} g^{ij} = h^{ij} \hspace{0.8cm} g^{vv} = g^{vi} =0
    \end{split}
\end{equation}
we find the following quantities
\begin{equation}\label{Qmuquantities}
    \begin{split}
        \widetilde{Q}^{rv} = 4 E^{rvrv}_R \, , \hspace{0.7cm} \widetilde{Q}^{ri} = 4 E^{rirv}_R \, ,  \\
        W^{ri}_v = - 2 E^{rirj}_R \omega_j + W^{rir} = \partial_v J^i_{(1)} \, . 
    \end{split}
\end{equation}
We also define the following quantities related to the extrinsic curvature of the horizon, as the final results have these structures:
\begin{equation}
    \begin{split}
        K_{ij} &= \dfrac{1}{2} \partial_v h_{ij} \hspace{0.8cm} K^{ij} = -\dfrac{1}{2} \partial_v h^{ij} \hspace{0.8cm} \Bar{K}_{ij} = \dfrac{1}{2} \partial_r h_{ij} \hspace{0.8cm} \Bar{K}^{ij} = - \dfrac{1}{2} \partial_r h^{ij} \\
        K &= \dfrac{1}{2} h^{ij} \partial_v h_{ij} = \dfrac{1}{\sqrt{h}} \partial_v \sqrt{h} \hspace{1.8cm} \Bar{K} = \dfrac{1}{2}h^{ij} \partial_r h_{ij} = \dfrac{1}{\sqrt{h}} \partial_r \sqrt{h}
    \end{split}
\end{equation}
The results of the explicit calculations of the Noether charge $Q^{\mu\nu}$ and $\Theta^{\mu}$ for examples considered in \citep{Bhattacharya:2019qal} are given in Tables \ref{table:symplecticpotential},\ref{table:noethercharge}.
\begin{table} [h!]
\centering
 \begin{tabular}{||c || m{6cm} || m{7cm} ||} 
 \hline \hline
 1 & \makecell{Einstein's theory: \\ $L=R$} & \makecell{$\Theta^{\mu} = g^{\mu\alpha}D^\sigma \delta g_{\alpha\sigma} - g^{\rho\sigma}D^\mu \delta g_{\rho\sigma}$} \\ 
 \hline
 2 & \makecell{ Ricci Scalar Squared Theory: \\ $L=R^2$} & \makecell{$\Theta^{\mu} = 2 \left[ R(g^{\mu\alpha}D^{\nu} \delta g_{\nu\alpha} - g^{\nu\alpha} D^{\mu} \delta g_{\nu\alpha}  ) \right.$ \\  $\quad\quad\left. -g^{\mu\alpha} (D^{\beta} R) \delta g_{\alpha\beta} + (D^{\mu}R) g^{\alpha\beta} \delta g_{\alpha\beta}  \right]$  } \\
 \hline
 3 & \makecell{Ricci Tensor Squared Theory: \\ $L=R_{\mu\nu}R^{\mu\nu} $} & \makecell{$\Theta^{\mu} = 2 g^{\mu\alpha} R^{\sigma\nu} D_{\sigma} \delta g_{\nu\alpha} - 2 (D^{\alpha} R^{\mu\nu}) \delta g_{\nu\alpha}$ \\$ - R^{\alpha\beta} D^{\mu} \delta g_{\alpha\beta} +  (D^{\mu}R^{\alpha\beta})\delta g_{\alpha\beta} $ \\ $\quad\quad - g^{\rho\sigma} R^{\mu\nu} D_{\nu} \delta g_{\rho\sigma} + \dfrac{1}{2} (D^{\mu}R) g^{\alpha\beta} \delta g_{\alpha\beta} $ } \\
 \hline
 4 & \makecell{Riemann Tensor Squared Theory: \\ $L=R_{\mu\nu\rho\sigma}R^{\mu\nu\rho\sigma}$} & \makecell{$\Theta^{\mu} = 4 R^{\mu\rho\nu\sigma} D_{\sigma} \delta g_{\nu\rho} - 4 (D_{\sigma} R^{\mu\nu\rho\sigma}) \delta g_{\nu\rho}$  } \\
 \hline\hline
\end{tabular}
 \caption{Symplectic Potential for Einstein gravity and other four derivative theories of gravity}
\label{table:symplecticpotential}
\end{table}

\begin{table} [h!]
\centering
 \begin{tabular}{||c || m{5.5cm} || m{7.5cm} ||} 
 \hline \hline
 1 & \makecell{Einstein's theory: \\ $L=R$} & \makecell{$Q^{\mu\nu} = ( D^{\nu} \xi^{\mu} - D^{\mu} \xi^{\nu} )$ } \\ 
 \hline
 2 & \makecell{ Ricci Scalar Squared Theory: \\ $L=R^2$} & \makecell{$Q^{\mu\nu} = [4 (D^{\mu}R) g^{\nu\alpha} - 4 (D^{\nu} R) g^{\mu\alpha}]\xi_{\alpha}$ \\ $+ 2 R ( D^{\nu} \xi^{\mu} - D^{\mu} \xi^{\nu})$ } \\
 \hline
 3 & \makecell{Ricci Tensor Squared Theory: \\ $L=R_{\mu\nu}R^{\mu\nu} $} & \makecell{$ Q^{\mu\nu} = \big( 2D^{\mu}R^{\nu\alpha} - 2D^{\nu}R^{\mu\alpha} + (D^{\mu}R) g^{\nu\alpha}$ \\ $\quad\quad- (D^{\nu} R) g^{\mu\alpha} \big)\xi_{\alpha} + 2 R^{\nu\beta} D_{\beta} \xi^{\mu} - 2 R^{\mu\beta} D_{\beta} \xi^{\nu}$ } \\
 \hline
 4 & \makecell{Riemann Tensor Squared Theory: \\ $L=R_{\mu\nu\rho\sigma}R^{\mu\nu\rho\sigma}$} & \makecell{$Q^{\mu\nu} = 4 R^{\mu\nu\alpha\sigma} D_{\sigma}\xi_{\alpha} - 8 (D_{\sigma} R^{\mu\nu\alpha\sigma}) \xi_{\alpha}$ } \\
 \hline\hline
\end{tabular}
\caption{Noether Charges for Einstein gravity and other four derivative theories of gravity}
\label{table:noethercharge}
\end{table}
Using the above expressions of Tables \ref{table:symplecticpotential},\ref{table:noethercharge}, it is straightforward to evaluate $\Theta^r$ and the various quantities of eq.\eqref{Qmuquantities} which are given in Table \ref{table:qmuquantities}.
\begin{table} [h!]
\centering
 \begin{tabular}{||c || c || l ||} 
 \hline \hline
 1 & \makecell{Einstein's theory: \\ $L=R$} & \makecell{$\Theta^r = (1 + v \partial_v) (-2 K)$ \\ $\widetilde{Q}^{rv} = -2 \hspace{0.7cm} \widetilde{Q}^{ri} = 0 \hspace{0.7cm} J^i_{(1)} = 0$ } \\ 
 \hline
 2 & \makecell{ Ricci Scalar Squared Theory: \\ $L=R^2$} & \makecell{$\Theta^r = (1 + v \partial_v )( -4 R K )$ \\ $\widetilde{Q}^{rv} = -4 R \hspace{0.7cm} \widetilde{Q}^{ri} = 0 \hspace{0.7cm} J^i_{(1)} = 0$ } \\
 \hline
 3 & \makecell{Ricci Tensor Squared Theory: \\ $L=R_{\mu\nu}R^{\mu\nu} $} & \makecell{$\Theta^r = (1+ v ~ \partial_v) [ -2 R^{ij}K_{ij} - 2 R_{vr}K + 2 K \partial_v \bar{K}$ \\ $- 2 \partial_v (K \bar{K}) ] + v~\partial^2_v [2 K \bar{K}]$ \\ $\widetilde{Q}^{rv} = -4 R_{rv}$ \\ $\widetilde{Q}^{ri} = h^{ij} \partial_v \omega_j - 2 \nabla_n K^{in} + 2 \nabla^i K + \omega^i K$ \\ $J^i_{(1)} = \omega^i K - h^{ij}\partial_v \omega_j + 2 \nabla_j K^{ij}  $ }  \\
 \hline
 4 & \makecell{Riemann Tensor Squared Theory: \\ $L=R_{\mu\nu\rho\sigma}R^{\mu\nu\rho\sigma}$} & \makecell{$\Theta^r = (1+v~\partial_v) [ 8 R_{vijr} K^{ij} + 8 K_{ij} \partial_v \bar{K}^{ij}$ \\ $- 8 \partial_v (\bar{K}^{ij}K_{ij}) ] + v~\partial^2_v (8 \bar{K}^{ij}K_{ij})$ \\ $\widetilde{Q}^{rv} = 8 R_{rvrv}$ \\ $\widetilde{Q}^{ri} = 4 h^{ij} \partial_v \omega_j + 4 \omega_k K^{ik}$ \\
            $J^i_{(1)} = -4h^{ij}\partial_v\omega_j + 4\omega_j K^{ij} + 8 \nabla_j K^{ij} $} \\
 \hline\hline
\end{tabular}
\caption{Summary of the various quantities needed to compute the Entropy currents}
\label{table:qmuquantities}
\end{table}
From the above results, it is clear that in examples considered, $\Theta^r$ is really of the form given in eq.\eqref{termfinal1}. Using the expressions of Table \ref{table:qmuquantities}, it is straightforward to evaluate the entropy currents $\mathcal{J}^v$,$\mathcal{J}^i$ which are summarized in Table \ref{table:entropycurrents}.
\begin{table} [h!]
\centering
 \begin{tabular}{||c || c || c ||} 
 \hline \hline
 1 & \makecell{Einstein's theory: \\ $L=R$} & \makecell{$\mathcal{J}^v = 1 \hspace{0.7cm} \mathcal{J}^i = 0$} \\ 
 \hline
 2 & \makecell{ Ricci Scalar Squared Theory: \\ $L=R^2$} & \makecell{$\mathcal{J}^v = 2R \hspace{0.7cm} \mathcal{J}^i = 0$} \\
 \hline
 3 & \makecell{Ricci Tensor Squared Theory: \\ $L=R_{\mu\nu}R^{\mu\nu} $} & \makecell{ $ \mathcal{J}^v = 2R_{rv} - K\bar{K} $ \\ $\mathcal{J}^i = -h^{ij}\partial_v \omega_j + 2 \nabla_j K^{ij} - \nabla^i K $ }  \\
 \hline
 4 & \makecell{Riemann Tensor Squared Theory: \\ $L=R_{\mu\nu\rho\sigma}R^{\mu\nu\rho\sigma}$} & \makecell{$ \mathcal{J}^v = -4 R_{rvrv} - 4 \bar{K}^{ij} K_{ij} $ \\ $\mathcal{J}^i = -4 h^{ij}\partial_v \omega_j + 4 \nabla_j K^{ij} $ } \\
 \hline\hline
\end{tabular}
\caption{Entropy currents for Einstein gravity and other four derivative theories of gravity}
\label{table:entropycurrents}
\end{table}

Finally, we provide the expressions for Gauss-Bonnet Theory: $L = R^2 - 4 R_{\mu\nu} R^{\mu\nu} + R_{\mu\nu\rho\sigma} R^{\mu\nu\rho\sigma}$. $\Theta^{\mu}$ is given by
\begin{equation}
    \begin{split}
        \Theta^{\mu} = 2R(g^{\mu\alpha} D^{\sigma}\delta g_{\alpha\sigma} - g^{\rho\sigma} D^{\mu}\delta g_{\rho\sigma}) - 2g^{\mu\nu}(D^{\alpha}R)\delta g_{\alpha\nu} \\ -8 g^{\mu\alpha}R^{\sigma\nu}D_{\sigma} \delta g_{\nu\alpha} + 4(D^{\alpha}R^{\mu\nu}) \delta g_{\nu\alpha} + 4R^{\alpha\beta}D^{\mu}\delta g_{\alpha\beta} \\ + 4g^{\rho\sigma} R^{\mu\nu}D_{\nu}\delta g_{\rho\sigma} + 4R^{\mu\nu\rho\sigma}D_{\sigma} \delta g_{\nu\rho}
    \end{split}
\end{equation}
The Noether charge is given by
\begin{equation}
    Q^{\mu\nu} = 2R(D^{\nu}\xi^{\mu} - D^{\mu}\xi^{\nu}) + 8( R^{\mu\beta}D_{\beta}\xi^{\nu} - R^{\nu\beta} D_{\beta} \xi^{\mu} ) + 4 R^{\mu\nu\alpha\sigma} D_{\sigma} \xi_{\alpha}
\end{equation}
Thus, $\Theta^r$ and the quantities in eq.\eqref{Qmuquantities} are straightforward to compute:
\begin{equation}
    \begin{split}
        \Theta^r &= (1+v \partial_v) [ -4 R K + 8 R^{ij}K_{ij} + 8 R_{rv} K - 8 K \partial_v \bar{K} \\ &+ 8 R_{vijr} K^{ij} + 8 K_{ij} \partial_v \bar{K}^{ij} + 8 \partial_v (K \bar{K} - 8 K_{ij} \bar{K}^{ij}) ] \\
        &+ v \partial^2_v [ -8 K \bar{K} + 8 \bar{K}^{ij} K_{ij} ]
    \end{split}
\end{equation}
\begin{equation}
    \begin{split}
        \widetilde{Q}^{rv} &= - 4R + 16 R_{rv} + 8 R_{rvrv} \\
        \widetilde{Q}^{ri} &= 8 \nabla_j K^{ij} - 8 \nabla^i K - 4 \omega^i K + 4 \omega_j K^{ij} \\
        J^i_{(1)} &= -4 \omega^i K + 4 \omega_j K^{ij}
    \end{split}
\end{equation}
The Entropy currents are therefore given by
\begin{equation}
    \begin{split}
        \mathcal{J}^v &= 2 R - 8R_{rv} - 4 R_{rvrv} + 4 K \bar{K} - 4 \bar{K}^{ij} K_{ij}  = 2 \mathcal{R} \\
        \mathcal{J}^i &= -4 \nabla_j K^{ij} + 4 \nabla^i K
    \end{split}
\end{equation}
where $\mathcal{R}$ is the intrinsic Ricci scalar of the horizon.

These results exactly match with the explicit results of \citep{Bhattacharya:2019qal}. To see this, we must compare the expressions written above for ${\cal J}^v$ and ${\cal J}^i$ as listed in Table-\ref{table:entropycurrents} with section-$3.1$ in \cite{Bhattacharya:2019qal}\footnote{It must be noted that there is an overall negative sign in the equations of motion used in the calculations above ((i.e.) $E_{vv}$ ) when compared to the expressions used in \citep{Bhattacharya:2019qal}. Due to an additional overall negative sign in definitions of the entropy currents, they exactly match with the results provided here.}. Thus, the results in this section are validation of the proof provided above.

\section{A proof of the physical process version of the first law} \label{sec:PPVFL_proof}
So far in this paper, our focus has been to construct an entropy current with the motivation to produce a proof of the linearized version of the second law of thermodynamics for dynamical black holes. However, a similar framework could be used to construct an alternative proof of a particular version of the first law, namely, the physical process version of it, \cite{Jacobson:1995uq, Gao:2001ut, Amsel:2007mh, Bhattacharjee:2014eea, Chakraborty:2017kob, Chatterjee:2011wj, Kolekar:2012tq}. Interestingly, one should note that in \cite{Wall:2015raa} the validity of this version of the first law was actually in a subtle way needed to prove the linearized version of the second law (see section-$(2.2.3)$ in \cite{Bhattacharya:2019qal} for a detailed discussion on this point). 

The purpose of this section is to reorganize the setup that we have already developed in this paper in order to devise an alternative proof of the physical process version of the first law applicable to the most general higher derivative theories of gravity. The main ingredient that we will use to achieve this, is essentially the fact that $\chi^\mu\xi^\nu E_{\mu\nu}$, \footnote{Recall $\chi = \chi^a\partial_a = \partial_v$ is the affinely parametrized null generator of the horizon and $\xi = \xi^\mu\partial_\mu = v\partial_v -r\partial_r$ is the Killing vector field.} could always be expressed in the form of eq.\eqref{eq:main1}. 

Here we are restricting ourselves to a situation where the dynamical fluctuation around a stationary horizon is of small amplitude. It is initiated by a small in-falling matter perturbing a black hole configuration at equilibrium. By small matter, what we mean is more precisely the introduction of a perturbation in the system of equations by a small matter stress tensor denoted by $\delta T_{\mu\nu}$. The perturbed horizon eventually settles down to a new stationary black hole configuration with slightly shifted asymptotic parameters (identified with the total energy, angular momentum, etc.) and slightly altered geometry around the horizon (identified with the total entropy). Since the same stress tensor component generates all such infinitesimal shifts, they are related, and this relation is nothing but the statement of the first law. 

Let us denote the change in entropy due to the evolution from the initial to the final equilibrium configuration as $\delta S$. The relevant component of the stress tensor which gets related to $\delta S$ can be identified with $\chi^\mu \xi^\nu \, \delta T_{\mu\nu}$ and that helps us writing down the first law more concretely as follows\footnote{A detailed derivation of this can be found in section $2$ of \cite{Jacobson:1995uq}. Also see section-(2.2.3) of \cite{Bhattacharya:2019qal} for more discussions regarding the same.}
\begin{equation}\label{physpro0}
\begin{split}
\delta S =& \, 2 \pi \int_{\cal H} dv \, d^{d-2} x \, \sqrt{h} \,  \chi^\mu \xi^\nu \delta T_{\mu\nu} = \,- 2 \pi\int_{-\infty}^{+\infty} dv \int_{{\cal H}_v}d^{d-2} x\, \sqrt{h} \,  v  \, E_{vv} \, 
\end{split}
\end{equation}
where $\cal H$ is the co-dimension one horizon at $r=0$, ${\cal H}_v$ is the constant $v$-slice of the horizon. Also for the last equality above we have used the fact that the equation of motion is given by $E_{\mu\nu} = -\delta T_{\mu\nu}$ (see eq.\eqref{EOM_Evv}), along with $\chi^v = 1$ and $\xi^v = v$ are the only non-zero components of $\chi^\mu, \, \xi^\nu$ in our set up. 

Next, let us recollect that the main result of \S\ref{sec:proof} was to obtain $E_{vv}$ in the form given in eq.\eqref{keyrel3}, which we write here again for convenience, 
\begin{equation}\label{keyrel_again}
\begin{split}
E_{vv} = & \, \partial_v\left({1\over\sqrt{h}}\partial_v \left(\sqrt{h}\, {\cal J}^v\right) +\nabla_i {\cal J}^i \right)  \, + {\cal O}(\epsilon^2),\\
\text{with} ~~ {\cal J}^v =& \, -\dfrac{1}{2} \left( \widetilde{Q}^{rv} + {\cal B}_{(0)} \right) \, , ~~ {\cal J}^i =  \, -\dfrac{1}{2} \left( \widetilde{Q}^{ri} - J^i_{(1)} \right) \, .
\end{split}
\end{equation}
In what follows, we will argue that the form of $E_{vv}$ as given in eq.\eqref{keyrel_again} is indeed consistent with eq.\eqref{physpro0}. Therefore, to show that eq.\eqref{keyrel_again} is true in arbitrary higher derivative theories of gravity is equivalent to verify the physical process version of the first law in such theories.

We first substitute for $E_{vv}$ from eq.\eqref{keyrel_again} in eq.\eqref{physpro0} followed by certain manipulations involving integrations by parts to move the overall $\partial_v$ on the RHS of eq.\eqref{keyrel_again} in $E_{vv}$ to obtain 
\begin{equation}\label{physpro1}
\begin{split}
\delta S
&=-\int_{-\infty}^{\infty}dv\int_{{\cal H}_v} d^{d-2} x~v\, \sqrt{h}~\partial_v\bigg({1\over\sqrt{h}}\partial_v\left(\sqrt{h}~ {\cal J}^v\right)+\nabla_i{\cal J}^i \bigg)\\
&=-\bigg[\int_{{\cal H}_v}d^{d-2} x ~v\, \bigg(\partial_v\left( \sqrt{h}~{\cal J}^v\right)+\partial_i\left(\sqrt{h} \, {\cal J}^i\right)\bigg)\bigg]_{v=-\infty}^{v=+\infty}\\
&~~+\int_{-\infty}^{\infty}dv\int_{{\cal H}_v}d^{d-2} x ~{\partial_v\left(v\sqrt{h}\right) \over \sqrt{h}}
\bigg(\partial_v\left(\sqrt{h}\,{\cal J}^v\right)+\partial_i\left(\sqrt{h} \, {\cal J}^i\right)\bigg) \, .
\end{split}
\end{equation}
Note that the combination ``$\partial_v\left( \sqrt{h}~{\cal J}^v\right)+\partial_i\left(\sqrt{h} \, {\cal J}^i\right)$" is of boost weight $(+1)$ and therefore must vanish when evaluated on any stationary configuration. Due to this and the fact that at $v=\pm\infty$ we have two stationary configurations at the beginning and the end of the dynamical process, the first term on the RHS of eq.\eqref{physpro1} (appearing on the second line) simply vanishes. 

Now, for the second term on the RHS of eq.\eqref{physpro1}(appearing on the third line), we note that it is a product of two factors, one of which is the same quantity with boost weight $(+1)$ that we mentioned above and therefore this is already linear in the amplitude of the dynamics at ${\cal O}(\epsilon)$. So, from the second factor, we have to consider only that piece which is boost-invariant, neglecting the other ${\cal O}(\epsilon)$ piece, i.e. 
$$\partial_v(v\sqrt{h}) =  \sqrt{h} +\underbrace{v \, \partial_v\sqrt{h}}_ {\sim {\cal O}(\epsilon)}\, ,$$
ensuring that the RHS of eq.\eqref{physpro1} is calculated up to ${\cal O}(\epsilon)$. Therefore we obtain 
\begin{equation}\label{physpro2}
\begin{split}
\delta S &\approx \int_{-\infty}^{\infty}dv\int_{{\cal H}_v} d^{d-2}x  \, 
\bigg(\partial_v\left(\sqrt{h}\,{\cal J}^v\right)+ \,\partial_i\left(\sqrt{h}\,{\cal J}^i\right) \bigg) \, + {\cal O}(\epsilon^2) \, .
\end{split}
\end{equation}
In eq.\eqref{physpro2} the second term on the RHS being a total spatial derivative, integrates to boundary terms on ${\cal H}_v$, the constant $v$ slices of the horizon. Therefore, it vanishes if we assume that the horizon is compact. Neglecting this term, we finally arrive at 
\begin{equation}\label{physpro3}
\begin{split}
\delta S &\approx \int_{-\infty}^{\infty}dv~\partial_v\left[\int_{{\cal H}_v}  d^{d-2}x \, \sqrt{h} \, {\cal J}^v \right] \, + {\cal O}(\epsilon^2)= \left[\int_{{\cal H}_v}  d^{d-2}x \, \sqrt{h} \, {\cal J}^v \right]_{v=-\infty}^{v=+\infty}\, + {\cal O}(\epsilon^2)
\end{split}
\end{equation}
From eq.\eqref{physpro3} it is clear that the change in entropy during this process is given by the change in the quantity ${\cal J}^v$ evaluated at the two stationary configurations at $v=-\infty$ and $v=+\infty$ respectively. We also obtain an expression that has to be identified with the total entropy on every constant $v$ slices, 
\begin{equation}\label{physpro4}
S \approx \int_{{\cal H}_v}  d^{d-2}x \, \sqrt{h} \, {\cal J}^v = -\dfrac{1}{2}\int_{{\cal H}_v}  d^{d-2}x \, \sqrt{h} \, \left(\widetilde{Q}^{rv} + {\cal B}_{(0)}\right) \, ,
\end{equation}
where we have substituted for the expression of ${\cal J}^v$ from eq.\eqref{keyrel_again}. 

From eq.\eqref{physpro4} we immediately recognize $\widetilde{Q}^{rv}$ as the Wald entropy density for equilibrium black holes, which we have already argued in the previous sections. The other term on the RHS of eq.\eqref{physpro4}, i.e., ${\cal B}_{(0)}$, has also previously been identified with the JKM ambiguities present in the out-of-equilibrium extensions of Wald entropy; see the discussion after eq.\eqref{termfinal1}. Since the JKM ambiguities vanish on stationary configurations, the term ${\cal B}_{(0)}$ will not contribute to the RHS of eq.\eqref{physpro3}, justifying that the physical process version of the first law is not affected by these terms. However, as we have already seen, they still play a crucial role in obtaining the linearized version of the second law.

It is worth highlighting that the crucial input for this derivation was the structure of $E_{vv}$ as predicted in eq.\eqref{eq:main1}. In other words, our proof of the second law within the approximation of linearized dynamics also allows us to construct the proof of the first law using the physical process version in the most general higher derivative theory of gravity.

Finally, we conclude this section with the following clarification regarding the implications of our derivation of this version of the first law. As it turns out, for this physical process version of the first law to be true in its most general form for an arbitrary higher derivative theory of gravity, the $vv$-component of the equations of motion, i.e. $E_{vv}$, has to satisfy the following restriction imposed on its off-shell structure, (see eq.(25) in \cite{Sarkar:2019xfd}),
\begin{equation}\label{condPPFL1}
\int_{-\infty}^{\infty}dv\int_{{\cal H}_v} d^{d-2}x  \,\sqrt{h} \, v \left(\partial_v^2 s_w -s_w \, R_{vv} + E^{\text{HD}}_{vv}\right) = {\cal O}(\epsilon^2) \, ,
\end{equation}
where $s_w = s^{\text{HD}}_w + s_{cor}$, such that $s^{\text{HD}}_w$ is the contribution to the Wald entropy density from higher derivative part of the Lagrangian defined in eq.\eqref{def_Wald_ent}, and $s_{cor}$ is the possible out-of-equilibrium extension or correction to  $s^{\text{HD}}_w$ defined in eq.\eqref{def_S_tot}. Also $E^{\text{HD}}_{vv}$ is the component of equations of motion getting contributions only from the higher derivative part of the Lagrangian; see eq.\eqref{EOM_Evv} \footnote{The condition written in eq.\eqref{condPPFL1} refers to the physical process version of the first law in a manner that is only determined by the geometry near the dynamical horizon and, most importantly, independent of the information at the asymptotic boundary of space-time.}. 
It has been shown that eq.\eqref{condPPFL1} is true for $f(R)$ theories \cite{Jacobson:1995uq}, and for Gauss-Bonnet or Lovelock theories \cite{Chatterjee:2011wj, Kolekar:2012tq, Sarkar:2013swa}. However, it is not yet known if this condition is true for arbitrary higher derivative theories of gravity \footnote{For the arguments to derive eq.\eqref{condPPFL1} and a detailed discussion on this we refer the reader to section-$2$ of \cite{Chakraborty:2017kob}. This is also discussed in section-$3$ of the review \cite{Sarkar:2019xfd}.}.

In this section, we have shown that the explicit form of $E_{vv}$ given in eq.\eqref{keyrel_again}  indeed satisfies this condition up to linearised order in the amplitude expansion. To see this, we should compute the $vv$-component of the Ricci Tensor, i.e., $R_{vv}$ in our metric gauge, eq.\eqref{met1}, which yields
\begin{equation} \label{relRvv}
R_{vv} = - \partial_v \left({1 \over \sqrt{h}}\partial_v (\sqrt{h} \,) \right)+ {\cal O}(\epsilon^2) \, .
\end{equation} 
Using eq.\eqref{relRvv} one can show that the terms inside the parenthesis in the integrand on the LHS of eq.\eqref{condPPFL1}, can be expressed as the following 
\begin{equation} \label{condPPFL2}
\partial_v^2 s_w -s_w \, R_{vv} + E^{\text{HD}}_{vv} = \partial_v \left({1 \over \sqrt{h}}\partial_v (\sqrt{h} \,s_w) \right)+ E^{\text{HD}}_{vv} + {\cal O}(\epsilon^2) \, .
\end{equation}
With eq.\eqref{condPPFL2} at hand, the restriction on $E_{vv}$ as written above in eq.\eqref{condPPFL1} takes the following form
\begin{equation}\label{PPFL3}
\int_{-\infty}^{\infty}dv\int_{{\cal H}_v} d^{d-2}x  \, \sqrt{h} \, v \left( \partial_v \left({1 \over \sqrt{h}}\partial_v (\sqrt{h} \,s_w) \right)+ E^{\text{HD}}_{vv} \right) = {\cal O}(\epsilon^2) \, ,
\end{equation}
It is now straightforward to check that eq.\eqref{PPFL3} follows trivially from eq.\eqref{cond2}, which we have already encountered in \S\ref{Evv_2nd_law}. Hence, it is also obvious that the form of $E_{vv}$ given in eq.\eqref{keyrel_again} with the specific identifications of ${\cal J}^v$ and ${\cal J}^i$, satisfies the condition in eq.\eqref{condPPFL1}. 

One crucial outcome of our analysis in this paper is to justify that certain total spatial derivatives in the form of $\nabla_i{\cal J}^i$ are generically present in the structure of $E_{vv}$ as in eq.\eqref{keyrel_again}. However, we now see that the presence of such total spatial divergences does not affect the arguments leading to this version of the first law. As we know, in the statement of the first law only the total entropies between two equilibrium configurations are compared and the expressions of this total entropy involve a spatial integration on the co-dimension two hyper-surface ${\cal H}_v$. Consequently, for compact horizons, it is easy to check that the spatial total derivatives integrate to zero,
\begin{equation}
\int_{-\infty}^{\infty}dv\int_{{\cal H}_v} d^{d-2}x  \,\sqrt{h} \, v\, \partial_v\left(\nabla_i{\cal J}^i\right) \sim {\cal O}(\epsilon^2) \, .
\end{equation}

\section{Discussions and future directions} \label{sec:conclusion}
In this paper, we have constructed a proof of the second law of black hole thermodynamics in arbitrary higher derivative theories of gravity working with a particular choice of coordinates. We have focussed on situations where the dynamical fluctuations connecting the two stationary black holes are of small amplitude such that a linearized amplitude expansion around the initial stationary background is possible. The dynamics are initiated by, e.g., throwing a small matter into the equilibrium black hole, and we have assumed a null energy condition for that matter stress tensor. In the course of designing the proof, we have also constructed an entropy current on the horizon. The divergence of this entropy current being non-negative signifies entropy production locally at each point of the space-time. The `time' component of this entropy current produces the correct expression for the Wald entropy density and also fixes the associated JKM ambiguities as a result of imposing the local second law. The spatial components of the entropy current provide us a mechanism of entropy equilibration via the spatial flow of entropy on the constant `time'-slice of the horizon. This, in turn, provides us with a possible out-of-equilibrium extension of the equilibrium Wald entropy function, maintaining the second law in its strongest possible form. As a by-product of our analysis, we have also completed an alternative proof of the first law using physical processes like in-falling matter initiating black hole dynamics in the most general higher derivative theories of gravity.

Let us now comment on some possible future directions of our work: 
\paragraph{Extending beyond linear order in amplitude expansion:} From our analysis in this paper, it might appear that our results up to linear order in amplitude expansion are enough to prove the second law completely in higher derivative gravitational theory up to all orders. In other words, there is apparently no need to worry about the non-linear terms as long as we are treating the higher derivative terms as perturbative corrections to the leading two derivative Einstein gravity. The reason for this is as follows.
When considering the terms that are non-linear in the amplitude of the dynamics, we could see entropy production at two derivative order itself. Furthermore, the higher derivative terms, being considered as a small perturbative correction, should always be suppressed compared to the two derivative Einstein-Hilbert terms in the gravity Lagrangian. Hence, they could never change the sign of the rate of entropy production. Therefore, one might conclude that the higher derivative corrections in the gravity Lagrangian have to be analyzed carefully only when the two derivative contributions to the entropy production vanish, namely at the linear order in the amplitude of the dynamics.
 
However, as it was shown in \cite{Bhattacharyya:2016xfs}, there are situations where such an argument does not work, at least locally, and one needs to add further corrections to the entropy density (and also presumably to the current) that are manifestly non-linear in the amplitude. In view of this observation, the immediate question is whether we could extend our proof to non-linear orders in amplitude.
\paragraph{Interpretation of the entropy current via fluid-gravity duality:} Another interesting future direction in our opinion is to explore the following question: what does this gravitational entropy density along with the spatial current translate to in the non-gravitational dual systems, whenever they exist, like fluid dynamics \cite{Bhattacharyya:2008xc} and membrane dynamics \cite{Dandekar:2019hyc, Saha:2020zho}. We already have algorithms to construct an entropy current on these dual systems that will maintain the second law locally even at nonlinear order in the amplitude of the dynamics \cite{Bhattacharyya:2013lha, Bhattacharyya:2014bha, Dandekar:2019hyc, Saha:2020zho}. If we can cleanly work out the map between the coordinates of this current paper and the different gauges or coordinates used to construct these dual fluid or membrane systems, it would presumably tell us a lot about the non-linear completion of the gravitational entropy current.
\paragraph{Exploring a ``covariant version" of the entropy current:} In the dual systems that we have mentioned above, in any given coordinate system, the space and time components of the entropy current are always related by the constraints imposed by the freedom of coordinate transformation. However, our proof relies heavily on the horizon adapted coordinate system that we have chosen, and our effective `time' is basically an affine parameter along a null direction. So naive mixing of time and space is not possible here. In other words, we have been able to justify the existence of an entropy current following the construction of \cite{Iyer:1994ys}, but only working in a particular choice of the space-time metric as in eq.\eqref{met1}. Nevertheless, we believe that the notion of non-negative divergence of an entropy current, if it exists, should be a robust feature in a covariant sense as well. It would be nice if we could obtain a proof supporting this in a more coordinate invariant fashion.

Additionally, the components of the entropy current that we have obtained in our analysis also crucially depend on the particular slicing of the null horizon. It would be interesting to understand how they transform under an arbitrary redefinition of the constant $v$-slices of the horizon, e.g., eq.\eqref{residual1}. Such an attempt might lead to a covariant formula also for the spatial entropy current. One possible direction going ahead would be to explore the problem with a viewpoint of covariant phase space analysis along the lines of \cite{Harlow:2019yfa}.

We hope to come back to these issues in our future works. 
\acknowledgments 
We want to thank Jyotirmoy Bhattacharya for initial collaboration, useful discussions, and important inputs. We are grateful to Aron Wall for his efforts in carefully reading our manuscript and providing us with his valuable suggestions and several significant comments. We would also like to thank Shamik Banerjee, Parthajit Biswas, Arunabha Saha for helpful discussions. M.P. would like to thank Shankha Singha Mahapatra for valuable discussions. P.D. would like to duly acknowledge the Council of Scientific and Industrial Research (CSIR), New Delhi for financial assistance through the Junior Research Fellowship (JRF) scheme.
 
\appendix
\section{Notations, conventions and useful definitions}\label{app:notations}
\begin{itemize}
\item We work in $d$-dimensional space-time with coordinates denoted by $x^\mu$.
\item Lower-case Greek indices: $\{\alpha, \beta, \cdots, \mu, \nu, \cdots, \alpha_1, \alpha_2, \cdots \}$, are used for space-time coordinates. 
\item The space-time metric is denoted by $g_{\mu\nu}$ and the associated covariant derivative is denoted by the operator $D_\mu$.
\item The horizon, a co-dimension one hyper-surface $\cal H$, is located at $r=0$. Here $r$ is the coordinate running along the affinely parametrized geodesics $\partial_r$ that takes us away from the horizon. 
\item The affinely parametrized null generator on the horizon is given by $\partial_v$ and $v$ is the coordinate along those null generators. 
\item Constant $v$-slices of the horizon is a co-dimension two hyper-surface ${\cal H}_v$ spanned by $(d-2)$ spatial coordinates $x^i$. 
\item Lower-case Latin indices: $\{i, j, \cdots, i_1, j_1, \cdots \}$ are used for coordinates on ${\cal H}_v$. 
\item The full space-time coordinates are $x^\mu =\{r,v,x^i\}$ and coordinates on the horizon $\cal H$ are $\{v, x^i\}$.
\item The intrinsic metric on the horizon is $h_{ij}$ and the associated covariant derivative is $\nabla_i$.
\end{itemize}
\section{More details regarding the boost weight of a covariant tensor}\label{app:BW}
In this appendix, we will provide additional details regarding some of the results which were used in the paper.
\subsection{Equivalence between different ways of evaluating the Boost weights}
We will first try to show the equivalence between the definitions of Boost Weight mentioned in \S\ref{KillSymBW}. We will collect them here for reference:
\begin{equation}
{\cal A} \rightarrow \widetilde{\cal A} = \lambda^w {\cal A} \, , ~~\text{under}~~ \{r\rightarrow\tilde r = \lambda ~r,~v\rightarrow\tilde v = \lambda^{-1} v\} ~~ \Rightarrow ~~ \text{boost weight of ${\cal A}$ is $w$} \, .
\end{equation}
\begin{equation}
\begin{split}
{\cal L}_\xi S_{\alpha_1 \alpha_2\cdots \alpha_k} =\,  \left[ w +(v\partial_v -r\partial_r) \right]S_{\alpha_1 \alpha_2\cdots \alpha_k} \, 
\end{split}
\end{equation}
where $\xi = v \partial_v - r \partial_r$ and we identify the quantity $w$ as the boost weight of the covariant tensor $S_{\alpha_1 \alpha_2\cdots \alpha_k}$ as 
\begin{equation}
\begin{split}
w\, \equiv \, &\text{net boost weight of a covariant tensor (with all indices lowered)}\\
= \, & \text{number of lower $v$ indices - number of lower $r$ indices.} 
\end{split}
\end{equation}
To prove the equivalence, we consider a generic tensor component with $m_v$ lower $v$ indices and $m_r$ lower $r$ indices : $S_{v \dots v r \dots r} (r,v,x^i) $ where we have explicitly indicated the functional dependence on the space-time variables. Let us consider an infinitesimal Boost transformation:
\begin{equation}
    \tilde{r} = (1 + \epsilon)r \hspace{1cm} \tilde{v} = (1 - \epsilon)v \hspace{1cm} \text{where} \, \, \, \epsilon \, \, \, \text{is infinitesimal}
\end{equation}
Under the above transformation, the covariant tensor component $S_{v \dots v r \dots r}$ transforms as follows:
\begin{equation}
    \begin{split}
        S_{\tilde{v} \dots \tilde{v} \tilde{r} \dots \tilde{r}} (\tilde{r},\tilde{v},x^i) &= \left(  \dfrac{\partial v}{\partial \tilde{v}} \right)^{m_v} \left( \dfrac{\partial r}{\partial \tilde{r}} \right)^{m_r} S_{v \dots v r \dots r}(r,v,x^i) \\
        &= (1+\epsilon)^{m_v} (1-\epsilon)^{m_r} S_{v \dots v r \dots r}(r,v,x^i) \\
        &= (1 + (m_v - m_r)\epsilon ) \, \, S_{v \dots v r \dots r}(r,v,x^i) + O(\epsilon^2)
    \end{split}
\end{equation}
Now we will extract the Lie Derivative along the Integral curves of this particular coordinate transformation: 
\begin{equation}
    \begin{split}
        &S_{\tilde{v} \dots \tilde{v} \tilde{r} \dots \tilde{r}} (r+\epsilon r,v - \epsilon v,x^i) = (1 + (m_v - m_r)\epsilon ) \, \, S_{v \dots v r \dots r}(r,v,x^i) + O(\epsilon^2) \\
        &S_{\tilde{v} \dots \tilde{v} \tilde{r} \dots \tilde{r}} (r,v,x^i) + \epsilon r \partial_r S_{\tilde{v} \dots \tilde{v} \tilde{r} \dots \tilde{r}}(r,v,x^i) - \epsilon v \partial_v S_{\tilde{v} \dots \tilde{v} \tilde{r} \dots \tilde{r}}(r,v,x^i) + O(\epsilon^2) \\ &= (1 + (m_v - m_r)\epsilon ) \, \, S_{v \dots v r \dots r}(r,v,x^i) + O(\epsilon^2) \\
        & S_{\tilde{v} \dots \tilde{v} \tilde{r} \dots \tilde{r}} (r,v,x^i) - S_{v \dots v r \dots r}(r,v,x^i) = (m_v - m_r) \epsilon S_{v \dots v r \dots r} (r,v,x^i) \\
        & \hspace{2cm} + \epsilon v \partial_v S_{v \dots v r \dots r}(r,v,x^i) - \epsilon r \partial_r S_{v \dots v r \dots r}(r,v,x^i) + O(\epsilon^2) \\
        & \implies \mathcal{L}_{\xi} S_{v \dots v r \dots r}(r,v,x^i) = (m_v - m_r) S_{v \dots v r \dots r} (r,v,x^i) \\
        & \hspace{2cm} + v \partial_v S_{v \dots v r \dots r}(r,v,x^i) - r \partial_r S_{v \dots v r \dots r}(r,v,x^i)
    \end{split}
\end{equation}
where in the final step we have defined $\xi^{\mu} \partial_{\mu} = v \partial_v - r \partial_r$ and the coordinate transformation is generated along the integral curves of this vector field. Thus, we have proved the equivalence between the two definitions of Boost Weight of the tensor component $S_{v \dots v r \dots r}$ which is given by $w = m_v - m_r$. The generalization of the above equivalence for tensors of the form $S_{v\dots v ij\dots}(r,v,x^i)$ and $S_{r\dots rij\dots}(r,v,x^i)$ is straightforward once we realise that the Boost transformation doesn't change the coordinates $x^i$. Thus, for $S_{v\dots v ij\dots}(r,v,x^i)$, we essentially have $m_r=0$ and for $S_{r\dots rij\dots}(r,v,x^i)$, we have $m_v=0$.
\subsection{Structure of covariant tensors in the stationary background solution}
We will now show that if $\xi$ is a Killing vector for our space-time, then the Lie derivative of any covariant tensor evaluated on that space-time will vanish. We will consider the above generic tensor component $S_{v \dots v r \dots r}$ which has $m_v$ number of lower $v$ indices and $m_r$ number of lower $r$ indices. Assume that once the tensor component is evaluated in our gauge, it has $m^{(1)}_v$ factors of $r$ and $m^{(2)}_v$ factors of $\partial_v$'s such that $m^{(1)}_v + m^{(2)}_v = m_v$. Thus, a Boost weight of $w = m_v - m_r$ implies that the tensor component takes the following form
\begin{equation}
    S_{v \dots v r \dots r} = r^{m^{(1)}_v} (\partial_v)^{m^{(2)}_v} (\partial_r)^{m_r} Y(rv)
\end{equation}
where $Y$ is a function of the product $rv$ so that it has Boost weight zero. Using this, we further simplify the tensor component:
\begin{equation}\label{finaleqm}
\begin{split}
S_{v \dots v r \dots r} &= r^{m_v^{(1)}} \left(\partial_v\right)^{m_v^{(2)} }\left(\partial_r\right)^{m_r} Y(rv)\\
&=r^{m_v^{(1)}} \left(\partial_v\right)^{m_v^{(2)} }\left(v^{m_r} \left[\left(d\over dz\right)^{m_r}Y(z)\right]_{z=rv}\right)\\
&=r^{m_v^{(1)}}\sum_{m=0}^{k}{^kC}_m\left[\left(\partial_v\right)^mv^{m_r}\right]\left(\partial_v\right)^{m_v^{(2)}-m}\left[\left(d\over dz\right)^{m_r}Y(z)\right]_{z=rv}\text{where}~ k = Min(m_r,m_v^{(2)})\\
&=r^{m_v^{(1)} + m_v^{(2)}}\sum_{m=0}^{k}{^kC}_m\left[\left(\partial_v\right)^mv^{m_r}\right]r^{-m}\left[\left(d\over dz\right)^{m_r +m_v^{(2)}-m}Y(z)\right]_{z=rv}\\
&=r^{m_v}\sum_{m=0}^{k}{^kC}_m\left[m_r!\over (m_r -m)!\right]v^{m_r - m}r^{-m}\left[\left(d\over dz\right)^{m_r +m_v^{(2)}-m}Y(z)\right]_{z=rv}\\
&=r^{m_v}v^{m_r}\sum_{m=0}^{k}{^kC}_m\left[m_r!\over (m_r -m)!\right](rv)^{-m}\left[\left(d\over dz\right)^{m_r +m_v^{(2)}-m}Y(z)\right]_{z=rv}\\
&=r^{m_v}v^{m_r}~\tilde Y(rv) 
\end{split}
\end{equation}
where
$$\tilde Y(rv) = \sum_{m=0}^{k}{^kC}_m\left[m_r!\over (m_r -m)!\right](rv)^{-m}\left[\left(d\over dz\right)^{m_r +m_v^{(2)}-m}Y(z)\right]_{z=rv},~k = Min(m_r,m_v^{(2)})
$$
We use the final result of \ref{finaleqm} to evaluate the Lie derivative of the tensor component:
\begin{equation}
    \begin{split}
        \mathcal{L}_{\xi} S_{v \dots v r \dots r} &= (m_v - m_r + v \partial_v - r \partial_r ) (r^{m_v} v^{m_r} \tilde{Y}(rv) ) \\
        & = (m_v - m_r + m_r - m_v ) r^{m_v} v^{m_r} \tilde{Y}(rv) = 0 
    \end{split}
\end{equation}
where in the final step we have used the fact that $(v \partial_v - r \partial_r)\tilde{Y}(rv) = 0$. This vanishing of the Lie derivative on our stationary space-time was possible only after we have explicitly worked out the $\partial_r$'s and $\partial_v$'s in terms of factors of $r$'s and $v$'s.

\section{A brief review of the Iyer-Wald procedure in component notation} \label{app:ReviewIyerWald}
We re-express the main results of \cite{Iyer:1994ys} in component notation. 

\begin{itemize}
\item \textbf{Lemma 2.1: Type of Lagrangians considered}

We consider Lagrangians which are diffeomorphism covariant in the sense of eq.\eqref{diffcovlag}. It has been proved that \citep{Iyer:1994ys} such Lagrangians can always be re-expressed in the following form
\begin{equation}
    L = L(g_{\mu\nu},R_{\mu\nu\rho\sigma},D_{\alpha_1}R_{\mu\nu\rho\sigma}, \dots , D_{(\alpha_1} \dots D_{\alpha_m)} R_{\mu\nu\rho\sigma},\phi,D_{\alpha_1}\phi, \dots , D_{(\alpha_1}\dots D_{\alpha_l)}\phi )
\end{equation}
where $\phi$ denotes the collection of all matter fields \footnote{We have suppressed the index structure of the matter sector.}. In \citep{Iyer:1994ys}, the Lagrangian is viewed as an $d$ form $\mathbf{L}$ given by $\mathbf{L}_{\alpha_1 \dots \alpha_d} = \epsilon_{\alpha_1 \dots \alpha_d} L$  \footnote{Here $\epsilon$ is canonical the volume form (i.e) $\epsilon_{b_1 \dots b_d} = \sqrt{-g} \varepsilon_{b_1 \dots b_d}$ where $\varepsilon_{b_1 \dots b_d}$ is the Levi Civita symbol }. The variation of this Lagrangian is given by
\begin{equation}
    \delta \mathbf{L} = \mathbf{E}^{\mu\nu} \delta g_{\mu\nu} + \mathbf{E}_{\phi} \delta \phi + d \mathbf{\Theta}
\end{equation}
where $\mathbf{E}^{\mu\nu}_{\alpha_1 \dots \alpha_d} = \epsilon_{\alpha_1 \dots \alpha_d} E^{\mu\nu} $ ($E^{\mu\nu}$ is the equation of motion for the metric field) and $\mathbf{E}_{\phi} = \epsilon_{\alpha_1 \dots \alpha_d} E_{\phi} $ ($E_{\phi}$ is the Equation of motion for the matter fields). $\mathbf{\Theta}$, an $d-1$ form is the total derivative term called the Symplectic Potential. By taking the Hodge dual we can revert to component notation using the following identity which can be proven straightforwardly
\begin{equation}
    d \mathbf{\Theta} = d (\epsilon_{\mu \alpha_2 \dots \alpha_d} \Theta^{\mu} ) = \epsilon_{\alpha_1 \dots \alpha_d} D_{\mu} \Theta^{\mu}
\end{equation}
Thus the variation of the Lagrangian becomes
\begin{equation}
    \begin{split}
        \delta (\epsilon_{\alpha_1 \dots \alpha_d} L ) = \epsilon_{\alpha_1 \dots \alpha_d} E^{\mu\nu} \delta g_{\mu\nu}+ \epsilon_{\alpha_1 \dots \alpha_d} E_{\phi} \delta \phi + \epsilon_{\alpha_1 \dots \alpha_d} D_{\mu} \Theta^{\mu} \\
       \implies \delta (\sqrt{-g} L) = \sqrt{-g} E^{\mu\nu} \delta g_{\mu\nu} + \sqrt{-g} E_{\phi}\delta \phi + \sqrt{-g} D_{\mu} \Theta^{\mu} 
    \end{split}
\end{equation}

\item \textbf{Lemma 3.1: General form of $\Theta^{\mu}$}

We can write Eqn (23) of \cite{Iyer:1994ys} explicitly as
\begin{equation}
    \mathbf{\Theta}_{\alpha_2 \dots \alpha_d} = 2 \left( E^{\mu \nu \rho \sigma}_R D_{\sigma}\delta g_{\nu \rho} \right) \epsilon_{\mu \alpha_2 \dots \alpha_d} + \mathbf{\Theta'}_{\alpha_2 \dots \alpha_d}
\end{equation}
where 
\begin{equation}\label{ERapp}
    E^{\mu\nu\rho\sigma}_R = \dfrac{\partial L}{\partial R_{\mu\nu\rho\sigma}} - D_{\alpha_1} \dfrac{\partial L}{\partial D_{\alpha_1} R_{\mu \nu \rho \sigma}} + \dots + (-1)^m D_{(\alpha_1} \dots D_{\alpha_m)} \dfrac{\partial L}{\partial D_{(\alpha_1} \dots D_{\alpha_m)} R_{\mu\nu\rho\sigma}}
\end{equation}
Writing $\mathbf{\Theta}$ and $\mathbf{\Theta'}$ which are $d-1$ forms as a Hodge duals of one-forms $\Theta_{\mu}$ and $\Theta'_{\mu}$ respectively,
\begin{equation}
    \epsilon_{\mu \alpha_2 \dots \alpha_d} \Theta^{\mu} = 2 \left( E^{\mu\nu\rho\sigma}_R D_{\sigma}\delta g_{\nu\rho} \right) \epsilon_{\mu \alpha_2 \dots \alpha_d} + \epsilon_{\mu \alpha_2 \dots \alpha_d} \Theta'^{\mu} 
\end{equation}
Now we can contract with $\epsilon^{\beta \alpha_2 \dots \alpha_d}$ \footnote{The contraction of two volume form tensors is given by ($\delta^{\alpha}_{\beta}$ is the Kronecker delta) $$
    \epsilon^{\mu_1 \mu_2 \dots \mu_p \alpha_1 \dots \alpha_{d-p}} \, \epsilon_{\mu_1 \mu_2 \dots \mu_p \beta_1 \dots \beta_{d-p}} = (-1)^s \,  p! \,  (d-p)!  \, \delta^{[\alpha_1}_{\beta_1} \dots \delta^{\alpha_{d-p}]}_{\beta_{d-p}}$$
} to get
\begin{equation}\label{app_theta}
    \Theta^{\beta} = 2 E^{\beta \nu \rho \sigma}_R D_{\sigma} \delta g_{\nu\rho} + \Theta'^{\beta}
\end{equation}
To find $\Theta'^{\beta}$, we write Eqn(24) of \cite{Iyer:1994ys} in component notation. In Eqn(24) of \cite{Iyer:1994ys}, since $\mathbf{\Theta'}$ is an $d-1$ form proportional to $\epsilon$, by inspection, it is clear that $\mathbf{S}^{\mu \nu}(\phi)$, $\mathbf{T_i}(\phi)^{\mu\nu\rho\sigma \alpha_1 \dots \alpha_i}$ and $\mathbf{U_i}(\phi)^{\alpha_1 \dots \alpha_i}$ are also $d-1$ forms proportional to $\epsilon$. Thus, we can write Eqn(24) of \cite{Iyer:1994ys} as
\begin{equation}
    \begin{split}
        \epsilon_{\mu \alpha_2 \dots \alpha_d} \Theta'^{\mu} = S^{\gamma \mu\nu}(\phi) \epsilon_{\gamma \alpha_2 \dots \alpha_d} \delta g_{\mu\nu} + \sum_{i=0}^{m-1} T_{i}(\phi)^{\gamma \mu \nu \rho \sigma \beta_1 \dots \beta_i} \epsilon_{\gamma \alpha_2 \dots \alpha_d} \delta D_{(\beta_1} \dots D_{\beta_i)} R_{\mu\nu\rho\sigma} \\
        +\sum_{i=0}^{l-1} U_i (\phi)^{\gamma \beta_1 \dots \beta_i} \epsilon_{\gamma \alpha_2 \dots \alpha_d} \delta D_{(\beta_1} \dots D_{\beta_i)} \phi 
    \end{split}
\end{equation}
Thus contracting with $\epsilon^{\beta \gamma_2 \dots \gamma_d}$,
\begin{equation}\label{app_theta_p}
    \begin{split}
        \Theta'^{\beta} = S^{\beta \mu \nu}(\phi) \delta g_{\mu\nu} &+ \sum_{i=0}^{m-1} T_{i}(\phi)^{\beta \mu \nu \rho \sigma \alpha_1 \dots \alpha_i} \delta D_{(\alpha_1} \dots D_{\alpha_i)} R_{\mu\nu\rho\sigma} \\ &+ \sum_{i=0}^{l-1} U_i (\phi)^{\beta \alpha_1 \dots \alpha_i} \delta D_{(\alpha_1} \dots D_{\alpha_i)} \phi 
    \end{split}
\end{equation}
which is exactly eq.(24) of \citep{Iyer:1994ys} written in Component Notation.

\item \textbf{The basic algorithm to compute the Noether charge}

First we compute the Noether current $d-1$ form associated to a vector field $\zeta^{\mu}$ which is defined as follows
\begin{equation}
    \mathbf{J}_{\alpha_2 \dots \alpha_d} = \mathbf{\Theta}_{\alpha_2 \dots \alpha_d} (\psi,\mathcal{L}_{\zeta}\psi) - \zeta \cdot \mathbf{L}_{\alpha_2 \dots \alpha_d} = \epsilon_{\beta \alpha_2 \dots \alpha_d} (\Theta^{\beta} - \zeta^{\beta} L)
\end{equation}
\footnote{Where $\psi$ denotes the collection of all fields.} Thus taking the Hodge dual we get the following expression for Noether current
\begin{equation}
    J^{\mu} = \Theta^{\mu}(\psi,\mathcal{L}_{\zeta}\psi) - \zeta^{\mu} L
\end{equation}
Given the Noether current $d-1$ form $\mathbf{J}$, we can define the Noether charge $d-2$ form as $\mathbf{J}= d \mathbf{Q}$ on-shell. We write this $d-2$ form as the Hodge dual of a $2$ form
\begin{equation}
    \mathbf{Q}_{\alpha_3 \dots \alpha_d} = \dfrac{1}{2} \epsilon_{\beta \gamma \alpha_3 \dots \alpha_d} Q^{\beta \gamma} 
\end{equation}
Using the relation
\begin{equation}
    d \left( \dfrac{1}{2} \epsilon_{\beta \gamma \alpha_3 \dots \alpha_d} Q^{\beta \gamma} \right) = D_{\gamma} Q^{\beta\gamma} \epsilon_{\beta \alpha_2 \dots \alpha_d}
\end{equation}
We can see that $\mathbf{J} =d \mathbf{Q} $ essentially amounts to
\begin{equation}
    J^{\mu} = \Theta^{\mu} - \zeta^{\mu} L = D_{\nu} Q^{\mu\nu} \hspace{1cm} (on-shell)
\end{equation}
To see how this works out in practice, let us consider a pure gravity theory with no matter sector. The variation of such a Lagrangian is given by
\begin{equation}
    \delta (\sqrt{-g}L) =\sqrt{-g} E^{\mu\nu} \delta g_{\mu\nu} + \sqrt{-g} D_{\mu} \Theta^{\mu} 
\end{equation}
We have performed a similar analysis in \S\ref{rel_Evv_Q}. Suppose we consider the variation under the diffeomorphism generated by $\zeta^{\mu}$ (i.e.) $\delta g_{\mu\nu} = \mathcal{L}_{\zeta} g_{\mu\nu} = D_{\mu}\zeta_{\nu}+D_{\nu}\zeta_{\mu}$, the variation of the Lagrangian is given by
\begin{equation}
    \delta (\sqrt{-g} L) = \sqrt{-g} D_{\mu} (\zeta^{\mu}L)
\end{equation}
due to the diffeomorphism covariance eq.\eqref{diffcovlag}. The variation of the Lagrangian thus provides
\begin{equation}
        \sqrt{-g}D_{\mu}(\zeta^{\mu}L) = \sqrt{-g} E^{\mu\nu} \mathcal{L}_{\zeta} g_{\mu\nu} + \sqrt{-g} D_{\mu}\Theta^{\mu}(\mathcal{L}_{\zeta}g_{\alpha\beta})
    \end{equation}
Thus we do some manipulations in the above expression as follows:
    \begin{equation}
        \begin{split}
            D_{\mu}(\zeta^{\mu}L) =& 2 E^{\mu\nu} D_{\mu}\zeta_{\nu} + D_{\mu}\Theta^{\mu} \\ 
            D_{\mu}(\zeta^{\mu}L) =& D_{\mu} (2 E^{\mu\nu} \zeta_{\nu} ) - 2 (D_{\mu} E^{\mu\nu}) \zeta_{\nu} + D_{\mu} \Theta^{\mu} \\
           D_{\mu}(\Theta^{\mu} - \zeta^{\mu}L)  =& D_{\mu} J^{\mu}= -D_{\mu} (2 E^{\mu\nu} \zeta_{\nu} )\\
             \implies J^{\mu} =& \Theta^{\mu} - \zeta^{\mu}L= -2E^{\mu\nu}\zeta_{\nu} + D_{\nu} Q^{\mu\nu}
        \end{split}
    \end{equation}
In the third step, we have used the Generalized Bianchi Identity $(D_{\mu} E^{\mu\nu} = 0)$. In the final step, when integrating out the covariant derivatives, it must be clear that there must be a piece $D_{\nu}Q^{\mu\nu}$ locally where $Q^{\mu\nu} = -Q^{\nu\mu}$ (we can trivially show that $D_{\mu}D_{\nu}Q^{\mu\nu} = 0$ just by going to a set of Local inertial coordinates). So we finally have 
    \begin{equation} \label{defDQ}
        \Theta^{\mu}(\mathcal{L}_{\zeta}g_{\alpha\beta}) - \zeta^{\mu}L+2E^{\mu\nu}\zeta_{\nu} = D_{\nu} Q^{\mu\nu}
    \end{equation}
Thus, the algorithm for extracting the Noether charge is to calculate $\Theta^{\mu}(\psi,\delta \psi)$ for $\delta \psi = \mathcal{L}_{\zeta} \psi$ and use the above equation to find $Q^{\mu\nu}$. This relation was derived in \S\ref{rel_Evv_Q}, see eq.\eqref{hijibiji-3}. 

\item \textbf{Proposition 4.1: General form of the Noether charge}

Proposition 4.1 of \citep{Iyer:1994ys} proves that the Noether charge $d-2$ form can always be expressed in the following form
\begin{equation}
    \mathbf{Q} = \mathbf{W}_{\rho} (\psi) \zeta^{\rho} + \mathbf{X}^{\rho\sigma}(\psi) D_{[\rho} \zeta_{\sigma]} + \mathbf{Y}(\psi,\mathcal{L}_{\zeta}\psi) + d \mathbf{Z}(\psi,\zeta) 
\end{equation}
where $\psi$ denotes the collection of all the fields of the theory. Since $\mathbf{Q}$ is an $d-2$ form, $\mathbf{W}_{\rho} (\psi),\mathbf{X}^{\rho\sigma}(\psi),\mathbf{Y}(\psi,\mathcal{L}_{\zeta} \psi)$ are $d-2$ forms. $\mathbf{Z}(\psi,\zeta)$ is an $d-3$ form. Thus, once again, we write the $d-2$ forms as Hodge duals of two forms:
\begin{equation}
    (\mathbf{W}_{\rho} (\psi))_{\alpha_3 \dots \alpha_d} = \dfrac{1}{2} \epsilon_{\mu\nu \alpha_3 \dots \alpha_d} W^{\mu\nu}_{\hspace{0.35cm}\rho} (\psi) 
\end{equation}
\begin{equation}
    \mathbf{Y}(\psi,\mathcal{L}_{\zeta}\psi)_{\alpha_3 \dots \alpha_d} = \dfrac{1}{2} \epsilon_{\mu\nu \alpha_3 \dots \alpha_d} Y^{\mu\nu}(\psi,\mathcal{L}_{\zeta} \psi)
\end{equation}
$\mathbf{X}^{\rho\sigma}$ can be chosen to be of the form
\begin{equation}
    (\mathbf{X}^{\rho\sigma}(\psi))_{\alpha_3 \dots \alpha_d} = - E^{\mu\nu\rho\sigma}_R \epsilon_{\mu\nu \alpha_3 \dots \alpha_d}
\end{equation}
Since $\mathbf{Z}$ is an $d-3$ form, we can write it as a Hodge dual of a $3-$form
\begin{equation}
    \mathbf{Z}_{\alpha_4 \dots \alpha_d} = \dfrac{1}{6} \epsilon^{\mu\nu\rho}_{\hspace{0.4cm} \alpha_4 \dots \alpha_d} Z_{\mu\nu\rho} 
\end{equation}
We now use the following result which can be proven straightforwardly:
\begin{equation}
    d \left( \dfrac{1}{6} \epsilon^{\mu\nu\rho}_{\hspace{0.4cm}\alpha_4 \dots \alpha_d} Z_{\mu\nu\rho} \right) = D_{\mu} Z^{\mu\nu\rho} \epsilon_{\nu\rho \alpha_3 \dots \alpha_d}
\end{equation}
Collecting all these results, we can rewrite Eqn(52) of \cite{Iyer:1994ys} as 
\begin{equation}
    \begin{split}
        \dfrac{1}{2} \epsilon_{\beta\gamma \alpha_3 \dots \alpha_d} Q^{\beta\gamma} = \dfrac{1}{2} \epsilon_{\beta\gamma \alpha_3 \dots \alpha_d} W^{\beta\gamma}_{\hspace{0.35cm}\rho} \zeta^{\rho} - E^{\beta\gamma\rho\sigma}_R \epsilon_{\beta\gamma \alpha_3 \dots \alpha_d} D_{[\rho} \zeta_{\sigma]} \\
        +\dfrac{1}{2} \epsilon_{\beta\gamma \alpha_3 \dots \alpha_d} Y^{\beta\gamma} + D_{\rho} Z^{\beta\gamma\rho} \epsilon_{\beta\gamma \alpha_3 \dots \alpha_d}
    \end{split}
\end{equation}
We can thus, contract with $\epsilon^{\mu\nu \alpha_3 \dots \alpha_d}$ to obtain
\begin{equation}
    Q^{\mu\nu} = W^{\mu\nu\rho} \zeta_{\rho} - 2E^{\mu\nu\rho\sigma}_R D_{[\rho}\zeta_{\sigma]} + Y^{\mu\nu} + 2D_{\rho} Z^{\mu\nu\rho}
\end{equation}
We can absorb the factor of 2 accompanying $Z^{\mu\nu\rho}$ into the definition of $Z^{\mu\nu\rho}$ since this ambiguity arises as $d \mathbf{Z}$ in $\mathbf{Q}$. Thus En(51) of \citep{Iyer:1994ys} explicitly written in component notation is given by
\begin{equation} \label{app_Q_1}
    Q^{\mu\nu} = W^{\mu\nu\rho}(\psi) \zeta_{\rho} - 2E^{\mu\nu\rho\sigma}(\psi) D_{[\rho}\zeta_{\sigma]} + Y^{\mu\nu}(\psi,\mathcal{L}_{\zeta}\psi) + D_{\rho}Z^{\mu\nu\rho}
\end{equation}
It should be noted that $Y^{\mu\nu}$ and $Z^{\mu\nu\rho}$ are ambiguities in the definition of the Noether current and charge respectively. The detailed algorithm mentioned above for calculating the Noether charge essentially results in the following general form for $Q^{\mu\nu}$
\begin{equation} \label{app_Q_2}
    Q^{\mu\nu} = W^{\mu\nu\rho}(\psi) \zeta_{\rho} - 2E^{\mu\nu\rho\sigma}_R D_{[\rho}\zeta_{\sigma]}  
\end{equation}
where $E^{\mu\nu\rho\sigma}_R$ is given by eq.\eqref{ERapp} and $W^{\mu\nu\rho} = W^{[\mu\nu]\rho}$. This Noether charge is then used to define an Entropy functional which satisfies the Equilibrium version of the first law of thermodynamics \citep{Iyer:1994ys}:
\begin{equation}
    S_W = -2 \pi \int_{\Sigma} d^{d-2} x^i \sqrt{h} \, \, E^{\mu\nu\rho\sigma}_R \epsilon_{\mu\rho} \epsilon_{\nu\sigma} \, ,
\end{equation}
where $x^i$ denote the coordinates induced on the horizon, $h= det (h_{ij})$ ($h_{ij}$ being the induced metric on the horizon), $\Sigma$ is a co-dimension $2$ surface on the Black hole horizon and $\epsilon_{\mu\rho}$ being the corresponding Bi-normal. $E^{\mu\nu\rho\sigma}_R$ is given by eq.\eqref{ERapp}. For $L = \dfrac{1}{16\pi G} R$, the entropy reduces to $S_W = \dfrac{A}{4G}$ which is the usual Bekenstein-Hawking Entropy.
\end{itemize}

\section{Arguments justifying eq.(\ref{eq:fstruct})} \label{App_eqShrtPrf}
In this appendix we provide some justifications in support of eq.\eqref{eq:fstruct}. According to the definition of $\Theta^\mu$ in a general higher derivative theory of gravity one can write it as a sum of two terms, i.e. $\Theta \sim \text{Term}_1 + \text{Term}_2$. This will be argued in the sub-section \S\ref{subsec:Theta_r} (see eq.\eqref{schemetheta}). One of them, named $\text{Term}_1$, has the following generic structure
$$\text{Term}_1 \sim \left[\text{Tensor-1}\right]~\mathcal{L}_\xi\left[\text{Tensor-2}\right] \, ,$$ with appropriate index contractions  being implied on the RHS above. On the other hand, 
$\text{Term}_2$ has a little bit different structure than $\text{Term}_1$, since it involves a covariant derivative of $L_\xi$:
$$\text{Term}_2  \sim D_\nu\left[ E^{r\mu\nu\rho}~\mathcal{L}_\xi g_{\mu\rho}\right]\vert_{r=0}.$$ 
However, if we explicitly evaluate this term at the horizon $(r=0)$ in our coordinate system we will find that it has the following form:  
$$\text{Term}_2 = E^{rijr}(-1 + v\partial_v)\left(\partial_r h_{ij}\right) + (1 + v\partial_v)\left[E^{rjkv}\left(\partial_v h_{jk}\right)\right].$$
A more careful derivation of this can be found in \S\ref{subsec:Theta_r} (see eq.\eqref{1stterm3}, eq.\eqref{termh} and eq.\eqref{termh2}). Although note that the term $\partial_r h_{ij}$ is not a covariant tensor component, the operator $$ \mathcal{L}_\xi= \text{boost weight} + v\partial_v $$  has a well-defined action on it and we could express this term as 
$E^{rijr}~\mathcal{L}_\xi\left(\partial_r h_{ij}\right)$. Since for our analysis we do not need any details other than boost weight and distribution of $\partial_v$s and $\partial_r$s, for our purpose, this term is also effectively of the same form as $\text{Term}_1$. We do not need to process the second term $(1 + v\partial_v)\left[E^{rjkv}\left(\partial_v h_{jk}\right)\right]$ since it is already of the structure of equation eq.\eqref{eq:fstruct} with ${\cal M}_{(0)} $ set to zero.

So finally we have to  analyze terms of the of the form of $\text{Term}_1$.
\begin{equation}\label{an1}
\begin{split}
&\text{A typical term in $\Theta^r$  generated from $\text{Term}_1$} ~~\sim~~ \tilde t_{(-k)}~ (k +1+ v~\partial_v) ~t_{(k+1)}\\
\Rightarrow~&\Theta_{(1)} \sim (k+1) ~\tilde t_{(-k)}t_{(k+1)},~~~\Theta_{(2)} = \tilde t_{(-k)}~\partial_v t_{(k+1)}
\end{split}
\end{equation}
For these terms, eq.\eqref{eq:fstruct} reduces to 
\begin{equation}\label{eq:fstruct2}
\tilde t_{(-k)}~\partial_v t_{(k+1)} - \partial_v\left[ (k+1)~\tilde t_{(-k)} ~t_{(k+1)}\right] =\partial_v^2~(\text {JKM type terms})
\end{equation}
Note that for $k=0$ this is trivially true with no JKM type terms.

Now extending the arguments in \cite{Wall:2015raa} (by transferring $\partial_v$ s to negative boost weight terms  after adding  total $\partial_v$ terms), we could show that a typical term of positive boost weight $(n+1)$ could always be expressed as (see eq.\eqref{sq01} in \S\ref{BW_cov_Ten} for more discussion on this)
\begin{equation}\label{r:R1}
\begin{split}
&\tilde T_{(-k)}~\partial_v^{n+1+k} T_{(0)} = \partial_v^{n+1}\left[\text{JKM type terms}\right] + (-1)^k  \left[^{k+n}C_n\right]\left[\tilde T_{(0)} ~\partial_v^{n+1}T_{(0)}\right]\\
\text{where} ~~&\tilde T_{(0)} = \partial_v^k ~\tilde T_{(-k)}
\end{split}
\end{equation}
Applying the result eq.\eqref{r:R1} to the two terms $\tilde t_{(-k)} ~t_{(k+1)}$ and $\tilde t_{(-k)} ~\partial_v ~t_{(k+1)}$ we find the following relations
\begin{equation}\label{eq:ddd1}
\begin{split}
\tilde t_{(-k)} ~t_{(k+1)} & \sim \tilde t_{(-k)} ~\partial_v^{(k+1)} t_{(0)} = \partial_v\left[\text{JKM type terms}\right] + (-1)^k\left[ \tilde t_{(0)} \partial_v t_{(0)}\right]
\end{split}
\end{equation}
by putting $(n=0)$ in  eq.\eqref{r:R1}, and similarly 
\begin{equation}\label{eq:ddd2}
\begin{split}
&\tilde t_{(-k)} ~\partial_v ~t_{(k+1)}\sim \tilde t_{(-k)} ~\partial_v^{(k+2)} t_{(0)} = \partial_v^2\left[\text{JKM type terms}\right] + (-1)^k(k+1)\left[ \tilde t_{(0)} \partial_v t_{(0)}\right]
\end{split}
\end{equation}
by putting $(n=1)$ in  eq.\eqref{r:R1}. 
Equation \eqref{eq:fstruct2} follows immediately from equation eq.\eqref{eq:ddd1} and  eq.\eqref{eq:ddd2}.

\section{Detailed proof of Result:\,$1$ in eq.(\ref{sq01}) }\label{app:pro1}
A typical component of a covariant tensor with boost weight $w = a+1>0$ will have the following structure when restricted to the horizon
\begin{equation}
t^{(k)}_{(a+1)}|_{r=0} = \tilde T_{(-k)} \partial_v^{(k+a +1)}T_{(0)}|_{r=0} + \mathcal{O}(\epsilon^2) \, .
\end{equation}
Here the subscripts denote the boost weight and as before, for convenience we have suppressed the index structure for $\tilde T$ and $T$.

Our aim is to prove Result:1 which states that $t^{(k)}_{(a+1)}$ could always be re-expressed as follows
\begin{equation}\label{sq0}
\begin{split}
&t^{(k)}_{(a+1)} = \tilde T_{(-k)} \partial_v^{(k+a +1)}T_{(0)} \\
&~= \partial_v^{a+1}\left[\sum_{m=0}^{k-1} (-1)^m\left[ {^{m+a}C}_{m}\right]\left(\tilde T_{(-k+m)}~\partial_v^{k-m}T_{(0)}\right)\right] + (-1)^k \left[ {^{k+a}C}_a\right]\tilde T_{(0)}~\partial_v^{a+1} T_{(0)}\\
&\text{where}~~\tilde T_{(-k +m)} \equiv \partial_v^m \tilde T_{(-k)}
 \end{split}
\end{equation}

\subsection{The Proof}
The basic idea in the proof is to carefully rearrange the $v$ derivatives and to read off the binomial coefficients appropriately. We use the method of induction to prove our desired result.

We check that eq.\eqref{sq0} is true for $a=0$:
\begin{equation}\label{sq1}
\begin{split}
t^{(k)}_{(1) } &=~ \tilde T_{(-k)} \partial_v^{(k +1)}T_{(0)}\\
&=~\partial_v\left[\tilde T_{(-k)} \partial_v^{(k)}T_{(0)}\right] - \tilde T_{(-k+1)} \partial_v^{(k )}T_{(0)}\\
&=~\partial_v\left[\tilde T_{(-k)} \partial_v^{(k)}T_{(0)}-\tilde T_{(-k+1)} \partial_v^{(k-1)}T_{(0)}\right] +\tilde T_{(-k+2)} \partial_v^{(k -1)}T_{(0)}\\
&=~\cdots\\
&=~\partial_v\left[\sum_{m=0}^{k-1}(-1)^m\tilde T_{(-k+m)} ~\partial_v^{k-m} T_0 \right] +(-1)^k \tilde T_{(0)} \partial_v T_{(0)}\\
&=~\partial_v\left[\sum_{m=0}^{k-1}(-1)^m\left[ {^{m+0}C}_0\right]\tilde T_{(-k+m)} ~\partial_v^{k-m} T_0 \right] +(-1)^k \left[ {^{k+0}C}_0\right] \tilde T_{(0)} \partial_v T_{(0)}
\end{split}
\end{equation}
where the fourth step indicates that we must extract all the $k$ derivatives in the form of total derivative. The final step indicates that $t^{(k)}_{(1)}$ is of the form predicted by eq.\eqref{sq0}. Next, we shall show that if the result eq.\eqref{sq0} is true for $a=n$, then it is also true for $a=n+1$.

The statement for $a=n$ which we assume to be true is given by:
\begin{equation}\label{sq2}
\begin{split}
&t^{(k)}_{(n+1)} = \tilde T_{(-k)} \partial_v^{(k+n +1)}T_{(0)} \\
&~= \partial_v^{n+1}\left[\sum_{m=0}^{k-1} (-1)^m\left[ {^{m+n}C}_n\right]\left(\tilde T_{(-k+m)}~\partial_v^{k-m}T_{(0)}\right)\right] +\left[ {^{k+n}C}_n\right]\left(\tilde T_{(0)}~\partial_v^{n+1} T_{(0)}\right)\\
&\text{where}~~\tilde T_{(-k +m)} = \partial_v^m \tilde T_{(-k)}
 \end{split}
\end{equation}
We shall now process $t^{(k)}_{(n+2)} = \tilde T_{(-k)} \partial_v^{(k+n +2)}T_{(0)}$
\begin{equation}\label{sq3}
\begin{split}
t^{(k)}_{(n+2)} =& \tilde T_{(-k)} \partial_v^{(k+n +2)}T_{(0)}\\
=&~\partial_v\left[\tilde T_{(-k)} \partial_v^{(k +n+1)}T_{(0)}\right] - \tilde T_{(-k+1)} \partial_v^{(k+n+1 )}T_{(0)}\\
=&~\cdots\\
=&~\partial_v\left[\sum_{m=0}^{k-1}(-1)^m~\tilde T_{(-k+m)} \partial_v^{(k +n+1-m)}T_{(0)}\right] + (-1)^k ~\tilde T_{(0)} ~\partial_v^{(n+2)}T_{(0)}
\end{split}
\end{equation}
In the first term of the final step, each term within the summation is of the form $t^{(k-m)}_{(n+1)}$. So we shall apply our previous assumption eq.\eqref{sq2} here for $a = n$.
\begin{equation}\label{sq4}
\begin{split}
&t^{(k)}_{(n+2)} = \tilde T_{(-k)} \partial_v^{(k+n +2)}T_{(0)}\\
=~&\partial_v\left[\sum_{m=0}^{k-1}(-1)^m~\tilde T_{(-k+m)} \partial_v^{(k +n+1-m)}T_{(0)}\right] + (-1)^k ~\tilde T_{(0)} ~\partial_v^{(n+2)}T_{(0)}\\
=~&\partial_v\sum_{m=0}^{k-1}(-1)^m\Bigg(\partial_v^{n+1}\left[\sum_{p=0}^{k-m-1}(-1)^p~\left(^{n+p}C_{n}\right)\tilde T_{(-k+m+p)} \partial_v^{(k -m-p)}T_{(0)}\right]\\
&~~~~~~~~~~~~~~~~~~~~~~ +(-1)^{k-m}\left(^{k-m+n}C_n\right)\tilde T_{(0)}\partial_v^{n+1} T_0\Bigg) + (-1)^k ~\tilde T_{(0)} ~\partial_v^{(n+2)}T_{(0)}\\
=~&\partial_v^{n+2}\bigg(\sum_{m=0}^{k-1}\sum_{p=0}^{k-m-1}(-1)^{m+p}\left(^{n+p}C_{n}\right)\left[\tilde T_{(-k+m+p)} \partial_v^{(k -m-p)}T_{(0)}\right]\bigg)\\
&+(-1)^k\left[1 +\sum_{m=0}^{k-1}\left(^{k-m+n}C_n\right)\right]\tilde T_{(0)}\partial_v^{n+2} T_0\\
=~&\partial_v^{n+2}\bigg(\sum_{m=0}^{k-1}\sum_{p=0}^{k-m-1}(-1)^{m+p}\left(^{n+p}C_{n}\right)\left[\tilde T_{(-k+m+p)} \partial_v^{(k -m-p)}T_{(0)}\right]\bigg)\\
&+(-1)^k\left[\sum_{m=0}^{k}\left(^{k-m+n}C_n\right)\right]\tilde T_{(0)}\partial_v^{n+2} T_0
\end{split}
\end{equation}
The final result of eq.\eqref{sq4} is almost of the desired form in terms of the structure of $\partial_v$ derivatives, but now we have to transform the coefficients to the desired form. For this purpose we use the following identity which is proved at the end of this section:
\begin{equation}\label{id}
\sum_{m=0}^{k}\left(^{k-m+n}C_n\right)= {^{k+n+1}C_{n+1}}
\end{equation}
Substituting in eq.\eqref{sq4} we find
\begin{equation}\label{sq5}
\begin{split}
&t^{(k)}_{(n+2)} = \tilde T_{(-k)} \partial_v^{(k+n +2)}T_{(0)}\\
=~&\partial_v^{n+2}\bigg(\sum_{m=0}^{k-1}\sum_{p=0}^{k-m-1}(-1)^{m+p}\left(^{n+p}C_{n}\right)\left[\tilde T_{(-k+m+p)} \partial_v^{(k -m-p)}T_{(0)}\right]\bigg)\\
&+(-1)^k\left[\sum_{m=0}^{k}\left(^{k-m+n}C_n\right)\right]\left(\tilde T_{(0)}\partial_v^{n+2} T_0\right)\\
=~&\partial_v^{n+2}\bigg(\sum_{m=0}^{k-1}\sum_{p=0}^{k-m-1}(-1)^{m+p}\left(^{n+p}C_{n}\right)\left[\tilde T_{(-k+m+p)} \partial_v^{(k -m-p)}T_{(0)}\right]\bigg)\\
&+(-1)^k\left[{^{k+n+1}C_{n+1}}\right]\tilde T_{(0)}\partial_v^{n+2} T_0
\end{split}
\end{equation}
So the second term in eq.\eqref{sq4} has been cast into the desired form. We will further process the double sum in the first term. Note that the sum has the following form
$\sim \sum_{m=0}^{k-1}\sum_{p=0}^{k-m-1} g(p)~ f(m+p)$,\\
where $g(p) \equiv\left(^{n+p}C_{n}\right)$ and $f(m+p) \equiv (-1)^{m+p}\left[\tilde T_{(-k+m+p)} ~\partial_v^{(k -m-p)}T_{(0)}\right]$. Thus, we have
\begin{equation}\label{id3}
\begin{split}
&\sum_{m=0}^{k-1}\sum_{p=0}^{k-m-1} g(p)~f(m+p)\\
=~&\sum_{m=0}^{k-1}\sum_{q=m}^{k-1}g(q-m)~f(q)~~~\text{$m$ has been changed to $q = (m+p)$}\\
=~&\sum_{q=0}^{k-1}\left[\sum_{m=0}^q~g(q-m)\right] f(q)
=\sum_{q=0}^{k-1}\left[\sum_{m=0}^q~\left(^{n+q-m}C_{n}\right)\right] f(q)\\
=~&\sum_{q=0}^{k-1}{^{n+q+1}C_{n+1} }~~f(q)
\end{split}
\end{equation}
Substituting eq.\eqref{id3} in eq.\eqref{sq5} we get the required result eq.\eqref{sq0} that we set out to prove
\begin{equation}\label{sqfinal}
\begin{split}
t^{(k)}_{(n+2)} = ~&\partial_v^{n+2}\bigg(\sum_{m=0}^{k-1}(-1)^{m}\left(^{n+1+m}C_{n+1}\right)\left[\tilde T_{(-k+m)} \partial_v^{(k -m)}T_{(0)}\right]\bigg)\\
&+(-1)^k\left[{^{k+n+1}C_{n+1}}\right]\left(\tilde T_{(0)}\partial_v^{n+2} T_0\right)
\end{split}
\end{equation}

\subsection{Proof of the identity eq.(\ref{id})}
Consider the finite sum 
\begin{equation}\label{id1}
\begin{split}
S(x) =~& \sum_{m=0}^k(1+x)^{k+n-m} = \sum_{m=0}^{k} \sum_{p=0}^{k+n-m}\left(^{k-m+n}C_p\right)x^p\\
 =~& (1+x)^n \sum_{m=0}^k(1+x)^{k-m}
  = (1+x)^n \left[(1+x)^{k+1} - 1\over x\right]\\
  =~& \left[(1+x)^{k+n+1} - (1 +x)^n\over x\right]
\end{split}
\end{equation}
Equating the coefficient of $x^n$ on both sides of eq.\eqref{id1}, we get
\begin{equation}\label{id2}
\begin{split}
\text{Coefficient of $x^n$ in $S(x)$}
=~&\text{Coefficient of $x^{n+1}$ in $\left[(1+x)^{k+n+1} - (1 +x)^n\right]$}\\
\Rightarrow~\sum_{m=0}^{k}\left(^{k-m+n}C_n\right)
=&~~ {^{k+n+1}C_{n+1}}
\end{split}
\end{equation}
Hence proved.


\section{Variation of a generic covariant tensor under diffeomorphism: a justification for eq.(\ref{diffeo_covT_rel})}\label{app:diffeo_covT}

Suppose we have a Diffeomorphism Covariant tensor $S_{\alpha_1 \dots \alpha_k}$ which is a functional of various dynamical fields $\{\psi\}$ (such as $g_{\mu\nu},R_{\mu\nu\rho\sigma},D_{\alpha_1} R_{\mu\nu\rho\sigma}, \dots, \phi, D_{\alpha_1} \phi, \dots $ ), the notion of Diffeomorphism covariance under the vector field $\xi^{\mu}\partial_{\mu}$ is expressed by the following equation:
\begin{equation}\label{diffeoproof}
    \delta S_{\alpha_1 \dots \alpha_k} [ \delta \psi \rightarrow \mathcal{L}_{\xi} \psi ] = \mathcal{L}_{\xi} S_{\alpha_1 \dots \alpha_k}
\end{equation}
where $\xi^{\mu}\partial_{\mu} = v \partial_v - r \partial_r$ which is a killing vector of our background Space-time eq.\eqref{metequl}. The reason for this specific choice will be clear in a moment. It should be emphasized that in the LHS of eq.(\eqref{diffeoproof}), the variation is induced by the variation of the dynamical fields. 
More precisely,
\begin{equation}
    \delta S_{\alpha_1 \dots \alpha_k} [ \delta \psi \rightarrow \mathcal{L}_{\xi} \psi ] = \lim_{\epsilon \to 0} \dfrac{S_{\alpha_1 \dots \alpha_k}[\psi + \epsilon \mathcal{L}_{\xi}\psi] - S_{\alpha_1 \dots \alpha_1}[\psi] }{\epsilon} 
\end{equation}
Whereas, in the RHS of eq.(\eqref{diffeoproof}), the variation is induced by a space-time Lie derivative or, in other words, the variation induced by a change of co-ordinate charts along the flow of the integral curves of $\xi^{\mu}\partial_{\mu}$ \citep{waldbook}.
Thus, the variation in the LHS affects only the dynamical fields, whereas the variation in the RHS affects both the dynamical and background fields if any \citep{Harlow:2019yfa}. For the equality to hold, we must make sure that the variations of all the background fields are set to zero.  Hence, the choice of $\xi^{\mu}\partial_{\mu}$ is clear: It is a symmetry of the only background field present in the problem eq.\eqref{metequl} (i.e.) $\mathcal{L}_{\xi} g^{eq}_{\mu\nu} = 0$ and hence the equality of eq.\eqref{diffeoproof} holds.
We provide a few examples which explicitly demonstrate this.

When one applies this notion of Diffeomorphism covariance to the Lagrangian of a theory, we get
\begin{equation} \label{diffcovlag} 
    \dfrac{\partial L}{\partial \psi} \mathcal{L}_{\xi} \psi = \mathcal{L}_{\xi} L [\psi] \, ,
\end{equation}
where the LHS is the usual variation of the Lagrangian with respect to the fields with $\delta \psi = \mathcal{L}_{\xi} \psi$. 
\paragraph{A scalar field coupled to non-dynamical gravity:} To illustrate this Diffeomorphism covariance, let us consider a simple scalar field Lagrangian:
\begin{equation}
    L = g^{\mu\nu} \partial_{\mu} \varphi \partial_{\nu} \varphi
\end{equation}
The Lie derivative is given by
\begin{equation}
    \mathcal{L}_{\xi} L = \xi^{\mu} \partial_{\mu} L
\end{equation}
The variation of the Lagrangian is however given by
\begin{equation}
        \delta L = \delta g^{\mu\nu} \partial_{\mu} \varphi \partial_{\nu} \varphi + 2 g^{\mu\nu} \partial_{\mu} \delta \varphi \partial_{\nu} \varphi 
\end{equation}
Substituting $\delta g_{\mu\nu} = \mathcal{L}_{\xi} g_{\mu\nu} = D_{\mu}\xi_{\nu}+D_{\nu}\xi_{\mu}$ and $\delta \varphi = \mathcal{L}_{\xi} \varphi = \xi^{\rho}\partial_{\rho} \varphi$, we get
\begin{equation}
    \begin{split}
        \delta L [ \delta \psi \rightarrow \mathcal{L}_{\xi} \psi ] &= - (D^{\mu}\xi^{\nu}+D^{\nu}\xi^{\mu}) \partial_{\mu} \varphi \partial_{\nu} \varphi + 2 g^{\mu\nu} \partial_{\mu} ( \xi^{\rho} \partial_{\rho} \varphi ) \partial_{\nu} \varphi \\
        & = - (\partial^{\mu}\xi^{\nu}+\partial^{\nu}\xi^{\mu} + g^{\mu\rho} \Gamma^{\nu}_{\rho\sigma} \xi^{\sigma} + g^{\nu\rho} \Gamma^{\mu}_{\rho\sigma} \xi^{\sigma} ) \partial_{\mu} \varphi \partial_{\nu} \varphi \\
        &\, \, \, \, \, \, + 2 g^{\mu\nu} \partial_{\mu} \xi^{\rho} \partial_{\rho} \varphi \partial_{\nu} \varphi + 2 g^{\mu\nu} \xi^{\rho} \partial_{\rho} (\partial_{\mu} \varphi) \partial_{\nu} \varphi \\
        & = g^{\mu\nu} \xi^{\rho} D_{\rho} ( \partial_{\mu} \varphi \partial_{\nu} \varphi ) \\
        & = \xi^{\rho} \partial_{\rho} L = \mathcal{L}_{\xi} L
    \end{split}
\end{equation}
where the end result is what you get by computing the Lie derivative directly. 

\paragraph{Einstein's theory of gravity:} We now consider a general diffeomorphism invariant theory. Assuming eq.\eqref{diffcovlag}, it has been shown \citep{Iyer:1994ys} that such a Lagrangian can be re-expressed as
\begin{equation}
    L = L(g_{\mu\nu},R_{\mu\nu\rho\sigma},D_{\alpha_1}R_{\mu\nu\rho\sigma}, \dots , D_{(\alpha_1} \dots D_{\alpha_m)} R_{\mu\nu\rho\sigma}, \phi, D_{\alpha_1} \phi , \dots , D_{(\alpha_1} \dots D_{\alpha_m)} \phi)
\end{equation}
where $\phi$ collectively denote the set of matter fields in the theory. Hence, with this choice of Lagrangian, the variation of the metric $g_{\mu\nu}$ and $R_{\mu\nu\rho\sigma}$ are treated independently. 

To again illustrate this diffeomorphism covariance for a more complicated theory, consider General relativity with $L = R$:
\begin{equation}
    \begin{split}
        \delta L &= 2 (\delta g^{\alpha\gamma} ) g^{\beta\delta} R_{\alpha\beta\gamma\delta} + g^{\alpha\gamma}g^{\beta\delta} \delta  R_{\alpha\beta\gamma\delta} \\
        \implies \delta L [ \delta g_{\mu\nu} \rightarrow \mathcal{L}_{\xi} g_{\mu\nu}] &= 2 (\mathcal{L}_{\xi} g^{\alpha\gamma} ) g^{\beta\delta} R_{\alpha\beta\gamma\delta} + g^{\alpha\gamma} g^{\beta\delta} \mathcal{L}_{\xi} R_{\alpha\beta\gamma\delta} \\
        &= \mathcal{L}_{\xi} (g^{\alpha\gamma}g^{\beta\delta}R_{\alpha\beta\gamma\delta}) = \mathcal{L}_{\xi} R
    \end{split}
\end{equation}
This Diffeomorphism covariance eq.\eqref{diffeoproof} has been used as an important definition \citep{Iyer:1994ys} to prove the first law of black hole thermodynamics for any diffeomorphism invariant theory.


\bibliographystyle{JHEP}
\bibliography{EntropyCurrent}
\end{document}